\documentclass[aps,showpacs,pre,twocolumn]{revtex4}
\usepackage{epsfig}
\usepackage{graphicx}
\usepackage{dcolumn}
\usepackage{bm}
\usepackage{amssymb}
\usepackage{amsmath}
\usepackage{amsthm}

\newcommand{\grad}{ {\mbox {\boldmath $\nabla$}} }
\begin{document}
\title{Imbibition in Disordered Media}
\author{Mikko Alava}
\affiliation{
Helsinki University of Techn., Lab. of Physics, 
P.O.Box 1100, 02015 HUT, Finland}
\affiliation{SMC-INFM, Dipartimento di Fisica,
Universit\`a ``La Sapienza'', P.le A. Moro 2
00185 Roma, Italy}
\author{Martin Dub\'e}
\affiliation{
CIPP, Universit\a'e du Qu\a'ebec \a`a Trois-Rivi\a`eres, 
C.P. 500, Trois-Rivi\`eres, Qu\a'ebec, G9A 5H7 Canada}
\author{Martin Rost}
\affiliation{
Abteilung Theoretische Biologie, IZMB,
Universit\"at Bonn, Kirschallee 1, 53115 Bonn, Germany}


\begin{abstract}
The physics of liquids in porous media gives rise to many
interesting phenomena, including imbibition where a viscous
fluid displaces a less viscous one. Here we discuss the
theoretical and experimental progress made in recent years
in this field. The emphasis is on an interfacial
description, akin to the focus of a statistical physics
approach. Coarse-grained equations of motion have been
recently presented in the literature. These contain terms
that take into account the pertinent features of imbibition:
non-locality and the quenched noise that arises from the
random environment, fluctuations of the fluid flow
and capillary forces.
The theoretical progress
has highlighted the presence of intrinsic length-scales
that invalidate scale invariance often assumed to be present
in kinetic roughening processes such as that of a two-phase
boundary in liquid penetration. Another important fact
is that the macroscopic fluid flow, the kinetic roughening
properties, and the effective noise in the problem are
all coupled. Many possible deviations from simple scaling behaviour exist,
and we outline the experimental evidence. Finally, prospects
for further work, both theoretical and experimental, are discussed.
\end{abstract}
\maketitle

\section{Introduction}
\begin{figure}[h]
\includegraphics[width=7cm]{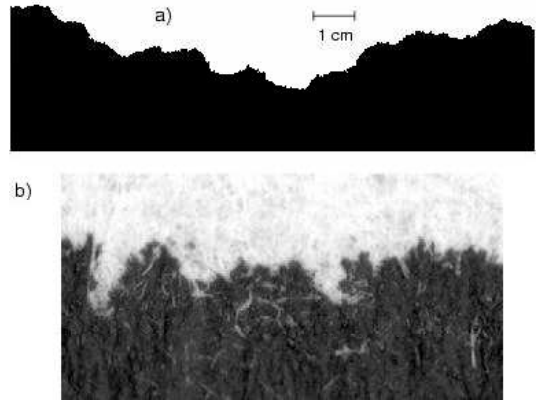}
\caption{Front of black ink sucked into a paper towel. a) digital photograph,
ca.\ 1200 pixels horizontal resolution, dark and light grey values were
enhanced to black and white. b) high resolution scan (1000 dots per cm) of a
small part (ca.\ 0.8 cm wide) in grey-scale. The structure of the medium, 
as the fibres on the paper top surface, and its effect on the fluid front 
becomes visible.}
\label{FigInkFront}
\end{figure}

It is easy to do qualitative observations on the physics of fluid
penetration in inhomogeneous media: a drop of coffee on a napkin or a
sugar cube held partly in the same coffee cup are enough to
demonstrate two fundamental facts. A moving interface is formed between
the wet and non-wet regions of the medium. It becomes apparent that
the disordered pore structure and uneven surface of the paper or the
structure of the sugar cube both influence its behaviour: the interface
is clearly rough. Furthermore, the dynamics of the phenomenon slows
down with time, meaning that the wetted area of the napkin or volume
of coffee absorbed by the cube increases more and more slowly. In fact,
the average position of the interface $H$ of the wet front usually
increases in time as $H(t) \sim t^{1/2}$. The coffee drop shows an
example of spontaneous imbibition, and it obeys what is known as
Washburn's law \cite{Washburn_1921}. The force driving the liquid from
the liquid reservoir to the front between the wet region and the air
in the medium has a weaker and weaker effect on the total flow as the
distance between these two gets larger. Such a naive first glance at
an imbibition experiment can be seen in Figure~\ref{FigInkFront}.

There are many similar examples of situations in which a liquid invades a
porous medium and pushes aside another viscous liquid or gas. They are often
of importance for technological applications or as ingredients in another
field than the physics of fluids. In Table \ref{imbiexamples} we list some
scenarios in which imbibition plays a role --- they range from oil recovery
(using water to displace it out of rock) to biology (water in living
organisms) and manufacturing processes. The flow of liquids through porous
media thus forms a very vast field which combines the porous structure of
the medium with the surface chemistry and physics of the liquids and/or
gases involved and is characterised by several parameters such as the
viscosity contrast of the fluids, their wettability and surface tension
as well as the displacement rate. The simplest realisation of imbibition
involves two immiscible fluids, one being displaced by the other, both
liquids being characterised by their viscosity $\eta$ and the surface
tension $\sigma$ of their interface.

\begin{figure}[h]
\includegraphics[width=8cm]{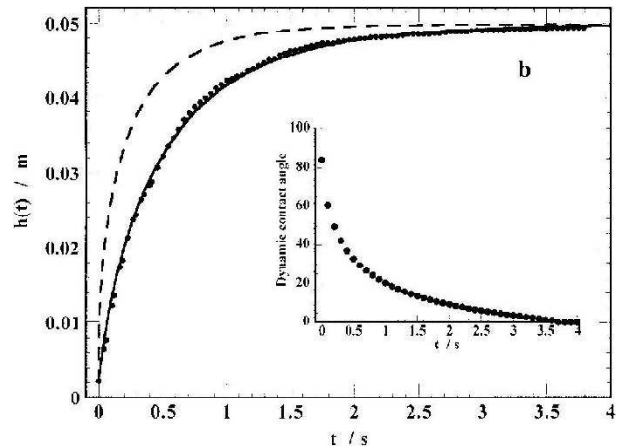}
\caption{Liquid drop in contact with a solid surface. The contact between
the surface and liquids 1 and 2 is characterised by surface tensions 
$\sigma_{s1}$ and $\sigma_{s2}$ respectively while the interface between
the two fluids has surface tension $\sigma$. If the contact angle 
$\theta < \pi/2$, fluid 1 wets the solid.}
\label{drop_solid}
\end{figure}

The solid matrix interacts with the fluids through their wetting
properties, as described by the Young-Dupr\a'e equation for a drop
a liquid in contact with a solid surface, shown schematically on
Fig.~\ref{drop_solid}: 
\begin{equation}
\sigma_{s2} - \sigma_{s1} = \sigma \cos \theta
\label{y_dupre}
\end{equation}
where $\sigma_{s1}$ and $\sigma_{s2}$ are the surface tensions of the solid
with fluids 1 and 2 respectively. The contact angle $\theta$ determines
whether liquid 1 is wetting ($\theta < \pi/2$) or non-wetting
($\theta > \pi/2$).

Imbibition means that a wetting fluid displaces the non-wetting one,
while the opposite case is called drainage. Spontaneous imbibition takes
place when the invading fluid does so under the sole influence of 
capillary forces, with no external pressure. Forced imbibition involves
a combination of capillary phenomena and an externally enforced
flow rate or pressure difference.

\begin{table}[htb]
\caption{Practical and experimental realisations of
imbibition}
\label{imbiexamples}
\begin{tabular}{p{3cm}p{5cm}}
Oil recovery & Displacement of a liquid by another,
possibly in presence of a third phase
\cite{Sorbie_1995,Cil_1996,Morrow_2001,Zhou_2002,Akin_2000,Dijke_2003b}
\\
\\
Printing processes & Ink penetration in paper \cite{Rosinski_1993};
Coating of paper \cite{Poulin_1997,Ridgway_2002b};
Absorbing materials \cite{Schuchardt_1991}
\\
\\
Food industry & Cooking \cite{Pinthus_1994}; Wine filtering \cite{Vernhet_1997}
\\
\\
Biological sciences & Fluid transport in plants or imbibition of
water into seeds (see Section \ref{biosection}); Water penetration into
soils \cite{Blunt_2002}; 
Medical applications \cite{Luner_2001}
\\
\\
Surface chemistry & Contact
angle measurements 
\cite{Chibowski_1993,Chibowski_1997,Marmur_1997,Labajos-Broncano_2002}; 
Droplets on surfaces \cite{Clarke_2002}
\\
\\
Composite materials & Invasion of voids
by a resin or a metal in filer or metal-metal composites.
\cite{Dopler_2000,Antonelli_2001,Michaud_2001,Michaud_2001b,Ambrosi_2000,Endruweit_2002,Michaud_2001c}
\\
\\
Textiles & Behaviour of garments in the presence of liquids 
\cite{Rajagopalan_2001,Kissa_1996,Hsieh_1992,Hsieh_1992b,Chen_2001a}
\\
\\
Construction & Water penetration into concrete or cement
pastes \cite{Leventis_2000,Ceballos-Ruano_2002}
\\
\end{tabular}
\end{table}

Although empirical relations for the flow of liquids through a porous
medium existed for a long time, an effort, in part inspired by statistical 
mechanics, to quantitatively understand
and predict the flow led to the study of 
{\em pattern formation} 
or the geometry of the regions occupied by the invading/receding 
fluids. At this level, the physics is
dependent on the combination of viscous and capillary forces
at the boundary between invaded and ``dry'' regions. The
viscous part arises due to the fluid flow in the (partly)
saturated pore space, and the relative importance of viscous 
and capillary
effects is described by the {\em capillary number} : 
\begin{equation}
C_{\rm a} = \frac{\eta \; v }{\sigma} 
\label{capillary_number_intro}
\end{equation}
where $\eta$ is the viscosity of the fluid, $v$ its average velocity and
$\sigma$ the interfacial surface tension.

\begin{figure}
\includegraphics[width=8cm]{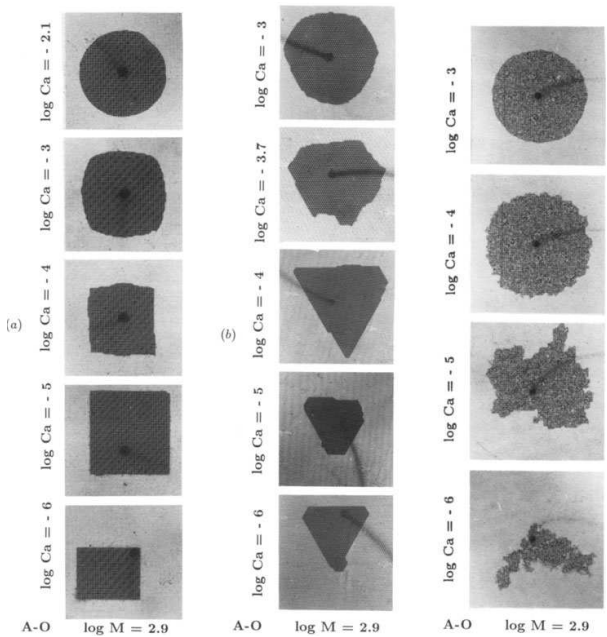}
\caption{Various cases of a fluid (oil) displacing air in a network,
with varying pore size distributions. From left to right: a square
network geometry, a triangular one, and finally a square one with very
wide pore sizes. As the Capillary number $C_a$ is varied, the effective
surface tension of the cluster of invaded pores changes. Note the
results for $\log C_{\rm a} \! = \! -6$ in particular (after Lenormand,
\cite{Lenormand_1990}).}
\label{lenormand5}
\end{figure}

The stability and existence of an interface was experimentally investigated
through the injection of liquid into various specially designed porous
networks by Lenormand \cite{Lenormand_1990} (see also 
\cite{Lenormand_1984,Wardlaw_1988,Lenormand_1988}) who
qualitatively summarised it as a phase diagram with three possible outcomes,
illustrated on Fig \ref{lenormand5}:
(i) discontinuous formation of wetted domains
due to {\em surface flow} in pores, (ii) formation of non-compact
branched structures due to weak surface tension, and finally
(iii) compact domains with well defined interfaces. 

A phase diagram in terms of both the capillary number and the viscosity ratio
of the two liquids (or liquid/gas) involved can then be established,
illustrating the competition of viscous and capillary forces.
The former stabilises the interface while the latter, if dominating,
leads to situations like (ii) above.

If imbibition dynamics are dominated by capillary forces and pore-level
invasion mechanisms like film flow, the effective surface tension of
an interface can vanish. Then percolation-like phenomena can ensue, which
means that the medium is apparently penetrated in disjoint clusters
of the imbibing liquid. On the pore scale, one thus has to deal with
either piston-like displacement or such gradual processes, leading finally
to ``snap-off'', as narrow pore throats become completely filled by the
invading liquid \cite{Roof_1970,Mohanty_1987}.

\begin{figure}
\includegraphics[width=8cm]{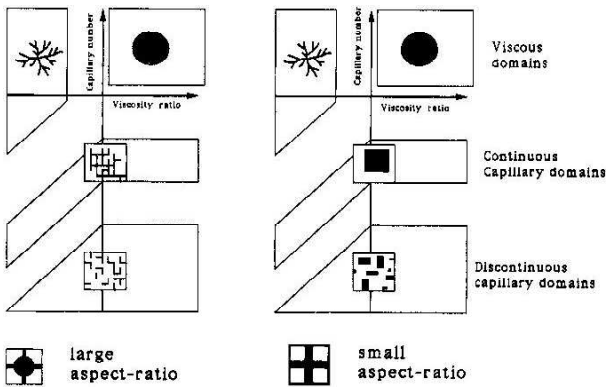}
\caption{A phase diagram for the geometry of imbibition as proposed by
Lenormand \cite{Lenormand_1990}, with the capillary number $C_{\rm a}$
and the viscosity $\nu$ being varied on one hand, and the pore-level
geometry on the other hand. Notice the possibilities of apparently
non-local (discontinuous) invasion, and of vanishing surface tension.}
\label{lenormand10}
\end{figure}

If the imbibing fluid is sufficiently more viscous than the defending one
the invasion front is compact. Typically, as in the example of the sugar cube
or the paper napkin, it is not flat but presents a random rough structure.
This additionally complicates any effective theory or equation of motion for
the invading fluid.

The main issues to be discussed in this review article are related to the
rough {\em interface} between the two phases involved in a compact
imbibition invasion. This is accomplished by {\em coarse-graining} from
the level of individual pores to the average position of the interface,
meaning that the micro-structure of the medium is averaged out when
possible. This is particularly complex in the case of spontaneous imbibition
and can fail, due to the slowing down of the fluid penetration and/or due
to the presence of competing mechanisms in the pore scale invasion dynamics.

This focus is also of fundamental interest to any imbibition process since
a correct interfacial description requires the knowledge of the 
time- and length-scales that control the entire process. For many
practical applications like soil mechanics or oil recovery 
it is of importance to understand this kind of ``upscaling'' 
\cite{Blunt_1990,Morrow_2001,Vogel_2003} in order to estimate 
remaining non-wetting fluid saturations or relative
permeabilities from laboratory
scale measurements or from microscopic simulations. 
Future advances can also be expected in more complicated scenarios,
as in the case of non-Newtonian fluids, in the presence of chemical
reactions between the fluids involved, as well as in the rapidly
developing field of microfluidics.

The process of imbibition is affected by {\em noise}, part of which
stems from ever-present thermal fluctuations, while the rest is due to
the quenched, frozen-in structure of the porous medium. The 
quenched nature of the noise becomes particularly 
important if the phase interface moves slowly
in avalanche-like behaviour. Then the dynamics consists of localised
bursts whose description directly couples the noise and
the dynamical fluctuations of the interface. 

An interesting theoretical question is then the universality of these
phenomena, i.e., whether the statistical description of the interface is 
similar to the roughness observed in fire fronts, cracks and rupture 
lines, domain walls in ferromagnets. The same question
arise with respect to the description of burst and avalanches in 
connection to the concept of Self-Organised Criticality. 
These questions are far from academic. In one part, the description
of roughening in non-equilibrium phenomena has received lots of 
attention from the theoretical side (witness the large number of publications
devoted to the Kardar-Parisi-Zhang (KPZ) equation \cite{Kardar_1986}), 
but it is only recently that
experimentalists were able to convincingly demonstrate the link between
the proposed theoretical models and the observed phenomena (see
\cite{Merikoski_2003} and references therein for KPZ behaviour in fire fronts). 
The same is true for roughening in imbibition.

Although it was examined in the early 1990's \cite{Buldyrev_1992a,Amaral_1994}
in connection with percolation theory and deviations from KPZ behaviour,
recent work \cite{Ganesan_1998,Dube_1999} pointing to the importance of
fluid conservation, has led to a flurry of new experimental results 
\cite{Hernandez_Machado_2001,Soriano_2002a,Soriano_2002b,Soriano_2002c,
Geromichalos_2002}
supported by further theoretical work \cite{Lam_2000,Paune_2002}. 
Even though many details remain obscure, we feel that the general theoretical
picture of {\em roughening in imbibition} is now well established. The goal of
this review is thus to highlight, from a statistical physics point of view,
the central aspects of imbibition that are understood, to relate them to
existing experimental results and to point to remaining gray zones that
deserve further studies. At the same time, we want to contrast this with
the array of experimental evidence and numerical simulation models used
for applied purposes and in connection to multi-phase flows. New
theoretical results on the dynamics of avalanches and on interface
roughening at constant flow and for columnar disorder are also presented
and related to the recent experiments
\cite{Hernandez_Machado_2001,Soriano_2002a,Soriano_2002b,Soriano_2002c}
in Sections \ref{spon_imbi}, \ref{constantflow} and \ref{imbi_columnar}.

The experimental situation is reviewed after that, with a brief
remainder of possible complicating effects and comments on non-Washburn-like
scaling in various setups. Next follows an account of various
experiments on front roughening, ranging from Hele-Shaw cells
to paper imbibition, divided between the forced fluid flow and
the spontaneous cases. In contrast to the recently focussed
theoretical progress, the experimental picture is much more
delicate and we discuss it in the light of the theoretical sections.
It turns out so that there are many confusing results, some of
which can be related to the theoretical ones, and some of which
clearly call for further work.

\begin{figure}[h]
\includegraphics[width=8cm]{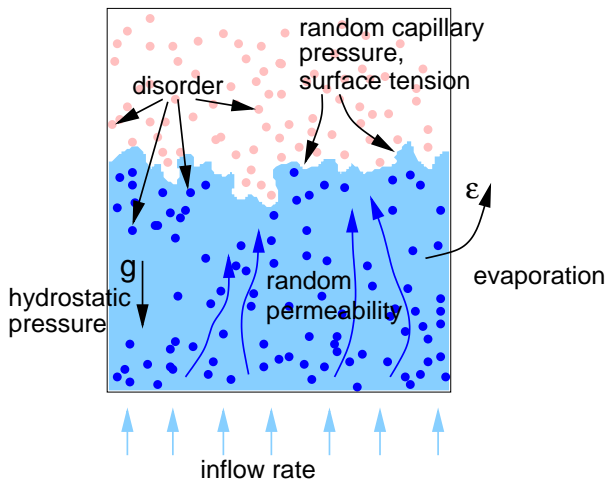}
\caption{Schematic representation of elementary processes
involved in quasi-two-dimensional imbibition.}
\label{imbielements}
\end{figure}

It is now clear that the physics of imbibition in disordered media
couples {\em randomness, kinetic roughening properties}, and
{\em interface velocity} in an complex fashion. This is illustrated
schematically in Figure~\ref{imbielements}.
The randomness acts as an effective noise that roughens the interface,
but the same randomness also influences the fluid flow in the bulk,
and thus the interface velocity and the net volume flow in the imbibition
process. Therefore, we must conclude that all these aspects are intimately
coupled and cannot be discussed separately. This creates intriguing
experimental and theoretical questions and brings engineering and
technological interests, surprisingly, close to those of fundamental
statistical mechanics.

At the basis of models and experiments dealing with imbibition interface
roughening lies the understanding of bulk fluid transport and saturation
in the invaded medium. In the section to follow next, an overview is given
from basic hydrodynamics in the medium to special effects related to
non-Newtonian fluids.

\section{Physics of Imbibition}

\subsection{Fluid flow in porous media}
\label{LucasWashburn}

The detailed description of liquid flow through a porous medium is
greatly complicated by the many time and length scales that are
involved. The flow takes place at the level of the pores (say
$1 \, \mu\mbox{m}$ or less) but the quantity of interest is the flow
averaged over the whole macroscopic sample.

Similarly, a pore may be invaded vary rapidly or may remain blocked
for the whole invasion time, which may be several hours or even days
in spontaneous imbibition. A description of the flow makes sense
only over a ``representative volume element'', loosely defined as the
minimal volume that can be defined such that the properties of the
flow (and of the porous medium itself) within it remain statistically
similar no matter where it is placed in the bulk of the medium
(see \cite{Zhang_2000,Melean_2003a} for some recent advances using
tomographic techniques). 

At this level, it was already empirically found long ago by 
Darcy that the flow is essentially proportional to the pressure gradient
across the medium, a result which can intuitively be explained by 
considering the flow of a liquid through a capillary tube.
A consequence of this result is that the height $H$ of a fluid column 
invading spontaneously a porous medium from a reservoir increases in time as 
$H(t) \sim t^{1/2}$, due to a combination of fluid conservation law (since
any amount of invading fluid must be transported from the reservoir) and 
capillary forces. Before going further into the concepts of interface
roughening, it is interesting to see how this result arises, and under
which conditions it can be expected to be valid. 

\subsubsection{The Lucas-Washburn description}
\begin{figure}[h]
\includegraphics[width=7cm]{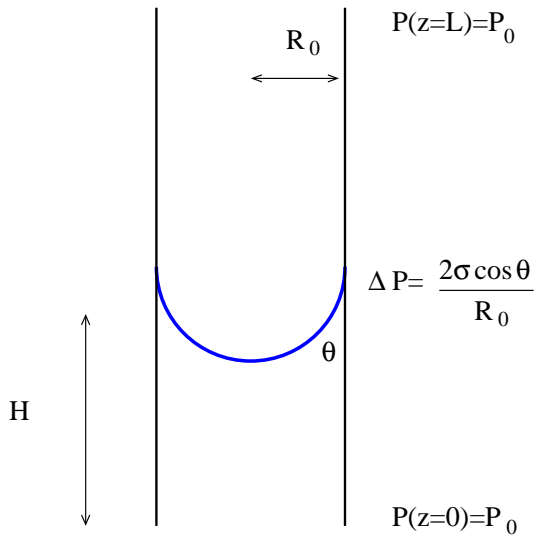}
\caption{Illustration of capillary rise, the fluid at $z=0$ is in
contact with a reservoir at atmospheric pressure $P_0$ and the top
of the tube is either closed or open. The fluid from the reservoir wets
the walls of the tube, which means that a curved interface, to which
corresponds a capillary pressure difference $p_c$ exists across the
interface.}
\label{cap_rise_fig}
\end{figure}

The simplest way to illustrate imbibition is the capillary rise. It
translates easily to the basic phenomenological description of flow in
porous media and it represents an important microscopic mode of flow
propagation. As illustrated in Fig.~\ref{cap_rise_fig}, we consider a
capillary tube of length $L$ and radius $R_0$, immersed into a reservoir
at ambient atmospheric pressure $P_0$. 

The two fluids are immiscible with an interface, of surface tension 
$\sigma$, located at a height $z \! = \! H$ above the reservoir and we 
assume that the fluid in the region $z \! < \! H$ wets the solid walls. A 
meniscus is formed at the interface, characterised at equilibrium by the
contact angle $\theta$ obtained from Eq.~(\ref{y_dupre}). The combination
of the curved interface with the surface tension creates a pressure
difference or capillary pressure $p_c \equiv \Delta P = 2 \sigma
\cos \theta /R_0$.  This Gibbs-Thomson or Laplace effect gives rise to
a thermodynamical force that will move the fluids. It also applies in more
general imbibition problems directly at the level of individual pores.

In its full generality the solution of the flow dynamics for this problem
needs to take into account the detailed flow right at the contact line
between the fluids and the solid  \cite{deGennes_1985,Joanny_1992} as
well as inertial effects at the entrance of the tube
\cite{Szekely_1971}. For now, these effects are neglected, but are
discussed in greater detail below. Neglecting the structure of the
meniscus means that the pressure field changes only along the length
of the tube, $P = P(z)$, with the interface described mathematically by
the height $H$. Likewise, neglecting inertia means that the fluids are
only described in terms of their viscosity $\eta$ 
using Stokes' Equation \cite{Landau_1986}. 
\begin{equation}
\eta \nabla^2 {\bf v} = \grad P 
\end{equation}
Three different cases can be then be easily examined: 

(i){\it Two incompressible Fluids:} The two fluids are described by
Stokes' equation, with a longitudinal velocity field depending only on the 
radial coordinate ${\bf v} = {\hat {\bf z}} \; v_z (r)$:
\begin{equation}
\eta_i \frac{1}{r} \frac{d}{dr} \left( r \; \frac{d v_{zi} (r)}{dr} 
\right) = \frac{d P (z)}{dz}
\end{equation}
where $i \! = \! w$ or $n$ denotes respectively the wetting ($z \! < \! H$)
and non-wetting ($z \! > \! H$) fluids. The tube is open at the top, to
allow the second fluid to leak out. The incompressibility condition leads to a  
Laplace equation $\nabla^2 P = 0$  for the pressure field, where the
instantaneous position of the interface appears only as a boundary condition
\begin{eqnarray}
P(z=0)=P_0 \nonumber \\
P(H^+)-P(H^-) = p_c \\
P(z=L) = P_0. \nonumber 
\end{eqnarray}
The Laplace equation for the pressure means that the pressure gradient
is constant, so that each fluid has a Poiseuille velocity field
\begin{equation}
v_{zi} = \frac{1}{4\eta_i} \left( r^2 - R_0^2 \right) 
\frac{d P}{dz} 
\end{equation}
and the velocity of the interface is associated with the average
flow velocity 
\begin{equation}
\pi R_0^2 \; \frac{dH}{dt} = \int \! d{\bf a} \, v_{zw} (r) 
 = \int \! d{\bf a} \, v_{zn} (r) \end{equation}
where $d{\bf a}$ is the area element of the tube, which leads to 
a pressure
\begin{eqnarray}
P(z \! < \! H) & = & - \frac{\eta_w}{\eta_w H + \eta_n (L \! - \! H)  } \; p_c
\; z + P_0 \nonumber \\
\\ 
P(z \! > \! H) & = & - \frac{\eta_n}{\eta_w H + \eta_n (L-H)  } p_c (z-L)+ P_0
\nonumber
\end{eqnarray}
and to an interfacial velocity 
\begin{equation}
\frac{dH}{dt} = \frac{R_0^2}{8} \frac{p_c}{\eta_w H + \eta_n (L-H)  }
\label{2inc}
\end{equation}
The wetting fluids thus continuously displaces the non-wetting fluid
until it occupies the whole length of the tube. Notice that at 
the beginning, it
does so as $dH/dt \sim p_c/H$, which implies an initial motion of 
the interface $H(t) \sim t^{1/2}$.

(ii){\it Second fluid is compressible:}
If the second fluid is a very compressible liquid or a gas, it 
simply adjusts itself to the pressure
of the wetting fluid and does not support any pressure gradient.
In the case of an open tube, the second fluid is 
the ambient gas, and its pressure is constantly at the atmospheric 
pressure $P_0$, such that 
\begin{eqnarray}
P(z \! < \! H) = - p_c \frac{z}{H} + P_0 \nonumber \\ 
  \\
P(z \! > \! H) = P_0 \nonumber
\end{eqnarray}
In this case, the classic result of Washburn \cite{Washburn_1921} 
and Rideal \cite{Rideal_1922} is obtained: 
the interface position is described in time as 
\begin{equation}
H^2(t)-H^2(t_0) = \frac{R_0^2}{8} \! \frac{p_c}{\eta_w} (t-t_0)
\label{t-half}
\end{equation}
where $H(t_0)$ is the initial height. 
This result is also obtained for two incompressible fluids 
(Eq.~(\ref{2inc})), as long as
$\eta_w H \gg \eta_n L$.
 
If the tube is closed at the top, so that the gas cannot escape, the 
pressure of the gas phase increases as the fluid occupies more and more
volume. If the gas was initially filling the whole tube with pressure 
$P_0$, then the pressure if the interface is at height $H$, 
\begin{equation}
P(z>H) = P_0 \left( \frac{L}{L-H} \right)^{\gamma}
\end{equation}
where $\gamma = C_p/C_v$, the ratio of the heat capacities at constant 
pressure and volume respectively. The flow field of the wetting fluid
is not modified, and the resulting interface equation 
\begin{equation}
\frac{dH}{dt} = \frac{R_0^2}{8} \frac{1}{\eta_w} 
\frac{P(H)-P_0 - p_c}{H}. 
\label{wash-trap} \end{equation}
Again, as long as $P(H)$ is not too far from the atmospheric pressure, 
the Washburn-Rideal result is obtained. However, in real porous media,
trapping of gas by the invading fluid may occur, in which case the 
propagation of the menisci, given by Eq.~(\ref{wash-trap}) may be 
markedly different from Washburn behaviour Eq.~(\ref{t-half}). 

(iii){\it Capillary rise and gravity:}
gravity acts on the fluids through their density $\rho$ as 
\begin{equation}
\eta_i \nabla^2 v_{zi} = \frac{d P (z)}{dz} - \rho_i \; g
\end{equation}
where $g$ is the gravitational constant. Assuming that the second
fluid is the ambient gas, with an open tube, the interfacial rise is
\begin{equation}
\frac{ d H}{d t} = \frac{\kappa}{\eta} \; \rho g 
\left( \frac{H_{eq}}{H(t)} - 1 \right),
\label{washburn}
\end{equation}
where we define the permeability $\kappa = R_0^2/8$, 
and the  equilibrium height $H_{\rm eq} = p_c/(\rho g)$, at which the 
hydrostatic and capillary pressures are balanced. 
The relevant parameter is the equilibration time $\tau_{\rm eq} \equiv H_{\rm eq}
\eta / \kappa g (\rho )^2$. For times $t \ll \tau_{\rm eq}$ (which corresponds
to $H \ll H_{\rm eq}$) or in absence of gravity (``horizontal imbibition'')
the time-dependence follows the square-root law, Eq.~(\ref{t-half}) 
seen earlier
At later times, $t \gg \tau_{\rm eq}$, the
finite equilibrium height $H_{\rm eq}$ (if
it exists) is approached exponentially
\begin{equation}
H (t) \; \sim \; H_{\rm eq} \; ( 1- e^{-t/\tau_{\rm eq}}),
\label{rise-eq}
\end{equation}
in a way that is simply related to the equilibrium quantities $H_{\rm eq}$
and $\tau_{\rm eq}$.
This simple law is 
interesting in many aspects since the relevant parameters 
(contact angle, surface tension) are easy to determine and the resulting 
easy-to-grasp predictions make it attractive and suitable in 
interpreting experiments. 

However, these results may be invalid for a number of reasons. 
For example, in a liquid-liquid
system Mumley et al.\ \cite{Mumley_1986},  
comparing dry and prewetted tubes, have shown that 
viscous dissipation at the contact line 
and precursor film dynamics 
\cite{deGennes_1985,Joanny_1992} may lead to 
both Washburnian or slower (in terms of the rise
exponent) behaviour depending on the contact angle and the condition of
the capillary rise. 

Inertial effects as Bosanquet-flow \cite{Bosanquet_1923} may also be
important in the early stage of pore invasion, before the dynamics
is described by Eq.~(\ref{t-half}) \cite{Szekely_1971}. Dimensionally, it is
easy to see that the characteristic time scale over which inertial effects
will be important $\tau_i \sim \rho R_0^2/\eta$, after which the usual 
Washburn dynamics follows. Before this time 
the fluid enters the capillary as 
\begin{equation}
H(t) \sim \left( \frac{\sigma}{\rho R_0} \right)^{ \! e/2}  \, t .
\label{inertia}
\end{equation} 
Although the time $\tau_i$ can be very small, the $R_{0}^{-1/2}$ dependence
of the rise on the radius of the capillary must be compared to the  
$R_{0}^{1/2}$ of Eq.~(\ref{washburn}). Although the effect of 
inertia lasts for a very short time in small pores, the capillary rise
can nevertheless be quite rapid. 
Experimentally inertial effects can give rise to effective initial height 
and time, $H_0$ and $t_0$ in Eq.~(\ref{washburn}) that must be 
taken into account in fitting measured data
\cite{Labajos-Broncano_1999,Labajos-Broncano_1999b}. 

\subsubsection{Macroscopic description of flow in porous media}
Equation (\ref{washburn}) is also often used 
in more general terms for porous media. Already at the level of Stokes'
equation, a simple dimensional analysis implies that 
$ {\bf v} \sim (\kappa/\eta) \grad P$ where $\kappa$ must have units 
of $\mbox{(length)}^2$. In fact, a common description of single-phase 
fluid flow in porous media is Darcy's equation 
\cite{Scheidegger_1974,Sahimi_1993},
\begin{equation}
\langle {\bf q} \rangle = - \frac{\kappa}{\eta} 
( {\mbox {\boldmath $\nabla$} } P - \rho \, {\bf g} ),
\label{darcy}
\end{equation}
where $ \langle {\bf q} \rangle$ is the average volume of fluid
transported per unit time per unit cross-section of the porous
medium, and $\kappa$ and $\eta$ are the {\em average} permeability and
viscosity respectively. This equation is obtained by coarse-graining
the structure of the porous medium, ignoring all fluctuation and pore
scale effects, using a line of reasoning based on ``representative
volume elements'' \cite{Scheidegger_1974,Bear_1990}.

The extension to two-phase or multi-phase flow is in principle 
straightforward, the dimensionless saturations of wetting and non-wetting
fluids $s_w$ and $s_n$ are defined as the ratio of fluid present with
respect to void space in a given representative volume element. 
Alternatively, the
concentrations are defined as $c_i = {\cal P} s_i$ where ${\cal P}$ is
the porosity. 
The dynamics of the concentration is then obtained through 
continuity equations to reflect the conservation of liquid. If 
the fluid densities and the porosity are constant, 
\begin{eqnarray} 
{\cal P} \frac{\partial s_w}{\partial t} + \grad \cdot {\bf q}_w = 0 
\nonumber \\
{\cal P} \frac{\partial s_n}{\partial t} + \grad \cdot {\bf q}_n = 0 
 \label{2-phase}
\end{eqnarray}
together with the constraint $s_w + s_n = 1$. The flux ${\bf q}_i$ of
each fluid obeys Eq.~(\ref{darcy}) and a {\it coarse-grained} capillary 
pressure is included as a closure relation between the pressure in the 
two fluids $P_w - P_n = p_c$. Since the permeability is only a geometric
factor, both fluids should have the identical values of $\kappa$ and the 
pressure gradient, ${\mbox {\boldmath $\nabla$} } P \sim {\bf \hat{z}}
\; p_c / H(t)$ then follows immediately.

If the imbibition front remains compact, the volume-flow per unit area
is nothing but the velocity of the interface, which leads to 
$\langle q \rangle = dH /dt$ and to the results described in
Eq.~(\ref{2inc}), and Eq.~(\ref{t-half}) if the second fluid cannot sustain
a pressure gradient. This however assumes that the fluid pushed
away has an easy escape, a situation denoted in the petrophysics
community as ``co-current flow''.

What kind of dynamical behaviour can now be expected in realistic 
porous media? Simplistic
analysis predicted by the above Washburn law, Eq.~(\ref{t-half}), can
be invalidated by many factors. For example, evaporation (to be examined
in more details in Section \ref{phasefield}) may slow down 
imbibition
when the quantity or volume of liquid needed to follow the typical
Washburn scaling is not available. 
Similar effects arise if the solid
phase matrix is permeable and acts as a sink. Likewise, for
three-phase flows, often important in industrial problems, the separation
of the two liquids may cause the introduction of new time-scales. 
In general, 
simply having several length-scales to describe the structure of the 
porous medium 
can lead to complications that obfuscate the Washburn scaling.
In the following, we briefly discuss when this scaling may fail, or 
what could be expected in various kind of porous media. 

Despite all these possible complications Washburn behaviour {\it is}
often observed experimentally. It is therefore an useful approach
to study roughening first under well controlled situations 
that lead to the $t^{1/2}$ progression of the average interface,
and then to look at possible deviations. We will therefore devote the
major part of this Review to Washburnian imbibition. The expected
roughening behaviour when $H(t) \sim t^{\delta}$ with
$\delta \neq 1/2$ is commented on at special occasions, in particular in
the concluding Section.

\subsubsection{Pore geometry and inertia}
At the microscopic level, it is clear that any simple picture of a
porous medium as a collection of capillary tubes with effective
permeability $\kappa$ is wrong, the pore geometry can be extremely
important, in particular when sharp corners or edges are present. 

It is already established that a compact (albeit rough) imbibition
front is formed if the viscosity of the invading fluid is sufficiently
greater than the viscosity of the defending fluid, as indicated on the
phase diagram proposed by Lenormand \cite{Lenormand_1990}, reproduced
in Fig.~\ref{lenormand10}. Thus, although containing interesting
physics at the level of pattern formation, the question of disconnected
invasion clusters is trivially not relevant in the context of interface
roughening (see e.g.\ refs.~\cite{Hirsch_1994,Berkowitz_1998} for
percolation approaches). The main exception to this rule is the
possibility of a ``dynamic'' transition between well-defined interfaces
and fingering instabilities in spontaneous imbibition due to the
slowing down of the interface \cite{Tzimas_1997,Hayashi_2001}. A
second possibility is the role of prewetting layers, whose presence
may again render an interfacial description meaningless
\cite{Lu_1994,Lu_1994b,Lu_1995}, since the essential dynamics are
ruled by the prewetting.

\begin{figure}
\includegraphics[width=8cm]{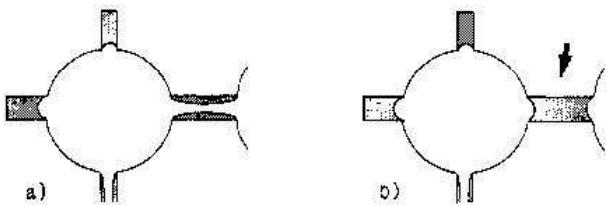}
\caption{In large pores or channels, the fluid can first cover the
surfaces and then collapse to fill the whole void \cite{Lenormand_1990}.}
\label{lenormand7}
\end{figure}

\begin{figure}
\includegraphics[width=8cm]{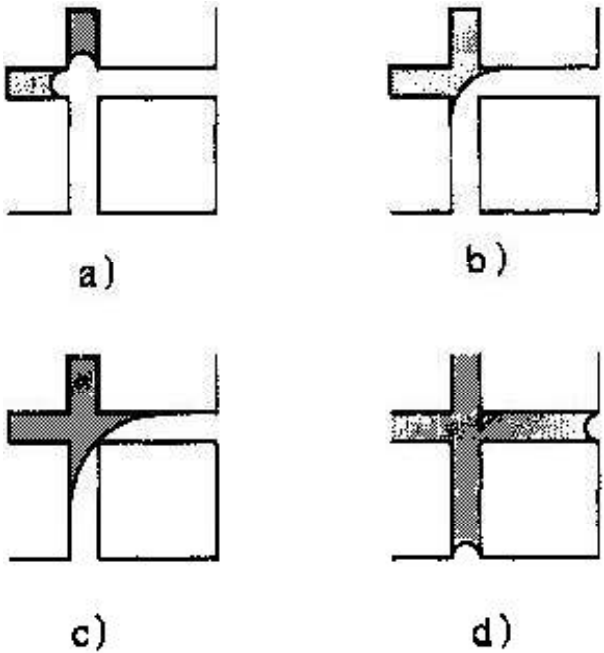}
\caption{Examples of pore invasion mechanisms \cite{Lenormand_1990}:
the progress is gradual from a) to b) as the meniscus radius increases,
and leading, at c), to a situation when the fluid suddenly invades the
channels.}
\label{lenormand6}
\end{figure}

In most cases it cannot be decided a priori whether the expected
macroscopic Washburn-Darcy behaviour is valid or not. The early studies
of Lenormand, later augmented by by the studies of Bernadiner, Knackstedt
et al.\ and others, have managed to shed further light into the question
of relevance or irrelevance of pore-scale physics
\cite{Bernadiner_1998,Senden_2000}. Three basic invasion processes were 
observed. A cylindrical pore may be invaded either as in the basic capillary
rise (piston flow) or by surface flow, followed by collapse (snap-off) of
the film, as shown in Fig.~\ref{lenormand7}. At the crossing of several 
cylindrical pores, there is a drastic change in the shape of the meniscus, 
followed by a rapid pinch-off leading to the invasion of the pore 
(Fig.~\ref{lenormand6}). Each of these modes of invasion are characterised
by different length scales, which can lead to a dynamical behaviour
that is very different from the one predicted by Eq.~(\ref{t-half}).
Blunt et al.\ \cite{Blunt_1992} argue that the width of a film
flow zone in networks (under forced imbibition) 
should scale as $w_{\rm ff} \sim C^{-1}_{\rm a} \; (w_t / R_0)$
where $w_t/R_0$ is a rough estimate of the ratio of tube wall roughness
to its diameter. Percolation-style
arguments for snap-off region sizes indicate that the associated
length-scale increases much more weakly with $1/C_{\rm a}$
\cite{Blunt_1992}. 

One question related to the pore-level description is whether
{\em correlations} matter. In various types of rocks the local permeability 
may be correlated over considerable length-scales due to geological
processes. The evidence from numerical modelling based on
microscopically faithful descriptions of pore-networks seems to
indicate, that multi-phase flow phenomena as imbibition in particular are much
more prone for correlation-induced effects than, e.g., simple one-phase
permeability 
\cite{Jeraud_1990,Bryant_1993,Ewing_1993,Oren_1998,Vidales_1998,Mani_1999,Tsakiroglou_2000,Knackstedt_2001,Dana_2002,Sok_2002,Arns_2003,Oren_2003}.
Typical quantities where this would be visible are relative
permeabilities for given saturations, and remaining saturations of
the non-wetting fluid. Note that there are no studies of the front dynamics
in imbibition in the presence of ``typical'' correlations
of pore structure, as arising in empirical contexts.

In addition, inertial effects \cite{Ridgway_2002,Schoelkopf_2000b} may 
be important in the early stage of pore invasion, where they can lead
to a preferential invasion of small pores (see Eq.~(\ref{inertia}))
\cite{Ridgway_2002,Schoelkopf_2000b}. But inertia can also be important
at all stages of the capillary invasion, whenever a sudden change in
the flow occurs, as in film snap-off, or pinch-off at a corner. In these
cases, the front position could advance as $t^k$ with $k \leq 1$
\cite{Quere_1997,Quere_1999}. On short time scales (smaller than a few
seconds), the front penetration may also proceed in pores first with the
establishment of a prewetting layer through {\it diffusion}
\cite{Eklund_1988,Handy_1960} or (precursor) film flow
\cite{Constantinides_2000}.

\subsubsection{Dynamical Saturation}
In general capillary flow occurs simultaneously with film flow and can
lead to a gradient in the concentration of the invading fluid: complete
saturation is established only over a certain distance. In addition,
{\em trapping} of the defending fluid in certain regions modifies the
flow of the imbibing fluid and may block some paths of progression,
leading to a reduced permeability for the invading fluid (e.g.,
Eq.~(\ref{wash-trap}) for the simplest case). It is also often possible
that the imbibing liquid interacts directly with the porous matrix,
changing directly the spatial structure of the porous medium and leading
to a slow change in the permeability (as observed in concrete with
water \cite{Lockington_2003}, or in ordinary paper \cite{Chen_1990}). 

These phenomena can, to some extent be taken into account by introducing
different permeabilities for each fluid in Eq.~(\ref{2-phase}):
${\bf q}_i = \kappa_i (\grad P_i -\rho_i g {\bf z} ) / \eta_i$ with
$i=w,n$. In addition, the permeabilities and capillary pressure will 
often be function of the saturation of the medium, $\kappa_i = \kappa_i(s_i)$
and $p_c = p_c (s_w)$.

However, even this basic set of equations has to be augmented to account
properly for surface tension and surface areas
\cite{Hilfer_1997,Hilfer_2000}. In fact, the two-fluid interfacial
area is one of the fundamental quantities --- it also partly defines the 
driving thermodynamic force in the case of spontaneous imbibition.
Many authours have addressed this question by network models and
experiments on one hand or by thermodynamical considerations 
\cite{Gray_1991,Reeves_1996,Saripalli_1997,Gladkikh_2003,Nordhaug_2003}.
Again the physics at pore level is complicated by the presence of precursor
films and the question of how entrapped volumes of the non-wetting
fluid behave.

If both permeabilities are equal and independent of saturations, and
likewise for the capillary pressure, then the ideal case of two
incompressible fluids in a capillary, Eq.~(\ref{2inc}), is recovered. 
As mentioned above, this implies that the non-wetting fluid can easily 
be pushed away by the
wetting liquid. This corresponds to ``co-current flow'' and ${\bf q}_w
= {\bf q}_n$. The opposite case is ``counter-current flow'', the
mass-flow of the expelled liquid takes place in the exactly opposite
direction from the invading one: ${\bf q}_w = -{\bf q}_n$ 
\cite{Li_2003}. In both 
cases, if the
geometric setup allows for a one-dimensional equation we arrive at
the effective diffusion equation 
\begin{equation}
\label{general_diffusion}
{\cal P} \frac{\partial s_w}{\partial t} + \frac{\partial}{\partial x}
\left( M_w \frac{\partial s_w}{\partial x} \right) = 0
\end{equation}
where $M_w (k_{rw},k_{rn},\partial p_c/\partial s_w)$
is an effective  mobility
and the subscript $r$ refers to relative permeabilities.
This has naturally a scaling solution with $x/\sqrt{t}$ as a scaling
parameter. Notice that this is also related to the piston-like fronts
in the Buckley-Leverett theory \cite{Buckley_1942,Dullien_2002}, and that
the implied scaling properties are of practical interest, as well
\cite{Zhang_1996,Ma_1997}.
If the non-wetting fluid cannot support a
pressure gradient (and we assume that it is free to escape), then
$P_n = P_0$ and Richard's Equation \cite{Richard_1931} appears as a
special case of  Eq.~(\ref{general_diffusion}). 

When dynamical saturation is present, the concept of an interface can 
easily become ill-defined  if there is no sharp jump in
concentration between the wet and dry parts of the medium. In many cases
the permeability is a rapidly increasing function of the 
saturation (e.g., $\kappa (s_w) \sim \exp( \beta (s_w -s_0) )$ 
\cite{Garder_1958,Mitkov_1998}) so that a sharp division exists between
the invaded and non-invaded part of the medium, which also leads to the 
basic scaling $H(t) \sim t^{1/2}$ for spontaneous imbibition. 

Hysteresis is also an omnipresent feature \cite{Beliaev_2001}. Partly it
occurs between what commonly is called ``primary'' and ``secondary''
imbibition. Such history effects arise if there is already a presence
(residual saturation) of the wetting fluid.

Saturation can be studied dynamically since the gradual increase in the
local fluid volume fraction can be followed by Nuclear Magnetic Resonance
(NMR) and by X-ray Computer Tomography (CT) techniques. These have
reached already accuracy levels in which individual pores can be monitored
as filled or not, and perhaps in the near future the interfacial features
will become accessible as well (see \cite{Wildenschild_2002} for a
hydrology example) to allow studies of interface dynamics in detail
\cite{Takahashi_1997,Bico_2001}). One of the important practical questions
is the dynamics of entrapped fluid. Though percolation-based descriptions
of such residual clusters have existed for quite some time
\cite{Wilkinson_1986}, the physics of such ``ganglions'' is still quite
open, often accessible only via numerical models.

Imaging techniques are sufficient to demonstrate phenomena like
{\em swelling} of the solid volume fraction \cite{Takahashi_1997}
in the case of disordered fibre networks (see also
\cite{Gupta_1988,Walinder_1999,Nederveen_1994,Hoyland_xxxx}).
Swelling, coupled to the fact that the porosity undergoes a simultaneous
decrease, leads to reduced fluid flow towards the interface and possibly
to deviations from Washburn behaviour. The problem of deformable
porous media is difficult to understand in general terms, and involves
from the theoretical viewpoint the solution of coupled elastic and
fluid mechanics problems (see e.g. \cite{Sommer_1996,Preziosi_1996}).
Other NMR studies have been recently performed in situations that are
of relevance for construction engineering (water imbibition into
cement pastes \cite{Ceballos-Ruano_2002,Leventis_2000}) and the
oil industry \cite{Rangel-German_2002,Zhou_2002,Akin_2000}. In
the latter case, the displacement of oil by water is an important
practical question, as is the achieved level of saturation. 
NMR imaging has now developed to a level, where one can consider the
saturation vs.\ pore sizes --- and distinguish between co- and
counter-flow due to the importance of film flow in the latter one
\cite{Chen_2003}.

By simply comparing the mass intake and the optical appearance of a paper
sample, it has been noted that the saturation and the actual visual
front may both display similar temporal scalings (Washburn-like). This
would define a widening ``saturation front'' as the partially
saturated volume behind the interface that drives the imbibition
process \cite{Bico_2003}. The principal question is whether the
saturation takes place in the wake of a ``front'' or whether
the process as a whole obeys diffusion into, e.g., small pores first
\cite{Handy_1960}. The first limit corresponds to piston-like advance
(a term using a one-dimensional analogy), and the second to a flat
saturation profile (which changes by the time-scale contained in
the scaling variable).

\begin{figure}[h]
\includegraphics[width=7cm]{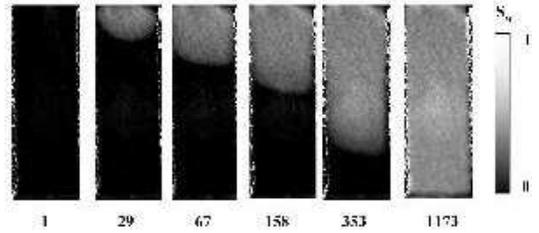}
\caption{X-ray computer tomography of spontaneous
counter-current imbibition of water into $n$-decane
in diatomite \cite{Zhou_2002}. Notice the slowing-down
of the sharp interface between dark and light (the
latter denoting water-saturated regions).}
\label{petrol}
\end{figure}

Recent imaging data \cite{Baldwin_2003,Melean_2003a,Melean_2003b} show
convincingly the capabilities of both CT and NMR approaches
in demonstrating both limits, and the cross-over in-between.
Figure~\ref{petrol} exhibits a clear example of an experimental front 
with only partial saturation \cite{Zhou_2002}. In the close future one
would expect that such techniques yield much more empirical evidence
\cite{Dong_1998,Dullien_2002,Schembre_2003} by looking at saturation
profiles and concomitantly capillary pressure curves as the saturation
is varied.

\begin{figure}[h]
\includegraphics[width=7cm]{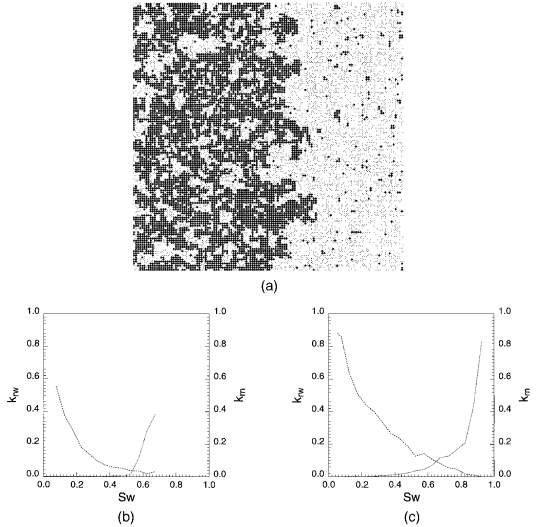}
\caption{Dynamic simulations of imbibition, showing
how a front can exist even in conditions of ``secondary
imbibition'' or initial wetting phase saturation.
\cite{Hughes_2000}. The capillary
number is $C_{\rm a} = 10^{-7}$, and the wetting
saturation 0.4. In the top panel, a snapshot,
while the two lower ones show the relative
permeabilities for two and three dimensional
systems ($128^2$ and $20^3$, the contact angle
$\theta =50$ degrees. Initial wetting phase
saturation was chosen as $0.063$.}
\label{blunt8}
\end{figure}

The transport of liquid into the less-saturated regimes can further be
complicated by the presence of an initial (residual) saturation
\cite{Jeraud_1990}. Figure \ref{blunt8} shows an example of how the
presence of wetting fluids is supposed to influence the {\em front}
geometry \cite{Hughes_2000}. The width of the front can spread in time
and exhibit what is called ``hyperdiffusivity'' via the dependence of the
$\kappa(s_w)$ on the local wetting fluid saturation. This is in fact
related to the question of diffusive hydrodynamic spreading of tracers,
and has been debated in the literature
\cite{deGennes_1985a,Bacri_1990,Davis_1990}. Recent imaging experiments
indicate that the hyper-dispersion can be related in a power-law fashion
to $s_w$ \cite{Melean_2003b,Melean_2003a}. This would result from
saturation dynamics following film behaviour in the pores.

Further such complications become evident if the description by
Richard's equation is modified by, e.g., a time-dependent  porosity
\cite{Lockington_2003}, or if anomalous diffusion is invoked
\cite{Kuntz_2001b}, both resulting presumably from liquid-porous matrix
interactions. The usual saturation dynamics, contained in simulation 
models \cite{Roels_2002,Thorenz_2002}, would indicate diffusive intake
--- which then would exhibit similar scaling as the spontaneous imbibition
front position \cite{Kuntz_2001}. The saturation dynamics can thus 
omplicate the dynamics at the front due to the conservation of the
liquid that a sample intakes per unit time \cite{Smiles_1998}, an
issue that has not been studied systematically to our knowledge in spite
of some ``snapshots'' in the literature, as Figure \ref{blunt8}.

\begin{figure}
\includegraphics[width=8cm]{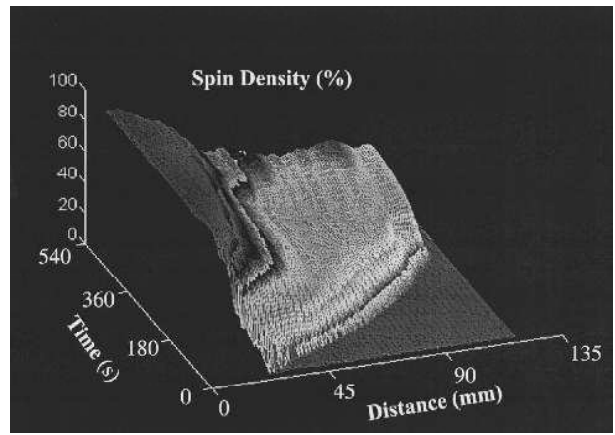}
\caption{Saturation profile of the liquid in a fibre
assembly (plug) \cite{Takahashi_1997}, as a function of time
and distance from the liquid reservoir.
}
\label{swede3}
\end{figure}
\begin{figure}
\includegraphics[width=8cm]{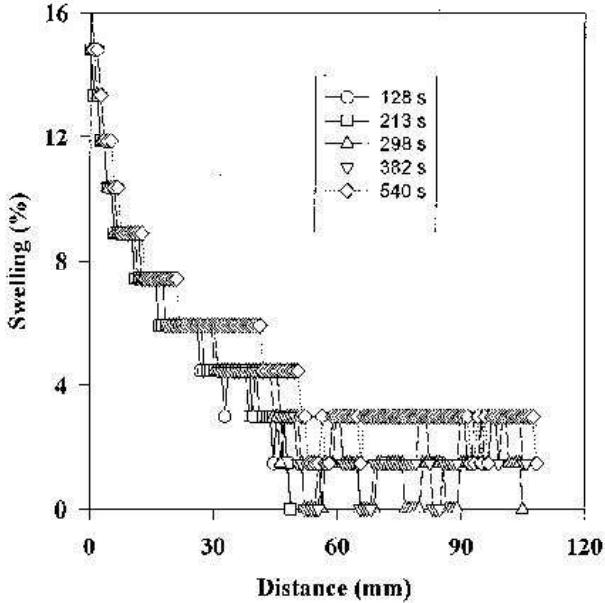}
\caption{The swelling of fibres, as a function of time and distance,
in a plug (\cite{Takahashi_1997}). A concomitant change in local
permeability, and thus fluid flow, is expected.
}
\label{swede6}
\end{figure}
\begin{figure}
\includegraphics[width=8cm]{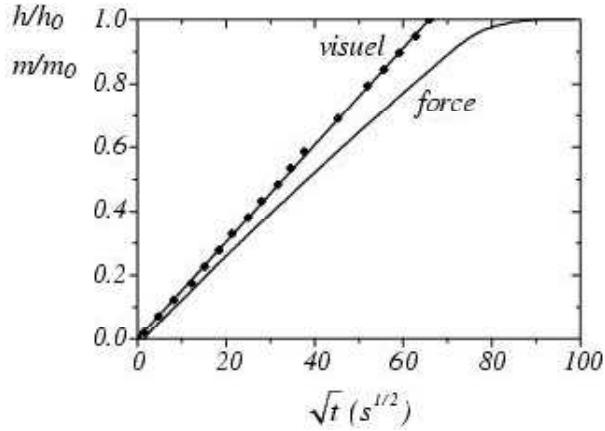}
\caption{The front and the mass (of the silicone fluid) in a paper
sample vs.\ time (Washburn-scaling, $\sim t^{1/2}$), according to
Ref.~\cite{Bico_2003}. The implication is that the saturation is
incomplete behind the front.
}
\label{Quere11}
\end{figure}

\subsubsection{Surfactants, additives and other non-Newtonian effects}

Finally, additives to a carrier fluid, such as dye, give rise to
phenomena that can cause a deviation from Washburn-like behaviour.
The dye particles diffuse as usual Brownian particles but are also
carried by the imbibing fluid, which gives rise to hydrodynamic
dispersion \cite{Sahimi_1993}. The dynamics of dispersion is of
fundamental interest in the imbibition context for two main reasons:
In some cases, the dye front is (maybe somewhat dangerously)
identified with the ``elastic interface'' for which theoretical
models are built. The dynamical contact angle at the advancing front
is generally dependent on the local dye concentration. This may
act as a surfactant, with a time-dependent concentration. Similar
behaviour can be observed for various mixtures of liquids: time-dependent
changes to the composition vary the viscosity and affect the time-scales
\cite{DeMeijer_2001}. Alternatively the absorption can be slow on
the time-scale of the dynamics \cite{Tiberg_2000}.
We do not want to deal extensively with the effects that these
complications may have on the reformulation of the interfacial
theories for imbibition, but list below only some possible effects
that one should keep in mind.

If the surfactant transport is diffusive, one gets accidentally
Washburn-like dynamics but anomalous diffusion can clearly change this
and lead to almost arbitrary temporal scalings. At the very least, the
expected length and time scales will change, sometimes drastically
\cite{Hodgson_1988,Chen_2001}. The effect of the surfactant (or
time-dependent composition) (e.g.\ \cite{Luokkala_2001,Rame_2001}) can
be understood based on, e.g., viscous dissipation at the contact line,
which changes the dynamical contact angle. This is a research field in
its own, related to the physics of thin films and droplet spreading
(see \cite{Davis_2000,Marsh_1993,Stoev_1998,Stoev_1999} among
others).

\begin{figure}
\includegraphics[width=8cm]{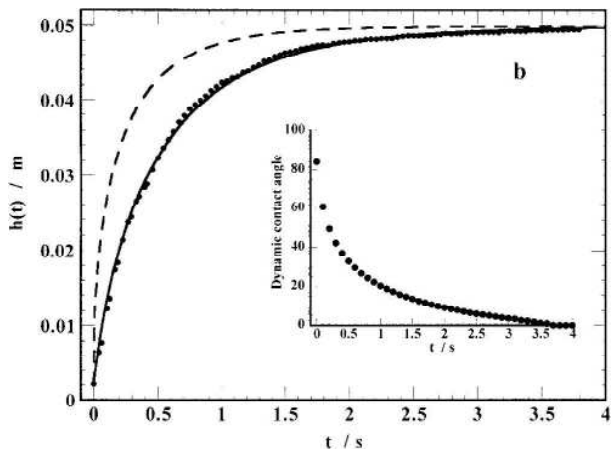}
\caption{
The rise of a column of water (and the effective dynamic contact
angle) vs.\ time in a capillary rise example. Notice the changing
angle and the deviations from Washburn's law (best fit with the dashed
line) \cite{Hamraoui_2000}.}
\label{6830a3}
\end{figure}

Generally, deviations from Washburn behaviour can easily exist. This is
particularly so at early times, due to the dissipation-induced changes
\cite{Quere_1997}. The fast movement of menisci, coming from microscopic
geometric considerations, coupled to the viscosity of the fluids involved,
leads to dissipation at the front (viscous stresses) and should persist
in all regimes and at all times in spontaneous imbibition
\cite{Tzimas_1997,Tzimas_2001}. Some progress in accommodating
dynamical phenomena can be made by modifying directly the contact
angle in Washburn-like effective equations \cite{Hamraoui_2002}, see
Fig.~\ref{6830a3}.

It should not be forgotten that the porosity can be also be changed
in a time-dependent way by the deposition of any particles carried by
the invading liquid \cite{Weir_1996,Frey_1999,Gagnon_2001}, which again
may lead to some erratic behaviour for the front. We note that the same
conclusions can coincidentally be obtained in the case of non-Newtonian
fluids \cite{Wu_1991,Chiu_1996,Pearson_2002}. Only recently there have
been some experimental developments on the scale of network model systems
\cite{Tsakiroglou_2003}. Similarl to three-phase systems
\cite{Fenwick_1998,Dijke_2003a} such work highlights the upscaling problems
when the pore-level behaviour varies widely. For shear-thinning fluids
the local fluid response inside a single pore depends on the flow rate,
i.e., on both the global front dynamics (due to fluid conservation) and
on pore details such as size or connectivity. A possibly general trend is a
less distinct front due to microscopic flows. Similarly, it should be
noticed that ink, often used in imbibition experiments behaves
rheologically as a non-Newtonian fluid \cite{Aspler_1993}

The treatment of microscale dynamics, such as inertia and film flow at
the pore-level, is very delicate at the level of hydrodynamic equations
and initial conditions \cite{Letelier_1979,Kornev_2001}.  However, the
central point is that these details enter the hydrodynamical description
in a coarse-grained way. This leads to a macroscopic front velocity which
has either a Washburn-like behaviour, but with effective permeability and
capillary pressure, or to a completely different scaling behaviour
\cite{Senden_2000}. Microscopic simulations of capillary imbibition imply
that modified Washburn-type equations ensue \cite{Martic_2002}. Nevertheless,
on a macroscopic scale, averaged over a history of penetration and a
representative volume, the situation is of course open.

\subsection{Non-equilibrium processes and roughening of fronts}
The front between the wet and dry regions roughens as it propagates.
This is the main emphasis in our review. The spatial and temporal
statistics of the interfaces is easily observed by the naked eye, as
in Fig.~\ref{FigInkFront}, and they remind, in general terms, of
other ``rough'' objects such as fractures in inhomogeneous
media, fronts in slow combustion, flux lines in disordered
superconductors, and Brownian paths in diffusion
\cite{HalpinHealy_1995,Meakin_1998,Barabasi_1995,Kardar_1998}.
The existence of so many analogies poses the fundamental 
question of universality. If the roughness is {\em scale-invariant},
it can be described by fluctuations and correlations that possess
power-law behaviour, reflected in typical quantities such as the spatial
and temporal two-point correlation functions, $G_2(r)$ and $C_2(t)$
respectively (see Eqs.~(\ref{G_vs._S}) and (\ref{Ct})).
These functions exhibit dynamical scaling that establishes
{\em critical exponents}, understood in the usual sense of statistical
mechanics. The exponents --- like those of the celebrated
Kardar-Parisi-Zhang (KPZ) equation \cite{Kardar_1986} ---
are related to a particular {\em universality class} with its
symmetries embedded in a Langevin-like equation for the interface. 
In Sections \ref{KinroughStart} to \ref{KinroughEnd} we review
the appropriate parts of the theory of kinetic roughening. 

In the late 1980's much effort was put into the physics of fluid fronts
(e.g., in disordered Hele-Shaw cells) to produce clear power-laws with
well-established exponents related to a particular universality class. 
The hope in the imbibition context was that the non-local nature of the
problem could be neglected in favour of a local description, partially
for simplicity but also because the complications involved would otherwise
seem unsurmountable \cite{Koplik_1985}. In reality, the experimental picture
turned out to be much more complicated and often not directly related to
any of the expected universality classes and the question of an effective
theory for imbibition front remains.

The main problems that a theoretical description must solve are the
nature of the noise acting on the interface and the effect of
non-locality induced by fluid conservation. At the simplest level the
fluid pushing the interface flows according to Darcy's law, which
already implies that the advances of sections of the fluid front are
coupled. Often a length scale $\xi_\times$ along the interface
emerges, which is related to the average motion of the interface and
sets the maximum extent of ``local fluctuations''.

This is intuitively clear for spontaneous imbibition since
fluctuations have a larger instantaneous velocity when behind the
front compared to when that are ahead of it. Consider an effective
surface tension $\gamma^*$, representing the macroscopic energy cost
of a curved interface. The curvature arising from fluctuations changes
the pressure by $\Delta P = \gamma^* W / \xi_\times^2$ (the Laplace
pressure or Gibbs-Thomson effect) where $W$ is the vertical and
$\xi_\times$ the lateral scale of the fluctuation. This pressure
change slows down advanced parts of the interface. Comparing it to the
overall pressure drop across the same vertical distance, $\Delta P =
p_c W / H$, yields a lateral length scale
\begin{equation}
\label{xi_cross_intro}
\xi_\times^2 \sim \gamma^* H/ p_c.
\end{equation}
This indicates that fluctuations on a scale larger than $\xi_\times$
are suppressed by the pressure field gradient $p_c/H$. For an
interface slowing down in spontaneous imbibition it follows directly
from Eq.~(\ref{t-half}) that $\xi_\times \sim H^{1/2} \sim t^{1/4}$
and for the transversal width of interface fluctuations $W \sim t^\beta$
with $\beta = \chi/4$ where $\chi$ is the roughness exponent of the
interface (see Section \ref{phasefield} for a more detailed discussion
of $\chi$ and the criticality of imbibition fronts).

Since the velocity of the interface $v = (\kappa p_c)/(\eta h)$, the
length scale $\xi_\times$ can be rewritten in term of the capillary
number $\xi_\times = (\kappa /C_{\rm a})^{1/2}$, which highlights the
importance of low capillary number experiments. Theoretical analysis,
to be presented in Section \ref{Theory} also shows that $\xi_\times$
is present in forced flow experiments, where it separates two
different spatial scaling regimes. 

The fluctuations of the interface have three sources: variations in the
permeability $\delta \kappa$ (the fluid has to flow to the interface but
does not do so uniformly), capillary forces $\delta p_c$ and volume
(the advancing fluid has various pores to fill, thus $\Delta h$ depends
on such fluctuations). It will be shown in Section \ref{Theory} that
disorder coming from saturation disorder can usually be neglected and that
capillary and ``mobility'' (arising from the random permeability) noise
act on different length scales, characterised by $\xi_{\rm mob} \sim
\kappa^2 \delta p_c / v \eta \delta \kappa$. On lengths $l \ll
\xi_{\rm mob}$, capillary noise is dominant, while mobility disorder
is relevant on length $l \gg \xi_{\rm mob}$. Of course, the permeability can
also be influenced by an interaction between the fluid and the medium and
complicate this picture.

It is also possible that for small interfacial velocities the physics
changes. Instead of the liquid front propagation via jumps or advancement
of the menisci, pores in contact may be partially wetted by film flow. From
a fundamental viewpoint this is the regime that, if described by local
interface equations, would exhibit {\em avalanche behaviour}: most of the
interface stays quiescent while only parts advance. The description
for such dynamics in usual kinetic roughening involves geometrical ideas
that draw parallels to directed percolation and its cousins in various
branching processes \cite{Marro_1999,Sornette_2000}.




\section{Theoretical approaches to imbibition}
\label{Theory}

The usual models, to analyse imbibition, can be divided into two main groups.
There are ones built in close connection to experimental studies, whose
goal is to further develop any details important for applications. Roughly
speaking these are related to the microscopic or small scale properties of
the imbibition systems. We shall also present some of these aspects in 
Chapter~\ref{Experiment} when reviewing experiments. Such models
are particularly useful when the microscopic physics at the front include
prewetting layers, film flow or when the saturation properties of the whole
medium, as a function of time, need to be addressed. In the latter case
one often needs to include detailed considerations about the behaviour
of residual pockets of the gas/liquid left behind the imbibition front.

On the other hand the models presented in this Chapter are more concerned with
questions of universality
in the morphology of the wetted region. Generally they include microscopic
details on a more abstract ground, incorporated, e.g., in the noise
terms included. They are meant to highlight the connection between 
the essential mechanisms, as ingredients into the model, and
the outcome in macroscopic morphology.

\subsection{Morphological phases in imbibition}
\label{morphologies}

The stability of the interface between the invading and the defending
liquid was one of the first issues to be studied \cite{Stokes_1986}.
A relation between the capillary number of the fluid invasion process
and the size of fingering structures in the interface has been found.
In the limit of very thin fingers the invading fluid forms a self-similar
fractal cluster which belongs to the universality class of invasion
percolation \cite{Koiller_1992}. In fact, there is a transition between
compact and fractal morphology of the invading fluid cluster
\cite{Ji_1991,Koiller_1992,Lenormand_1990}. The capillary number $C_{\rm a}$
in Eq.\ (\ref{capillary_number_intro}) relating viscous and capillary
forces is a control parameter, e.g., in an experimentally obtained
phase diagram and examples of morphologies from \cite{Lenormand_1990}
are shown in Figures~\ref{lenormand5} and \ref{lenormand10}. If the
porous medium is easily wetted the cluster tends to be compact. The
wetting properties, measured in terms of the static contact angle
$\theta$, are related to the fingering width of the interface. With
increasing wetting tendency it increases and finally diverges at a
critical contact angle $\theta_c$, below which the invading fluid
cluster remains compact and forms a well defined albeit rough front
\cite{Koiller_1992}.

Compact invasion clusters with self-affine rough fronts have already
been reported earlier \cite{Rubio_1989}. At the same time theoretical
description and modelling as well as experimental studies of roughening
interfaces were intensively studied and have thus created a lot of interest
in imbibition. The central objective of this Review are rough fronts
in imbibition, and in this Chapter we present different model approaches
to them and their theoretical evaluation. Before that we give a phenomenological
introduction to the elementary processes in imbibition (Section~\ref{pheno}),
as well as a brief recall on kinetic roughening in 
Section~\ref{kinrough}.

\subsection{Phenomenological approach to imbibition with rough fronts}
\label{pheno}

The interest on front roughening in imbibition was first and foremost
motivated by the assumed connection to the KPZ equation \cite{Kardar_1986} and its
underlying theoretical relation to experimental systems.  Such spatially
local equations were considered after the ``full'' 
problem was proven to be very complicated and they hope to capture the 
essential features by focusing on the interface only. 
This description is valuable, perhaps, for (nearly) pinned
interfaces, although also here interesting physical complications arise 
near the pinning transition which are the subject of 
Section~\ref{avalanches_sect}.  

It was however quickly realised that the detailed nature of fluid
invasion could not be ignored \cite{He_1992}. The first theoretical efforts in 
front roughening with a {\it global} conservation law were done 
by Krug and Meakin \cite{Krug_1991}. They considered the roughening of a
Laplacian front as discussed below, using a linear analysis.
This is similar to the problem of
Saffman and Taylor \cite{Saffman_1958}, for the stable case 
of a high viscosity fluid entering a low viscosity one. 

Since the Laplacian one is a basis for later theories as well, it is worth 
considering
it in some details. The starting point of such an analysis is based on the 
phenomenological law of Darcy for flow in porous media, 
\begin{equation}
{\bf q} = - \frac{\kappa}{\eta} \, \grad P 
\label{darcy_th}
\end{equation}
where ${\bf q}$ is the flux of liquid, $\eta$ the viscosity, and
$\kappa$ the permeability of the medium, essentially dependent of the
size of the channels through which the liquid flows (c.f.\ Equations
(\ref{washburn}) to (\ref{t-half}) in the Introduction). The
incompressibility condition $\nabla \cdot {\bf q} = 0$ leads to a Laplace
Equation for the pressure
\begin{equation}
\nabla^2 P = 0.
\end{equation}
The front then propagates because of mass transport by the flow ${\bf q}$,
whose component normal to the front determines the front velocity
\begin{equation}
v_n = - \frac{\kappa}{\eta} \, \partial_n P,
\end{equation}
where $\partial_n$ represents the normal derivative to the front.

\begin{figure}[ht]
\includegraphics[width=8cm]{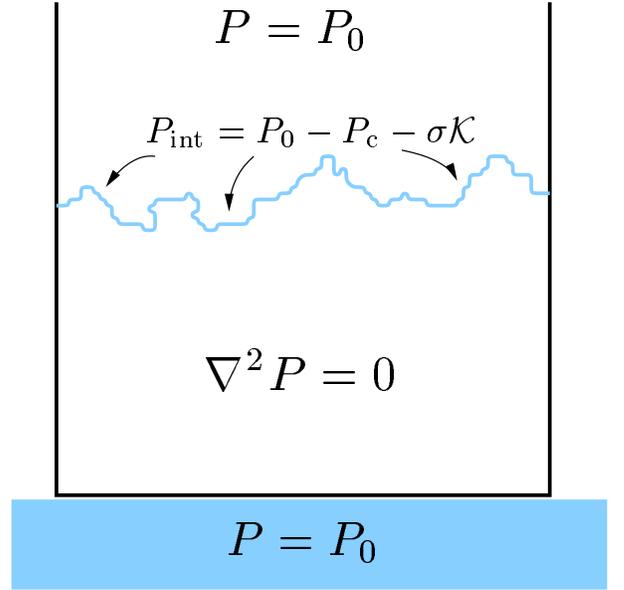}
\caption{Theoretical concepts of modelling imbibition. A rough front is situated
at the average height $H$. A pressure field solves Laplace's equation in the
bulk, boundary conditions are given at the contact line to the liquid reservoir
and at the rough interface. Pressure gradient is proportional to flux, and
influx at the interface proportional to interface displacement velocity.}
\label{th_setup}
\end{figure}

We consider the situation described in Fig.\ \ref{th_setup}.
The interface, of average height $H$, is described by a single-valued function
$y = h(x,t)$ with Fourier decomposition
\begin{equation}
h(x,t) = H + \sum_{k\neq 0} h_k(t) \, e^{ikx}.
\end{equation}
At the level $y = 0$ the medium is in contact with an outer reservoir of
liquid and the pressure is supposed to be controlled to $P(x,0;t) = {\cal
P}_0 (t)$. At the interface the boundary condition for the pressure is
\begin{equation}
P_{int} = P(x,h(x,t);t) = P_0 - p_c - \sigma {\cal K} 
\label{p_boundary}
\end{equation}
where $p_c$ is the coarse-grained capillary pressure, introduced in Section
\ref{LucasWashburn}, $P_0$ is the atmospheric pressure of the ambient gas, and
${\cal K}$ is the curvature of the front. Up to linear order in the interface 
fluctuations $\delta h (x,t)= h(x,t) - H$, the pressure field is given by
\begin{equation}
p(x,y;t) = {\cal P}_0 - \frac{\eta v}{\kappa} y 
- \sum_{k \neq 0}  e^{ikx} \left( \frac{\eta v}{\kappa} + \sigma k^2 \right) 
\; \frac{\sinh ky}{\sinh kH} \; h_k
\end{equation}
where the average displacement of the interface
\begin{equation}
 v = \frac{d H}{dt} = \frac{\kappa}{\eta} \; \frac{{\cal P}_0 -P_0 +p_c}{H}.
\end{equation}
The dynamical equation for the interface fluctuation \cite{Krug_1991} is then 
\begin{equation}
\frac{d h_k (t) }{dt} = - \frac{|k|}{\tanh |k| H}
\left( v + \frac{\kappa}{\eta} \; \sigma k^2 \right) h_k (t).
\label{basic_linear}
\end{equation}

Several important conclusions can already be drawn from this analysis:

\noindent
{\bf (i)} The average behaviour of the interface is determined from the
pressure at the bottom of the porous medium. A constant pressure gradient
results in a constant average velocity of the interface $v$.
On the other hand, if the porous medium is simply immersed into a reservoir
at atmospheric pressure $P_0$, the average interface velocity
decreases with the average height
\begin{equation}
v = \frac{\kappa}{\eta} \; \frac{p_c}{H}.
\end{equation}
Solving for $\partial_t H = v$ makes the Washburn-Darcy behaviour apparent,
\begin{equation}
H^2 (t) - H^2 (t_0) = 2 \; \frac{\kappa}{\eta} \; p_c \, (t-t_0).
\end{equation}

\noindent
{\bf (ii)} There exists an intrinsic length scale 
\begin{equation}
\xi_{\times}^2 = \frac{\kappa}{\eta v} \; \sigma
\label{cross} \end{equation} 
that separates two different modes for the relaxation of the interface
fluctuations. If this interface moves at a constant velocity, this 
length is fixed and static, while it is dynamical ($\xi_{\times} \sim
v^{-1/2} \sim t^{1/4}$) for spontaneous imbibition. Note that the capillary
number $C_a = \eta v /\sigma$, relating the viscous and capillary forces,
can be used to rewrite $\xi_{\times} = (\kappa / C_a )^{1/2}$.

\noindent
{\bf (iii)} There are odd powers in momentum, more precisely a dependence
on $|k|$ in the dynamical Equation (\ref{basic_linear}), which reflects
non-locality in space arising from the conservation law. However, the truly
non-local $|k|$-dependence cancels for wave-vectors $|k| \ll 1/H$, meaning that the
interface must be far enough from the reservoir of liquid to start to feel
the effects of conservation. The surface dynamics is then local 
in the sense that points with mutual
distance greater than $H$ are essentially independent of each other. 

\noindent
{\bf (iv)} 
This equation is in principle readily applicable to the case of fluid
propagation in Hele-Shaw cells filled with glass beads \cite{Martys_1991}, 
but the nature of the disorder remains
ambiguous. Krug and Meakin \cite{Krug_1991} chose for their application to use
non-conserved thermal noise, $\langle \eta_k (t) \eta_{k'} (t') \rangle
\sim \delta_{k,-k'}\delta (t-t')$, which gives rise to asymptotic logarithmic
roughness, not observed experimentally in imbibition. Nevertheless, this equation
already gives some hint about the influence of the conservation law. 

The simultaneous treatment of quenched disorder and liquid conservation
was first attempted by Brenner and Ganesan \cite{Ganesan_1998}. They assumed
that the capillary pressure was dependent on the spatial location ${\bf x}$
and random, $p_c = p_c ({\bf x})$, as well as a constant mobility $\kappa$. 
Their end result is similar to that of Eq.\ (\ref{basic_linear}), 
although it includes non-linearities and the quenched nature of disorder, 
\begin{widetext}
\begin{equation}
\frac{d h_k (t) }{dt} = |k| \; \left( \frac{\kappa p_c}{H} + \kappa \sigma \;
k^2 \right) h_k (t)
- \frac{\kappa p_c}{H} \int \! dk' \; k' \; [k-k'] \; h_{k-k'} h_{k'} 
- \kappa \sigma \int \! dk' \; |k'|^3 \; [k-k'] \; h_{k-k'} h_{k'} + \eta_k .
\label{imbi_non_linear}
\end{equation}
\end{widetext}
The noise term $\eta_k$ is obtained by assuming that the boundary
condition Eq.\ (\ref{p_boundary}) can be modified by adding a random
part $P_{\rm int} \rightarrow P_{\rm int} + \eta (x,h)$ with so-called
random field correlations $\langle \eta(x,h(x,t)) \; \eta(x',h(x',t')
\rangle = \Delta  \; \delta(x \! - \! x') \; \delta(h(x,t) \! - \!
h(x',t'))$. These would amount to the fluctuating part of the
capillary force at the interface. Note that the non-linearities also
appear non-locally in space, contrary to the usual non-linear term
introduced in the KPZ equation outlined in the next subsection. The
noise term also enters the interface equation in a non-local way. It
is however difficult to extract any results from this method, since
Eq.\ (\ref{imbi_non_linear}) is not very well suited even for direct
numerical integration. A Flory-type analysis (term-by-term comparison
of typical magnitudes of the terms) yields a roughness
exponent $\chi = 3/4$ but this result is questionable since it is well
known that a similar analysis gives wrong results in the simpler case
of the so called quenched Edwards-Wilkinson (QEW) equation discussed
also below.

This analysis was pushed further by Paun\'e and Casademunt
\cite{Paune_2002}. They considered the specific problem of fluid flow
in a Hele-Shaw cell, where the only disorder is through variation in
the thickness of the cell $b$. This implies that both the capillary
pressure
\begin{equation}
p_c = \sigma \cos \theta \left( \kappa + \frac{1}{b} \right)
\end{equation}
and the permeability $\kappa \propto b^2/\eta$ become random
quantities. They then proceeded to derive a phenomenological equation
that combines Eq.~(\ref{imbi_non_linear}) and earlier equations
derived by Hern\'andez-Machado {\it et al.} and Dub\a'e {\it et al.}
(these are discussed in more detail later). This work shows that
disorder in the capillary pressure and in the mobility act on very
different length scales depending on the velocity of the average
interface, as will be discussed in details in Section
(\ref{constantflow}). At the simplest level the main source of
disorder will come from variation of thickness $b \rightarrow b +
\delta b$, which affects the capillary pressure $p_c \sim p_c + \delta
p_c$ and the permeability $\kappa \sim \kappa + \delta \kappa$,
\begin{equation}
\frac{\delta p_c}{p_c} = - \frac{\delta r}{r_0} \; \; \; \; \; \;
\mbox{and} \; \; \; \; \; \; \frac{\delta \kappa}{\kappa} =
\frac{\delta r}{r_0}. 
\end{equation}
Based on a simple Darcy analysis, this would suggest that variations
in the velocity due to permeability disorder $\delta v_{\kappa}$ are
of the same order of magnitude as those due to variation in the
capillary pressure $\delta v_{p_c}$
\begin{equation}
\left| \frac{\delta v_{p_c}}{\delta v_{\kappa}} \right| \sim
\left| \frac{\kappa \; \delta p_{c}}{p_c \; \delta k} \right| = -1,
\end{equation}
which would imply that both sources of disorder are equally important.
This may be right for the specific case of a Hele-Shaw cell with varying
thickness but not necessarily for more general random media such as 
paper. 

\subsection{Theoretical description of rough interfaces}
\label{kinrough}
\subsubsection{The concept of roughness}
\label{KinroughStart}
\begin{figure}[h]
\vspace{10cm}
\includegraphics{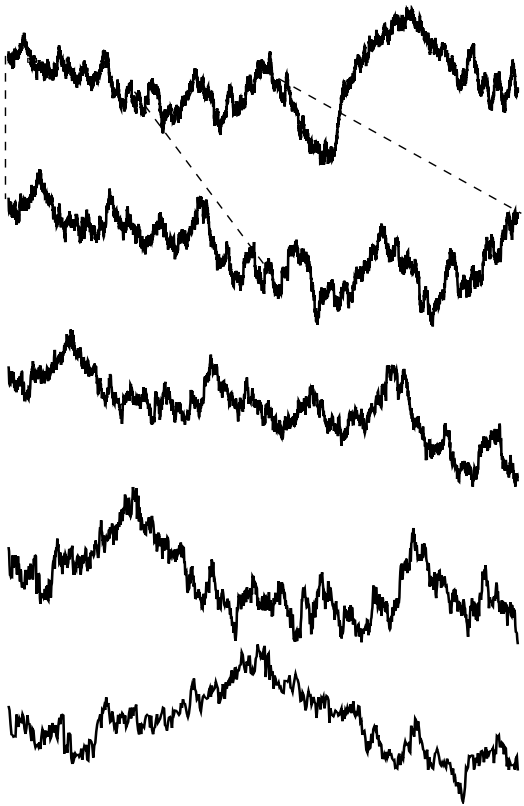}
\caption{Different aspects of an interface, magnified by a factor 2 in each
  step from top to bottom, as indicated by the dashed lines between the first
  two panels. No inherent length scale becomes visible, all surface pieces
  ``look the same''}
\label{statscale}
\end{figure}

The exact shape of a rough front or interface in a given dynamical
process depends on the particular realization of randomness
encountered and is unpredictable. The statistical properties of its
shape fluctuations however can be described in a controlled way. Quite
often one observes statistical scale invariance in certain ranges of
time and space. Naively this means that the interface ``looks the
same'' when its parallel and perpendicular length scales are ``blown
up'' or ``shrunk'' by certain factors. In this Section, as well as in
the entire Article, we consider interfaces with translational
invariance, meaning that they ``look the same'' no matter which part
of them you consider. For an illustration see Figure \ref{statscale}.

Roughness need not be a static phenomenon but changes its properties
in time. Typically, a growing or moving interface will increase its
roughness with time, as single fluctuations are accumulated. We will
also see scale invariance in time, i.e., an interface ``behaves the
same way'' when time is speeded up or slowed down with an according
spatial rescaling.

The relation to critical phenomena is evident. Statistical scale invariance
over a certain range of length scales reflects itself in power law
shapes of correlation functions. As with critical phenomena, one
observes different types of power law behaviour of interfaces, which
are therefore referred to as {\em universality classes} as well
\cite{Krug_1992,Barabasi_1995,Marsili_1996}.

\subsubsection{Formal description of roughness}
\label{kinroughformal}
Let us try to formalise the intuitive picture of shrinking and enlarging 
the aspect of an interface. Generally, an interface $\mathcal{S}$ is an 
orientable $d$-dimensional object in $(d+1)$-dimensional space. A crystal 
surface is an example for $d=2$ while a step separating two terraces on 
it is one for $d=1$. Since the interface is orientable, we can define an 
indicator funtion $\varphi : \mathbb{R}^{d+1} \to \mathbb{R}$ which takes 
the value $1$ on one side of the interface, $-1$ on the other, and $0$ 
right on top of it. In the example of a crystal one can choose  $\varphi = 1$
inside and $\varphi = -1$ outside.

\begin{figure}[h]
\vspace{4cm}
\includegraphics{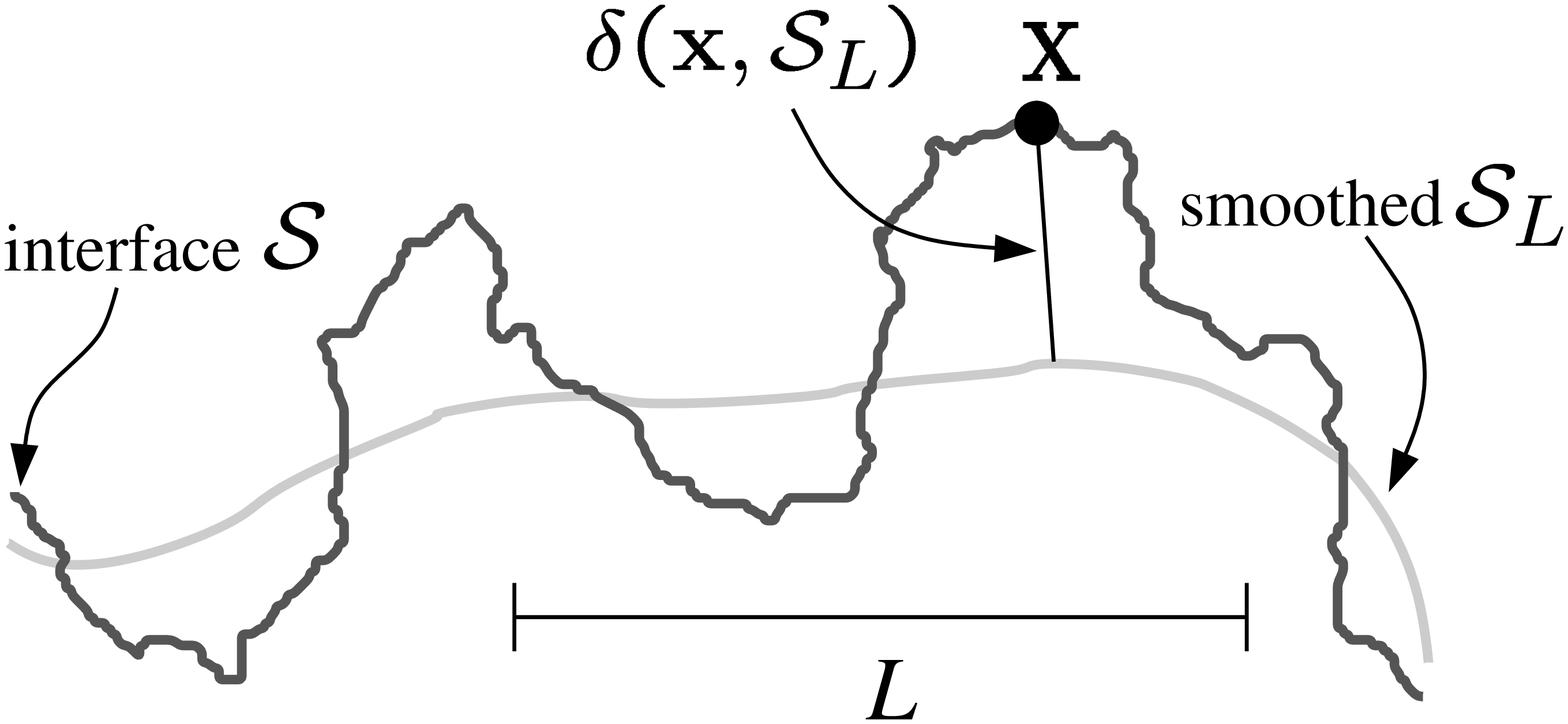}
\caption{Example for construction of $\mathcal{S}_L$ and $\delta({\bf
    x},\mathcal{S}_L)$ in $d \! + \! 1 = 2$ dimensions.} 
\label{roughdef}
\end{figure}

To examine the roughness on various scales we use a smoothing kernel
\begin{equation}
f_{d+1}^L ({\bf x}) \equiv \frac{1}{B_{d+1} L^{d+1}} \; f 
\left(\frac{|{\bf x}|}{L} \right)
\end{equation}
for some monotonously decreasing function $f(r)$ with the property 
$\int_0^\infty dr \;r^d\; f(r) = 1$, and the $d$-dimensional surface
volume of the $(d+1)$-dimensional unit sphere
$B_{d+1} = \frac{ 2 \pi^{(d+1)/2}}{(d + 1) \Gamma((d + 1)/2)}$
By convolution one can obtain via the smoothed profile function
\begin{equation}
\varphi_L ({\bf x}) \equiv \int \! d^{d + \! 1} \! y \; f_{d+1}^L \! ({\bf x
 \! - \! y}) \; \varphi ({\bf y})
\end{equation}
the surface smoothed to a scale $L$
\begin{equation}
\mathcal{S}_L \equiv \bigl\{ {\bf x} | \varphi_L ({\bf x}) = 0 \bigr\}.
\end{equation}
The profile function $\varphi_L ({\bf x})$ contains information about how 
far a point is away from $\mathcal{S}_L$: If $\mathcal{S}$ is a 
hyperplane, $\mathcal{S}_L$ remains identical to it and $\varphi_L ({\bf 
x}) = F(\delta/L)$ where $\delta$ is the distance between ${\bf x}$ and 
$\mathcal{S}$, and $F(r) = \int_0^r ds \; s^d\;f(s)$. So, it is natural to 
define
\begin{equation}
\label{def_delta}
\delta({\bf x},\mathcal{S}_L) \equiv L \; F^{-1}(\varphi_L ({\bf x}))
\end{equation}
as a measure for the distance of a point ${\bf x}$ from $\mathcal{S}_L$.
When  ${\bf x}$ is taken on the original interface $\mathcal{S}$ Equation 
(\ref{def_delta}) gives an access to its roughness fluctuations on scales 
smaller than $L$. For an illustration see Figure \ref{roughdef}

Generally one observes power law scaling of the various moments of roughness,
\begin{equation}
\label{w_q_def}
w_q(L,t) \equiv \langle \langle \delta({\bf x},\mathcal{S}_L)^q \rangle_{{\bf
  x}  \in \mathcal{S}} \rangle_\mathcal{S}^{1/q} \sim L^{\chi_q} \mbox{ for
  } L < \xi(t)
\end{equation}
saturating to a constant for $L$ larger than the so called {\em correlation
length} $\xi(t)$. In this definition $\langle \cdot \rangle_{{\bf x} \in
\mathcal{S}}$ takes the average over all points in $\mathcal{S}$. The outer
brackets stand for the ensemble average over all possible interface
configurations. The measure for these averages is inferred from the area of the
interface and is straightforward to construct in most cases; $\chi_q$ is
called roughness exponent of the $q$th moment.

%

In the case of multiscaling the $\chi_q$ differ from each other. A
classical illustration of this is given by the concept of ``turbulent
interfaces'' introduced by Krug in \cite{Krug_1994}. Its point is
best illustrated on a discrete crystal lattice: The height fluctuations
may be related to the {\em largest} step heights between two neighbouring
points on the discrete lattice. This introduces a new length scale
$\Delta h = |h_{x+1} - h_x|$ which then makes the scaling more
complicated. The different momenta of $w_q$ would show a relation to
the largest values of $\Delta h$ in $L$. $\Delta h$ can then depending
on the model exhibit various behaviours: E.g., its probability
distribution can be directly a function of the system size
\cite{Krug_1994,Seppala_1998}, or at least have a cut-off imposed by
it \cite{Asikainen_2002,Mitchell_2002}. Many physical processes lead
instead to (regular) or normal self-affine scaling with $\chi_q \equiv
\chi = \mbox{const}$, to which we restrict ourselves for most of this
article, omitting the index $q$ and mentioning possible cases of
multi-scaling by remarks when necessary.

In a roughening process the correlation length $\xi(t)$ increases with time,
because physical coupling of different points in the interface spreads
fluctuations. Again, in general one observes power law behaviour, which for
historical reasons related to the theory of critical phenomena is written
\begin{equation}
\label{xi_t_def}
\xi(t) \sim t^{1/z}
\end{equation}
with $z$ the dynamical exponent. The maximal extent of interface
fluctuations is
\begin{equation}
\label{wt_scaling}
w(t) \equiv \lim_{L \to \infty} w(L,t) \sim \xi(t)^\chi \sim
t^{\chi/z} \equiv t^\beta,
\end{equation}
which connects the relations (\ref{w_q_def}) and (\ref{xi_t_def})
and defines the scaling exponent $\beta$.

All these relations can be combined into a scaling form
\begin{equation}
\label{scaleform}
w(L,t) = \alpha(t) \; \xi(t)^\chi \; \mathcal{W} \left( \frac{L}{\xi (t)}
\right).
\end{equation}
with the scaling function $\mathcal{W}(x)
\sim x^\chi$ for $x < 1$ and approaches a constant for $x \gg 1$.
In many cases the amplitude $\alpha(t)$ is constant, which is referred
to as normal scaling. It may however increase with time, causing one sort of
so-called anomalous scaling \cite{Lopez_1997}. This behaviour occurs for forced
flow imbibition with columnar disorder, analysed theoretically in section 
\ref{imbi_columnar} and experimentally in Section \ref{Experiment}.

In most cases $0 \le \chi \le 1$, but it need not be restricted to that
range. Interfaces with $\chi > 1$ sometimes are called ``super-rough''
\cite{Lopez_1997,Lopez_1999,Dube_1999}. One-dimensional interfaces in disordered two-dimensional
media are examples for super-roughness \cite{Jost_1996,Leschhorn_1993a,Leschhorn_1993b,Leschhorn_1994,Leschhorn_1996} and of
particular interest for imbibition in effectively two-dimensional media such
as paper or thin Hele-Shaw cells \cite{Dube_1999, Dube_2000a,Dube_2000b}.

\subsubsection{Height fields without overhangs}
\begin{figure}[h]
\vspace{5cm}
\includegraphics{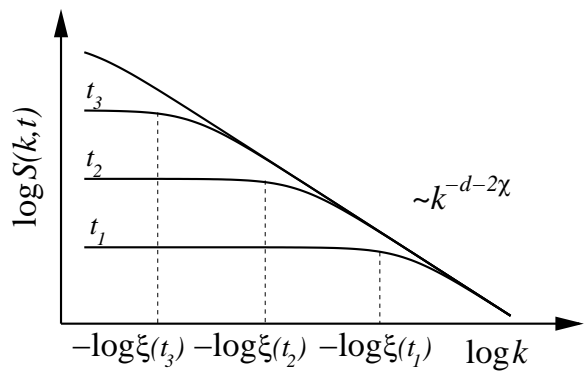}
\caption{Schematic plot of structure factors $S(k,t)$ of an interface at times
  $t_1 < t_2 < t_3$ in double logarithmic scale. The correlation length
  $\xi(t)$, as well as the intensity of long wavelength fluctuations $S(k \to
  0,t)$ increase with time.}
\label{sk_scheme}
\end{figure}

The quantitative description of interfaces can be simplified a good deal when
overhangs can be neglected. An interface is then represented by a height field
$h({\bf x},t)$ moving in time, $t \in \mathbb{R}$, over some $d$-dimensional
substrate, whose points are ${\bf x} \in \mathbb{R}^d$. The roughness exponent 
$\chi$ is then linked to the {\em  structure factor} or spatial power spectrum
\begin{equation}
\label{structfact}
S({\bf k},t) \equiv \langle | h({\bf k},t) |^2 \rangle \sim
\Biggl\{
\begin{array}{ll}
\! k^{-(2 \chi + d)} & \mbox{for }k \gg 1/\xi(t) \\
\! \xi(t)^{2 \chi + d} & \mbox{for }k \le 1/\xi(t)
\end{array}
\end{equation}
where $h({\bf k},t)$ denotes the spatial ($d$-dimensional) Fourier transform
of $h({\bf x},t)$. The scaling behaviour of Equation (\ref{structfact}) is
illustrated schematically in Figure \ref{sk_scheme}. From this schematic
representation the scaling form
\begin{equation}
\label{sk_scaleform}
S({\bf k},t) = \xi(t)^{2 \chi + d} \; \mathcal{S} \left( \xi(t) |{\bf k}|
\right)
\end{equation}
of the power spectrum becomes evident. As in Equation (\ref{structfact}), we
have $\mathcal{S}(\kappa) \sim \kappa^{-2\chi-d}$ for $\kappa \gg 1$, and
$\mathcal{S}(\kappa) \equiv \mbox{const}$ in the opposite limit.

Often the height difference correlation function
\begin{equation}
\label{G_vs._S}
G_2({\bf x},t) \equiv \langle | h({\bf x},t) - h(0,t)|^2 \rangle =
\int_{\bf k} S({\bf k},t) \left( 1 - \cos({\bf k \cdot x}) \right)
\end{equation}
is measured. It is self-averaging (unlike $S({\bf k},t)$), so ``smooth
curves'' are easily obtained from spatial averaging, but one tends to neglect
that the initial power law increase for $| {\bf x} | \ll \xi(t)$ reflects only
the {\em local} roughness exponent $\chi_{\rm loc}$ which due to convergence
of the integral in Eq.\ (\ref{G_vs._S}) is restricted to values $\le 1$
\cite{Leschhorn_1993b,Krug_1997}. 
Thus, to study a super-rough interface with $\chi > 1$, the power law
decrease of $S({\bf k},t)$ should be analysed instead.

\begin{figure}[h]
\vspace{10cm}
\includegraphics{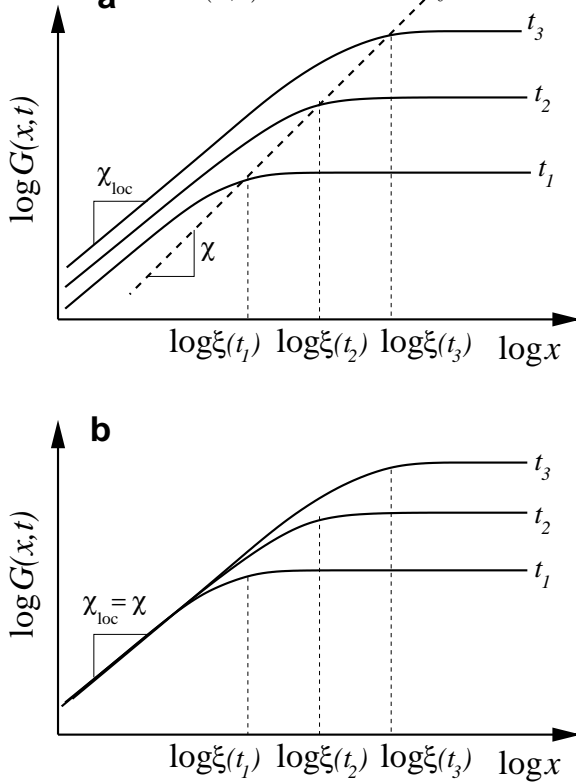}
\caption{Schematic plot of height difference correlation functions $G_2(x,t)$
  of an interface at times $t_1 < t_2 < t_3$ in double logarithmic scale. Part
  {\bf a} illustrates anomalous scaling of a super-rough interface with $\chi >
  \chi_{\rm loc} = 1$, part {\bf b} normal scaling with $\chi = \chi_{\rm
  loc}$.}
\label{gx_scheme}
\end{figure}

Figure \ref{gx_scheme} contains a schematic representation of $G_2({\bf x},t)$
for a super-rough (part {\bf a}) and a normally rough (part {\bf b})
interface. In part {\bf a} it becomes clear why measuring the initial power
law increase does not yield the global roughness exponent $\chi$.

\subsubsection{Temporal interface fluctuations}
In the same way as in the spatial case, interface fluctuations can be studied
in the course of time. We have already seen the increase of the interface 
width with time, $w(t) \sim t^\beta$ with $\beta \equiv \chi/z$ in Eq.\
(\ref{wt_scaling}), valid for early times as long as $\xi(t)$ is smaller than
the lateral size of the system containing the interface.

Analogous to $S({\bf k},t)$ and $G({\bf x},t)$ one can construct quantities
reflecting the temporal fluctuation of the interface height above a substrate
point ${\bf x}$. These are conveniently defined for stationary processes at
late times in a finite system, i.e., when $\xi(t) = L_{\rm system} < \infty$,
and contain, e.g., the power spectrum
\begin{equation}
\label{PowerSpectrum}
S({\bf x},\omega) \equiv \langle | h({\bf x},\omega) |^2 \rangle
\end{equation}
of the Fourier transform of the height field in time ($t \to \omega$), or the
correlation function of the $q$th moment of height ``jumps'' over a temporal
distance $t$
\begin{equation}
\label{Ct}
C_q({\bf x},t) \equiv \langle | h({\bf x},t+s) - h({\bf x},s)|^q \rangle^{1/q}.
\end{equation}

Generally one finds an increase $C_q({\bf x},t) \sim t^{\beta_q}$ 
over ``short''
time distances $t$. In the case of {\em (conventional) scaling} it is related
in a simple fashion to the early time increase of the width in Eq.\
(\ref{wt_scaling}), i.e., for all $q$-moments $\beta_q \equiv \beta =
\chi/z$. For practical purposes $C_q({\bf x},t)$ is easier to measure than
$w(t)$ at early times and therefore this relation is often used.

This simple picture does not hold if height fluctuations are intermittent,
e.g., if the interface advances in avalanches with a non-trivial size
distribution. In the experimental context this is known as Barkhausen noise,
originally measured for the motion of boundaries between domains of different
magnetisation in a solid with disorder \cite{Barkhausen_1919}. It is not surprising
that imbibition into a disordered medium also exhibits multi-scaling in time,
and it will play a role in several parts of this Review.

\subsubsection{Brief history of roughness}
\label{KinroughEnd}
Roughness of interfaces and surfaces has been studied in many contexts and on
all scales, and the interest in the roughness of imbibition fronts was based
on these previous studies to a large extent. To name a few examples: cloud
fronts, mountain rims, shore lines, rivers, forest ranges, sandpiles,
biofilms, cell colonies, cancer, grain boundaries in polycrystals, crystal
surfaces, crystal terrace steps, deposited thin films, flux lines in
superconductors, fracture fronts \dots  (see again e.g.
\cite{HalpinHealy_1995,Barabasi_1995,Meakin_1998,Sornette_2000}).

The theoretical interest arose from the observation that many completely
different roughness phenomena exhibit the same (or similar) scaling exponents,
and therefore be understood as belonging to the same ``universality
class''. Consequently a main theoretical emphasis has been to find the
essential mechanisms behind the different universality classes by constructing
the simplest possible models with the same roughness properties as a given
observed process in nature.

To use a simple ``classification'' there are two kinds of models: Continuum
equations, in most cases stochastic partial differential equations, are more
suitable to an analytical approach, either by exact solution or by
renormalisation and perturbative approaches. On the other hand, lattice or
particle based models are well suited for simulations, often on large scales.

Among the continuum equations we name only the most classic ones. First, the
linear Gaussian model of Edwards and Wilkinson (EW equation)
for a surface relaxing by
surface tension or downhill surface current and perturbed by noisy forces
\cite{Edwards_1982}, which can in the stationary state be said to
be in equilibrium. Second,
introducing a preferential direction of growth to the surface, Kardar, Parisi,
and Zhang (KPZ) constructed a true non-equilibrium model. It (as they noted)
is equivalent to Burger's equation for an vortex-free, compressible, randomly
driven fluid \cite{Kardar_1986,Burgers_1974}. It reads 
\begin{equation}
\label{kpz}
\partial_t h({\bf x},t) = \sigma \nabla^2 h({\bf x},t) 
+ \lambda (\nabla h({\bf x},t))^2 + \eta(x,t),
\end{equation}
where $\sigma$ is a surface tension, and $\lambda$ measures the 
strength of the celebrated KPZ nonlinearity, proportional to
the squared local slope of interface. In the case $\lambda\equiv0$
one obtains the EW case. $\eta(x,t)$ denotes here a simple
delta-correlated ``thermal'' noise field, that roughens the interface.
One has thus the correlator  $\langle \eta(x,t)) \; \eta(x',t') \rangle = 
2 T \delta(x \! - \! x') \delta(t\! - \!t')$.
The main ingredient of Eq.~(\ref{kpz}) is that the interface always
has a non-zero normal component of the local growth velocity, with
respect to the local interface orientation. Also, the coupling between
the noise field $\eta$ and the non-linearity gives rise to the
much-studied complex dynamics and morphology of the KPZ class. As it
is in some sense the most fundamental non-equilibrium field theory this
equation has gained much interest and was often thought to be the
correct large scale description for most experimental and model
systems for surface roughening, among them for imbibition ---
erroneously as we shall see in this review.

Similarly, among the discrete models we briefly mention the family of
(driven) lattice gases and equivalent deposition  or solid-on-solid models for
crystal surfaces, with evaporation or surface diffusion. Discrete in the
number of deposited particles, but continuous in space and time are, e.g.,
ballistic deposition models. A pedagogical overview of growth models can be
found in the lecture notes of Krug and Spohn \cite{Krug_1992}. Covering only
the earlier years of studies on kinetic roughening it nevertheless presents
the types of models which were relevant for the approach of non-equilibrium
statistical mechanics to imbibition and which are presented in this Chapter.

\subsection{Interfaces in disordered media and avalanches}
\label{avalanches_sect}

As already stated, there are several cases where the interface separates
regions with common characteristics, distinguished by the field $\phi$
introduced in Section~\ref{kinrough}. The example of a step separating 
two terraces was used, but interfaces will also arise between 
magnetic domains, or in paper around regions that are either burned or
wet; in some alloys, it can be the boundary between grains with
different orientation of the crystal structure, or even with
completely different structure, such as in martensitic materials.

These interfaces can move either spontaneously, under thermodynamic
forces, or else through an applied force, such as a magnetic field or
a pressure difference. The essential point however is the thermal
influence of the environment is often negligible; fluctuations in the
interface arise only from the presence of impurities or defects in the
material structure itself. These impurities are quenched, i.e., static
in time and the disorder felt by the interface is thus intrinsically
dependent on its {\it position}  itself. For example, the KPZ
equation, Eq.\ (\ref{kpz}), is modified to 
\begin{equation}
\label{quenched_kpz}
\partial_t h({\bf x},t) = \nu \nabla^2 h({\bf x},t)
+ \lambda (\nabla h({\bf x},t))^2 + F + \eta(x,h(x,t)),
\end{equation}
where $F$ is the force acting on the interface and the quenched nature
of the disorder is reflected in the properties of the noise
correlator. Two broad classes are usually distinguished: Random Bond (RB)
disorder, where the correlator of $\eta$ arises essentially from
the fluctuations of the surface tension (like in an Ising magnet with
random coupling constants). More important is  Random Field (RF)
disorder, also since in the QEW case discussed below (Eq.~(\ref{qew_eq}))
it can be explicitly shown that the Random Bond and Random Field
correlators renormalise to the same under rescaling.

In the RF case, the effectively felt noise depends directly on the
position of the interface and has correlations
\begin{equation}
\langle \eta(x,h(x,t)) \eta(x',h(x',t')) \rangle = \Delta
(h(x,t)-h(x',t')) \delta(x-x')
\end{equation}
where $\Delta(u)$ is a function which is strongly peaked around $u=0$, 
often taken to be a delta function
\begin{equation}
\Delta (h(x,t)-h(x',t')) \rightarrow \delta (h(x,t)-h(x',t'))
\label{quenched_random_field} \end{equation}

For an interface moving at some velocity $v$, we can separate the average
motion from the fluctuations $h(x,t) = vt + \delta h(x,t)$. For length scales
$l_{\rm th} > (vt)^{1/\chi}$, the quenched noise correlator
\begin{equation}
\delta (h(x,t)-h(x',t')) \rightarrow \delta(v(t-t')+\delta h) 
\rightarrow \frac{1}{v} \delta(t-t')
\label{crossover}
\end{equation}
and the interface effectively feels an annealed thermal noise with 
effective strength $\tilde{\Delta} = \Delta_0 / v$. 

\subsubsection{Avalanches}
\label{avalanche}

On length scales below $l_{\rm th}$, the quenched nature of the noise
is fully felt and a finite threshold force $F_c$ is needed to 
get the interface in motion. The velocity then shows critical 
behaviour $v \sim (F-F_c)^{\theta}$ until $F \gg F_c$ for which
$v \propto F$. Equation~(\ref{quenched_kpz}) is then just one example
of a system which shows a {\em depinning transition} when a control
parameter, in this case the force  $F$, is tuned. 

The physics in this case is best discussed in terms of avalanches, and
it is useful to first consider
the limit of vanishing nonlinearity $\lambda \rightarrow 0$ of
Eq.\ (\ref{quenched_kpz}), the
so-called quenched Edwards-Wilkinson equation (QEW),
\begin{equation}
\partial_t h(x,t) = \nu \nabla^2 h(x,t) + F + \eta(x,h(x,t)),
\label{qew_eq}
\end{equation}
which is appropriate if the anisotropy ($\lambda$) in the dynamics has
a ``kinematic'' origin and vanishes in the limit $v \rightarrow
0$. See also the discussion in subsection~\ref{cellular} about the
origins of such behaviour and Figure~\ref{alb98} in
Section~\ref{Experiment}.

The case $\lambda=0$ has received lots of attention in the disordered
systems community since it presents a stereotype for the
renormalisation group treatments of such systems and associated
problems. The QEW-equation is also an interesting candidate for a
local description of liquid propagation in porous media since it
contains three very important ingredients of the problem: surface
tension, driving force, and variations in the local (capillary) force,
e.g., due to pore structure fluctuations. This was indeed noticed in
the mid-eighties by Koplik and Levine \cite{Koplik_1985} who extended
ideas of domain wall dynamics in random field Ising magnets
\cite{Bruinsma_1984} (see also \cite{Kessler_1991}). And indeed, we
shall show below that within the length scale $\xi_{\times}$
(Eq.\ (\ref{cross})), the physics of imbibition is very similar to the
physics that comes from Eq.\ (\ref{qew_eq}).

It is also worth keeping in mind that spontaneous
imbibition exhibits something akin to a velocity-dependence: 
an interface that slows down experiences different average velocities
depending on the distance to the reservoir. Thus, tilts
in the interface should affect the local average velocity,
again. 

For forces just slightly larger than $F_c$, 
the interface dynamics in the QEW model is characterised by
{\em avalanches}. This implies typically that large regions
of the interface stay quiescent, while only parts move in
a coherent manner. This is described
by  the correlation length $\xi_\parallel = (F-F_c)^{-\nu}$, 
which is related to the fluctuations of the interface 
via a roughness exponent such that $\xi_\perp \sim
\xi_\parallel^\chi$. Temporal scaling, $\xi_\parallel \sim t^{1/z}$,
is also assumed, and the length- and time scales are related via the
relation 
\begin{equation}
\theta=\nu(z-\chi),
\end{equation}
defining the order parameter (velocity) exponent \cite{Leschhorn_1997,Nattermann_1992,Narayan_1993}.

Since the disorder that the interface ``experiences''
has a history effect, it is clear that it is non-trivial to compute
the correlator of the disorder in a coarse-grained
QEW, as already was noticed in the
beginnings \cite{Koplik_1985}. Later functional 
renormalisation group calculations have resolved
this issue by pushing perturbation theory up to second order.
For uncorrelated disorder
the original one-loop functional RG results
read $\chi = (4-d)/3$, and $z = 2 - (4-d)/9$,
with $\nu = 1/(2-\chi)$. These,  
together with the $\theta$-exponent, manifest 
the fact that there is only one temporal and one spatial
scale at the critical point.  Later work by Chauve, LeDoussal,
and Wiese on much more tedious higher-order
RG expansions \cite{Chauve_2001,LeDoussal_2002}
have yielded corrected expressions for the exponents.

For example with $\epsilon \equiv 4-d$ the
original $\zeta =\frac{\epsilon}{3}$ changes to
$\zeta =\frac{\epsilon}{3} (1 +0.14331 \epsilon +\dots)$.
In particular in $d=1$, this means that the RG beyond
one loop is better able to adhere to the numerical results 
\cite{Leschhorn_1993a,Leschhorn_1997,Nattermann_1992,Rosso_2001}. 
These imply that $\chi_{{\rm QEW},d=1} \sim 1.2 \dots 1.25$, i.e.,
the pinned interface is super-rough. This approximate
value makes for an interesting piece of numerology, since the exponent
agrees with the phase field model result for the 
global roughness exponent as discussed below.

The usual discussion of depinning in the QEW
takes place by considering the limit $F \rightarrow F_c^+$
and concerns the values of such critical exponents.
One may of course ask questions about the behaviour 
for a slightly larger driving force, or
in particular consider the {\em avalanches} right at
the critical point. For $F>F_c$ one has a cross-over
to the thermal EW on large enough scales, Eq.\ (\ref{crossover}), 
and for the avalanches it holds that they start to overlap,
eventually, signalling such a cross-over. It is possible
that one can define here effective exponents \cite{Koplik_1985},
perhaps analogous to imbibition in many ways. For $F\leq F_c$
in a finite system the velocity goes to zero, into an 
``absorbing state''.

Avalanches can be observed in the vicinity of
a critical point (for a general discussion about
phase transitions with an absorbing state see 
\cite{Marro_1999}, and for an overview of 
avalanches in similar, critical systems \cite{Paczuski_1996}) 
by imposing a suitable ensemble.
With the caveat from above (overlap of avalanches)
there are two set-ups that give rise to well-defined
avalanches: 
\begin{itemize}
\item
{\em constant velocity} ensemble, resembling
forced fluid flow imbibition,
\item
{\em self-organised} critical ensemble, induced
by a combination of slowly increasing $F$ and
boundaries that pin the interface.
\end{itemize}
The former is useful in this context since it
exhibits translational invariance; one may thus
proceed to study in this ensemble the statistics of avalanches
and the character of the critical point, as e.g.\
\cite{Narayan_2000,Lacombe_2001}. SOC, self-organised
criticality was coined originally by Bak, Tang
and Wiesenfeld \cite{Bak_1987}, in a context
that is actually directly related to an interface
description or model of charge-density waves. 
Recent work has expanded and clarified much the
picture of SOC as an interface depinning transition
\cite{Alava_2001,Alava_2002} or as an absorbing
state phase transition as such \cite{Dickman_2000}.

The geometric picture of ``avalanches'' considers
a generalised situation where the interface
has ``pinning paths'', akin to paths on the backbone of a percolation
cluster. Assuming that there is a ``punctuation event'' 
if the interface is pushed at any particular spot, 
the interface is released and propagates locally till
another connected pinning path is encountered.
The geometric properties of avalanches set now
the size distribution of avalanches: it is determined
by the usual critical exponents and (except perhaps
in the SOC ensemble) does not involve other physics.
The important quantity is the scaling of voids, between
two such paths that pin the interface, or their
relation of size vs.\ area, which immediately
involves the roughness exponent. The size distribution
of avalanches is then expressed as
\begin{equation}
P(s) \simeq s^{-\tau_s} f(s/L^D).
\label{ps}
\end{equation}
where, if {\em translational invariance} exists (see above),
$D = d + \chi$ and it usually holds that $\tau_s=2-2/D$ \cite{Alava_2003}.
 
In the DPD case \cite{Huber_1995} (see next Section and
Fig.~\ref{dpd_fig}), it is possible to derive the avalanche size
exponent $\tau_s$ in $d = 1 \! + \! 1$ dimensions as 
\begin{equation}
\tau_{s} = 1 + (1/(1+\chi))(1-1/\nu).
\end{equation}
Similar scalings can be attempted for the SOC case.

Note, finally, that such scaling laws as Eq.~(\ref{ps}) can be
established for the duration and support as well, not only for the
size or area/volume. The presence of long-range interactions instead
of the Laplacian surface tension in Eq.~(\ref{qew_eq}) might present 
some interesting twists \cite{Tanguy_1999}.

\subsection{Cellular automata for imbibition}
\label{cellular}

Since the apparent failure of the usual kinetic roughening
framework to account for the large value
of the roughening exponent was partially due to the thermal nature 
of the disorder, specific models were then developed to acccount for the
influence of {\em quenched} disorder. In particular, the
``Directed Percolation Depinning'' (DPD) \cite{Buldyrev_1992a,Amaral_1994} 
model completely ignores all effects of liquid conservation to focus only on 
the quenched nature of disorder. The algorithm that advances the front in
this model is essentially similar to the sandpile (and forest fire) models
\cite{Bak_1987} and has three main ingredients, also illustrated in Figure
\ref{dpd_fig}:

\begin{figure}[ht] 
\includegraphics[width=8cm]{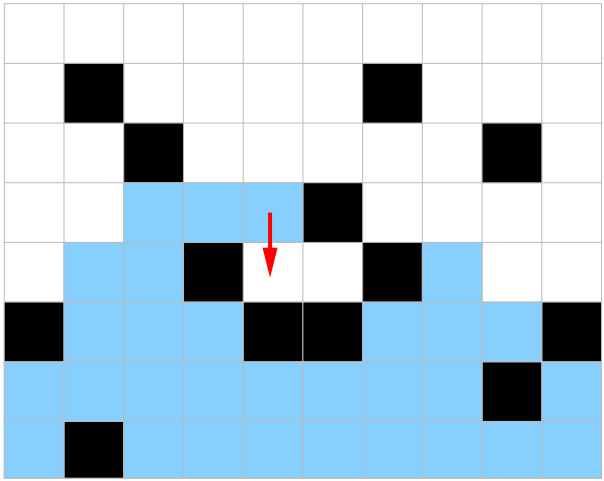}
\caption{Schematic representation of the DPD model after
\cite{Buldyrev_1992a,Amaral_1994}. The medium is divided into regular cells.
Blocked cells are black, invaded (wet) cells grey, invadable (dry) cells white.
Any dry cell adjacent to a wet one can be invaded in the next time step. The
arrow indicates a cell which gets wet immediately in order to erase an
overhang.}
\label{dpd_fig}
\end{figure}

\noindent
{\bf (i)} The porous medium is divided into cells that are either
blocked or open for fluid flow. This model was first applied to
imbibition of ink into paper, the ink pigment then being instrumental
in blocking the flow progression. 

\noindent
{\bf (ii)} The progression of the front takes place only in open cells, 
with overhangs being erased instantaneously. A later development of the
model introduced a linear gradient in the density of blocked cells to
phenomenologically mimic the effects of evaporations through an increasing
difficulty of propagation \cite{Amaral_1994}.

\noindent
{\bf (iii)} The progression of the front stops when no neighbouring open cell
is available, i.e., when a percolating line of blocked cells is encountered.
If $p$ represents the probability of a cell being blocked, interface pinning
occurs at the percolating threshold $p_c \approx 0.47$.

The statistical properties of the interface are related to the
properties of the percolating path. Close to $p_c$ the directed
percolating path is characterised by the parallel and perpendicular
length scales $\xi_{\|} \sim | p- p_c|^{-\nu_{\|}}$ and $\xi_{\bot}
\sim | p- p_c| ^{-\nu_{\bot}}$, with values $\nu_{\|} \approx 1.7$ and
$\nu_{\bot} \approx 1.09$. The saturated width of the interface then
is $ W \sim \xi_{\bot} \sim \xi_{\|}^{\nu_{\bot}/\nu_{\|}}$, which
defines a roughness exponent $\chi = \nu_{\bot}/\nu_{\|} \approx
0.63$. It was argued that the right continuum description of this
automaton is the quenched KPZ equation, given above in Eq.\
(\ref{quenched_kpz}).

The physical justification to the main feature of the DPD model, the
erosion of overhangs has already been approached by experimentalists
\cite{Horvath_1995}. It is plausible that propagation of the fluid in
a direction parallel to the reservoir, if it has a chance to occur, will be
favoured over the motion of the fluid away from it, since this does
not result in any change in the average capillary pressure. In that
sense, any overhangs will eventually be removed, although the question
of the different time  scales involved in such dynamics is certainly
extremely complex. As shall be seen below, phase-field models, which
include this possible effect do not show any kind of DPD roughening.
Other specific predictions of the DPD or quenched KPZ class would
be that of the DPD critical point, a front moving with constant velocity has 
an effective roughness exponent $\chi = 0.7$ and a dynamical exponent $z = 1$.
Leschhorn \cite{Leschhorn_1996} has moreover shown numerically that the DPD 
model consistently gives rise to multi-scaling.

A variation of the DPD idea, introduced by Sneppen \cite{Sneppen_1992},
allows invasion always at sites of lowest resistance. An interesting
aspect of this model is temporal multi-scaling \cite{Sneppen_1993}, due
to avalanche motion \cite{Olami_1994,Leschhorn_1994}. Avalanches are
also present in the process of spontaneous imbibition, although
the conservation law imposes a natural cutoff on their size and
distribution \cite{Dougherty_1998}. It is nevertheless
interesting that a similar lack of temporal scaling is also seen
in the phase field model of imbibition, thus surviving the
introduction of a conservation law. We refer to that in Section 
\ref{phasefield} below.

The DPD model in its simplest form may or may not describe well the
statistical properties of a pinned interface. However, this 
model cannot describe the whole dynamical motion of the interface,
since it does not account for liquid conservation. It already fails by
predicting a {\em constant} average interface velocity, contrary to
Washburn's law. This also holds for dyed liquids, for which the DPD model
was originally developed, since the conservation law governing the motion
of the fluid must be reflected on the motion of the dye particles.
We conclude that any such {\em local} interface equation, as for instance the
QKPZ one, is not appropriate for imbibition. Note that this extends
to interface equations where the coupling between the interface
points is non-local \cite{Mukherji_1997,Tanguy_1999,Kechagia_2001}.

\subsection{Microscopic simulations} 
\label{Network}
The normal alternative approach to imbibition is based on pore network models.
The porous medium is represented by an assembly of pores and throats
connecting pairs of pores. A simplified case is a network (in 2d) of 
horizontal
and vertical capillary tubes with a stochastic distribution of
cross-sections. The nodes between the tubes are structureless in order
to avoid all the possible complications described in Chapter~\ref{Experiment}
or by Lenormand in \cite{Lenormand_1990}. With this
simple stucture, the flow within each capillary is of Poiseuille type
and the progression of the overall front is dictated by the global
pressure difference and local fluid conservation at each node of the
network. 

Note that there are recent lattice-Boltzmann simulations
of multiphase flows that have also tackled flow in porous media
and imbibition.
The main advantage of the method are easy-to-implement
boundary conditions. Other phases such as surfactant are easily
added and questions such as Onsager transport coefficients can be
analysed from a first-principle viewpoint. Actual examples both in 2d 
and 3d however highlight
the technical challenges \cite{Maillet_2000,Love_2001} associated with
the method. A natural idea is to study the front properties exactly
as for coarse-grained models and the authours present
some examples of rough interfaces with unfortunately little
analysis that would compare to usual kinetic roughening measures. 
However, it seems to
us that the capacities of such models have to be augmented 
considerably before one can expect ``realistic'' results.

A recently much practised approach is to construct network geometries that 
attempt
to mimic the statistics of the porous media more faithfully via
statistical laws for the pore sizes, and fluctuating connectivities
(pore throats) for neighbouring pores. Other dynamical
aspects (behaviour of residual gas pockets related to saturation, 
prewetting layers, Bosanquet-flow --- see Section~\ref{Experiment})
can also be added
\cite{Bernadiner_1998,Mogensen_1998,Blunt_2001,Carmeliet_1999,Constantinides_2000,Ridgway_2002a,Ridgway_2002,Schoelkopf_2002,Chang_2002}.
Simulations can be either quasi-static --- the dynamics is applied at 
the ``weakest link'', and no other processes are allowed ---
or dynamic so that the pressure balance is maintained
dynamically at the single pore level 
\cite{Koplik_1985,Dias_1986,Chan_1988,Mogensen_1998,Blunt_1990,Blunt_1992,Dahle_1999,Hughes_2000}.
Such simulations are closer to describing the reality of flow through porous media but
are naturally heavier computationally. In addition, quasi-static simulations 
have no time scale, which poses a genuine limitation for 
many comparisons with experiments and makes studies of kinetic
roughening impossible. Only for dynamical simulations are 
time-dependent mass flows included explicitly using
Darcy's laws and possibly more complicated effects on
the pore level \cite{Hughes_2000,Blunt_2001}.

Both kinds of models have already been used extensively to study the forced
and slow displacement of a wetting liquid by a non-wetting liquid,
often in connection with oil recovery techniques. Generally there are
difficulties for such methods to study interfacial dynamics since the
wetting front is a very non-compact object - one has to pay attention
to the definition of the interface. Questions that are of relevance
from the interface viewpoint and that have been addressed include
pore-scale phenomena as broad pore size distributions
\cite{Schoelkopf_2000a} and the role of the fluid viscosity ratio of
the two fluids in controlling micro-fingering for low $C_a$ \cite{Vizika_1994},
studies of relative permeablities \cite{Maximenko_2000}. 
Of major interest is the use of micro-structural parameters
(from tomography or from other experimental sources) and
the analysis of correlations in the pore scale
\cite{Jeraud_1990,Bryant_1993,Oren_1998,Mani_1999,Tsakiroglou_2000,Knackstedt_2001,Sok_2002,Arns_2003,Hilpert_2003,Oren_2003}.
Typical quantities considered
are relative permeabilities and capillary pressures as well as saturation
curves. These all tend to highlight the fact that {\em in the presence
of correlations} the imbibition-related quantities change
much more readily than, e.g., single-phase permeability.

Concerning ``interfacial aspects'' it is of course true
that any mass flux into the system in imbibition arises
due to and is controlled by the interfacial area between the two different
phases. Thus it is a natural quantity to compute, and
consider \cite{Gray_1991}. A related problem is the
presence of initial saturations \cite{Jeraud_1990}, 
and some network simulations have been done to address
exactly this question \cite{Reeves_1996,Gladkikh_2003,Nordhaug_2003}.

\begin{figure}[h]
\includegraphics[width=7cm]{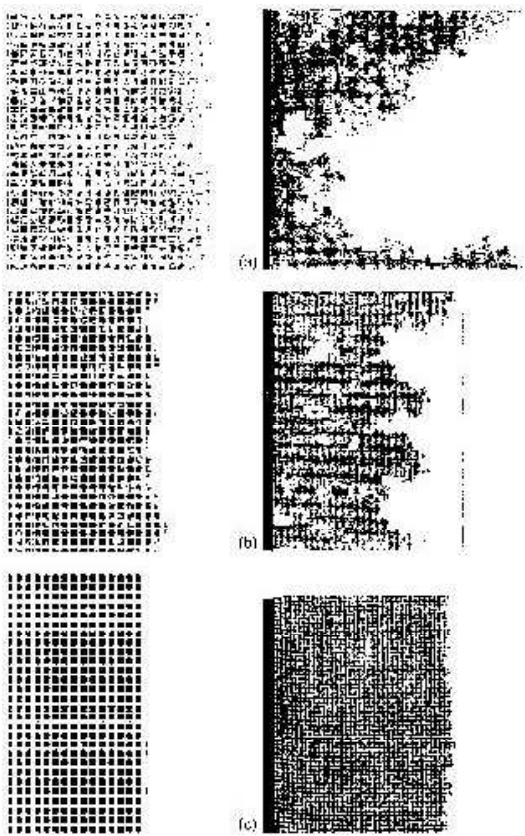}
\caption{Simulations of a dynamic network model vs.\
experimental results from a laboratory channel network
\cite{Hughes_2000,Lenormand_1984}. The capillary
numbers are $C_a = 6 \times 10^{-7}$,
$C_a = 1.4 \times 10^{-5}$, and $C_a = 3 \times 10^{-4}$,
in the simulations, from top to bottom. Notice the transition
to a compact front with increasing capillary number.
The ratio of average pore throat conductivity to
that of wetting layers is about 2000, controlling
the ``percolation'' of the fronts.}
\label{dynamfront}
\end{figure}

Direct interfacial
studies in network models \cite{Blunt_1992,Hughes_2000}.
Given complicated --- or realistic enough --- rules the 
boundary between completely non-wet (occupied by the
non-wetting fluid) and partially saturated regions
can be quite complex due to the effects discussed already
in detail (film flow, initial presence of wetting fluid etc.).
This is of course the strength of network models. Figure~\ref{dynamfront}
illustrates the problems that ensue even in qualitative
comparisons with simplistic laboratory setups.
The generic features in the Figure show the agreement between 
dynamical network simulation results and the experimental snapshots
as the fronts get more and more complex with decreasing $C_a$.
Nevertheless, it is clear that the similarity is mostly
superficial (given the many free choices in the details of
such simulations as here). It would seem to be of both practical and
theoretical interest to apply such network models to at least
two tasks at hand.

The first concerns analysing the detailed
dynamics of the saturation in the presence of relatively
well defined fronts --- what is the actual coarse-grained driving
mechanism here, if the mass flux is used as the ``order parameter''?
Second, the morphology of such fronts, the noise affecting
them and the relation to coarse-grained models as the phase
field approach seems like an interesting avenue, in particular in
the presence of pore-level details as snap-offs and film flow. We underline
that the first of these suggestions has to do with the very
basic physics of imbibition, while the second would bridge
the gap between any analytical theory and models that contain
much of the microscopic effects actually in action.

\begin{figure}[ht] 
\includegraphics[width=8cm]{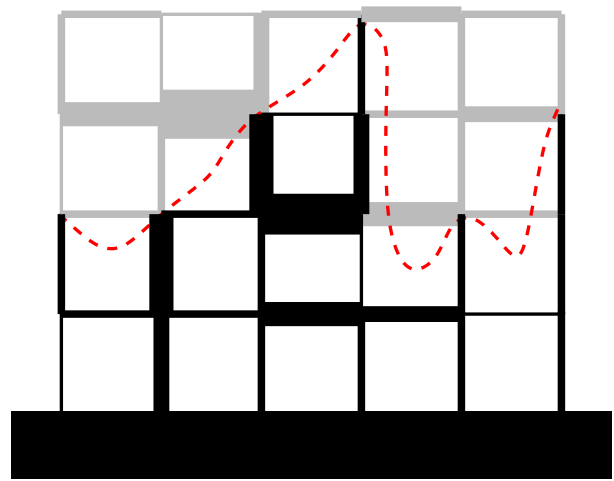}
\caption{Schematic representation of the pipe network model of
Lam \& Horv\'ath \cite{Lam_2000}. A regular grid of pipes is invaded
by a fluid. Pipe diameter is random and independent for each pipe, determining
capillary pressure at invasion and conductivity for transport when invaded.
Dark pipes are invaded, grey ones dry. The dashed line highlights
the ``interface''.}
\label{lam_fig}
\end{figure}

To circumvent the difficulty of obtaining well-defined interfaces,
conditional wetting rules, depending on the state of nearest and
next-nearest neighbours may be introduced
\cite{Lam_2000,Blunt_1995}. These produce a more compact front so that
a single-valued interface may be defined. This approach is discussed
below, as it was used by
Lam and Horv\'ath \cite{Lam_2000}, with the goal of reproducing the
experimental results of Horv\'ath and Stanley \cite{Horvath_1995}
discussed in Chapter~\ref{Experiment}. This is a priori not obvious
since one of the main characteristics of this experiment is
progression of the interface different from Washburn's law, $v \sim
H^\Omega$ with $\Omega \approx 1.6 \neq 1$. It was found that
a very specific distribution of pore radii would give rise to the desired
relation of average front height and velocity, although no plausible
physical reason was given. In particular, it is not clear, that what
appears to be some power law, is in fact transient behaviour.  

The scaling behaviour of the temporal fluctuations was found to be 
similar to the experimental results. The temporal correlation function
(defined in Eq.~(\ref{Ct}), here the moment $q \! = \! 2$ is
considered) could be described by the scaling function
\begin{equation}
C(t) = v^{-\kappa} \; f \left(t v^{(\theta_t+\kappa)/\beta} \right)
\label{LH_ct_scaling}
\end{equation}
with exponents $\kappa = 0.49$, $\theta_t = 0.38$, and $\beta = 0.63$.

Spatially, the roughening process shows clear anomalous scaling. The
stationary two-point correlation function (cf.\ Eq.~(\ref{G_vs._S})) 
had the scaling behaviour
\begin{equation}
G_2 (r) = v^{-\kappa} g(r v^{(\theta_l+\kappa)/\chi} )
\label{LH_ge_scaling}
\end{equation}
where $g(u) \sim u^{\chi}$ for small $u$ tending to a constant at
larger values. The values $\theta_l = -0.25$ and $\chi = 0.63$ were
reported. Based on the assumption that a single time scale controls the
roughening process, Lam and Horv\'ath propose the exponent identity
\begin{equation}
\beta = \frac{\Omega}{\Omega+1} ( \theta_t + \kappa ) 
\label{exp_ident}
\end{equation}
which, for both theory and experiment, agrees with about 15 \% error 
margin. 

These results raise several questions from a theoretical point of view.
First, it is unclear why such a particular
choice of the distribution of radii could give rise to universal scaling
behaviour. The exponent $\Omega$ does not seem to depend in any controllable
way on the disorder, so a long transient towards an asymptotic Washburn
behaviour is not to be excluded.

The value $\chi = 0.63$ for the local roughness exponent can be
interesting since it could indicate that the imbibition process belongs
to the DPD class of roughening. The anomalous nature of roughening and
the continuously slowing down of the interface of course do not support
this view, and only a measurement of the structure factor could
would give some further insight.

In fact, the most questionable result is Eq.\ (\ref{exp_ident}), that relates
the different scaling exponents. We shall show below how
this identity conflicts with the picture of imbibition that emerges
from the phase field models, and how the results of Lam and Horv\'ath
can be interpreted within this approach. 

\subsection{Phase-Field Models}
\label{phasefield}

\subsubsection{General concept}
A continuum approach different from interface equations and conceptually
related to the reasoning of Sections~\ref{pheno} and \ref{Network} was
taken by Dub\a'e et al.\ and Hern\'andez-Machado et al.\
\cite{Dube_1999,Dube_2000b,Dube_2001,Hernandez_Machado_2001}. This
approach concentrates on the bulk of the porous medium rather than on
a hypothetical interface equation and presents the following
advantages:

\noindent
{\bf (i)} Interface equations for particular models arise naturally 
from the sharp interface limit of the model. Simulations using the 
phase field approach thus include automatically the relevant interface
equation, with all possible non-linearities. Equally, non-local interactions
between different parts of the interface arise naturally.

\noindent
{\bf (ii)} Disorder is directly introduced at the level
of the bulk of the porous medium and thus does not need to be
postulated in the interface equation. 

The simplest phase field approach to imbibition starts from the 
observation that all coarse-grained descriptions of flow in 
porous media, based on Darcy's Law Eq.\ (\ref{darcy_th}), 
are essentially diffusive. All inertial effects of hydrodynamics are 
neglected and only pressure determines the flux of liquid ${\bf q}$.
The dynamical equation for the fluid fluid concentration $c$ is then defined
through a continuity equation \cite{Scheidegger_1974} 
\begin{equation} 
\frac{\partial c}{\partial t} + \grad \cdot \, {\bf q} = 0
\end{equation}
where the concentration is related to the saturation through the 
porosity $c = s {\cal P}$. The concentration refers to both the wetting
and non-wetting fluids. The two phases are differentiated by the 
field $\tilde{\phi}({\bf r},t)$, defined over
all space and related to the saturation concentration of either liquid or
air as $\phi_e$. The total concentration field over the porous medium is then 
\begin{equation}
c ({\bf x},t)= \frac{1}{2} c_w \left( 1
+ \frac{\tilde \phi({\bf x},r)}{\phi_e} \right) 
+ \frac{1}{2} c_{n} 
\left( 1- \frac{\tilde \phi({\bf x},t)}{\phi_e} \right) 
\end{equation}
where $c_w$ and $c_n$ are the respective concentrations of the 
wetting and non-wetting fluids and  the values
$\tilde{\phi} = +\phi_e (-\phi_e)$ are
associated to the wetting (non-wetting) phases. 

The field is locally conserved, which is reflected in a continuity equation that
controls the dynamical evolution of the field
\begin{equation}
\frac{\partial \tilde{\phi} ({\bf x},t)}{\partial t} + \nabla \cdot 
{\bf j} = 0 
\label{mob_b_basic} \end{equation}
where the current is taken to be proportional to the gradient of a chemical
potential, ${\bf j} = -M({\bf x},\tilde{\phi}) \nabla \mu$, with a
mobility that may be a function of both space and concentration.
The chemical potential has a 
thermodynamic origin, $\mu = \delta F / \delta \tilde{\phi}$, 
with the free energy functional
\begin{equation}
F = \int d{\bf r}  \left[ 
V(\tilde{\phi}) 
+ \frac{c}{2} (\nabla \tilde{\phi})^2
-(\tilde{\alpha}({\bf x})-\alpha_0) \, \tilde{\phi} \right] \, ,
\label{free_energy_alpha}
\end{equation}
and the potential
\begin{equation}
V(\tilde{\phi}) = - \frac{r}{2} \tilde{\phi}^2
+ \frac{u}{4} \tilde{\phi}^4.
\label{2w-potential}
\end{equation}
For $\tilde{\alpha} = 0$, this free energy represents two liquid phases,  
described by the values $\phi_e^2 = r/u$ at chemical potential 
$\mu=\alpha_0$. The double-well nature of the free energy, 
combined with the square gradient term ensures 
that a well defined interface (of width $\zeta \sim (c/r)^{1/2})$ 
exists between the regions with $\phi = \pm \phi_e$. 
The fluids described by Eq.\ (\ref{2w-potential}) have a slight 
compressibility ($\sim  r^{-1}$) but since the model does not 
consider hydrodynamical effects explicitly, this represents only a minor
correction to the overall dynamics of the interface.

Notice that the ``saturation dynamics'' described by the
phase field model is diffusive. Attempts have been made to look at this
aspect separately by Mitkov et al.\ \cite{Mitkov_1998} (see also
refs.\ \cite{Gray_2000} and \cite{Mitkov_2000}). This is similar to
the approach based on Richard's equation that incorporates a 
a non-linear (and non-trivial) mobility  $M_w$ (see Eq. (\ref{general_diffusion}))
usually determined empirically for each particular case \cite{Bear_1990}.

The model is defined in terms of the chemical potential, so as to
make close contact with theories of critical dynamical phenomena 
with conserved field (Model B in the nomenclature of Halperin and
Hohenberg \cite{Halperin_1977}. See \cite{Gunton_1983,Bray_1994} 
for general reviews on phase fields), used to study binary fluids as well as
the liquid-gas transition. The correspondence between pressure $p$
and chemical potential $\mu$ is easily accomplished in this limit through 
$p \sim \mu \phi_e$. Notice that Eq.~(\ref{free_energy_alpha}) is
also a continuum limit of a random field Ising lattice gas with
locally conserved magnetisation. 

Disorder enters the problem naturally though the transport properties
of the porous medium and we can identify three main sources:  

\noindent
{\bf (i) Capillary disorder.} 
The term $\tilde{\alpha} -\alpha_0$, sets the local value of the
chemical potential and thus effectively represents the pressure
difference due to capillary forces,
\begin{equation}
p_c = \phi_e \tilde{\alpha},
\label{alpha-cap-pressure}
\end{equation}
If this quantity is taken to be random it is natural to assume the
following correlations
\begin{eqnarray}
\langle \tilde \alpha({\bf x}) \rangle & = & \bar\alpha > 0 \nonumber \\ 
\langle \tilde \alpha({\bf x}) \alpha({\bf x'})\rangle & - & 
\bar \alpha^2  =  (\Delta \alpha)^2 \delta({\bf x} \! - \! {\bf x'}).
\end{eqnarray}
(If one assumes a Gaussian distribution for $\tilde{\alpha}$, as one usually does
when defining a random variable by its 1st and 2nd moment, one has to exclude
negative values by hand. However, the model results do not depend on the
precise shape of distribution, only on the moments quoted above.)

\noindent
{\bf (ii) Mobility disorder.} 
The local mobility $M({\bf x})$ (neglecting any dependence on the
fluid density $\tilde \phi$), which controls the ease of flow of the
liquid, is related to the permeability of a porous medium through the
relation
\begin{equation}
\frac{M}{\phi_e^2} \equiv \frac{\kappa}{\eta}. 
\label{mobility-permeability}
\end{equation}
and is random and characterised by the correlations (again the
precise shape of distribution does not matter)
\begin{eqnarray}
\langle M ({\bf x}) \rangle & = & M_0 > 0 \nonumber \\ 
\langle M({\bf x}) M({\bf x'})\rangle & - & M_0^2  = 
(\Delta M)^2 \delta({\bf x} - {\bf x'}).
\end{eqnarray}
There is a priori no reason to expect any simple relation between the
two fluctuating quantities and they are kept independent. 
 

Since the chemical potential is analogous to pressure, it also enters
the model in the boundary conditions. We next outline the choice in
two spatial dimensions, the only one studied so far in the
literature. In the case of spontaneous imbibition the chemical
potential obeys the boundary conditions $\mu (x,y \! = \! 0) =
\alpha_0$ (equivalent to atmospheric pressure) and $\partial_y \mu
(x,y \! = \! L) = 0$ (closed at the upper end) \cite{Dube_1999}. The
boundary conditions on the phase field $\partial_y \phi (x, y \! = \!
L) = 0$ and at $y =0$, the field's value is the solution to
\begin{equation}
- \tilde{\alpha} - \tilde{\phi} + \tilde{\phi}^3 = 0.
\label{compressibility} 
\end{equation}
Since the term $\alpha_0$ simply corresponds to a constant shift in the
chemical potential, we consider $\alpha_0 = 0$ from now on. 
In order to model imbibition situations at constant flow rate,
as in \cite{Soriano_2002a}, constant flux boundary conditions
$M \partial_y \mu = \gamma_0$ are imposed. Both situations, spontaneous
imbibition and constant flux shall be examined in details below. 

When the effects of evaporation or gravity are sufficiently strong,
they act in a way that stops the imbibition front and produce a pinned
interface. Both effects can be taken into account in a generalised
imbibition model of the form \cite{Dube_2001}
\begin{equation}
\frac{\partial  \tilde{\phi} ({\bf r},t)}{d t} - \tilde{G}
\frac{\partial \tilde{\phi} ({\bf r},t)}{\partial y} =
\nabla ( M(\tilde{\phi},{\bf x}) \nabla \mu)
 - \frac{1}{2} \tilde{\epsilon} \left( \phi_e+\tilde{\phi}({\bf r},\tau) \right)
\label{pf_gen}
\end{equation}
The convective term is included to describe
gravity. Although gravity can only be introduced properly through an
hydrodynamical field \cite{Kawasaki_1982,Jasnow_1996}, one can argue
that this term reproduces the correct equation of motion for the
average position of the imbibition front. The connection to Darcy's
law is established through
\begin{equation}
G \equiv g \; \frac{\kappa}{\eta}. 
\end{equation}
When $\tilde \epsilon \neq 0$, a non-conserving term is introduced into the
equation of motion, which, to a first approximation, describes
evaporation of liquid (the $\tilde{\phi} = \phi_e$ phase), at a rate
$\phi_e \tilde \epsilon$ and proportional to the total area covered by the
fluid. 
Note, that in this case one thinks of a thin medium from which the fluid
can evaporate on the sides, as in the case of a wet paper sheet.

Taking the mobility to be $M(\tilde \phi,{\bf x}) = M_0 \, m ({\bf x})$
(here we have neglected explicit dependence on the field value), 
the basic scales for length, energy, and time of the problem are 
\begin{eqnarray}
\zeta = \left( c/r \right)^{1/2}, \nonumber \\
\mu_0 = \left( r^3/u \right)^{1/2}, \\
\tau_M = \frac{ \phi_e \zeta^2}{M_0 \mu_0}.  \nonumber 
\end{eqnarray}
Equation~(\ref{pf_gen}) can then be put in a dimensionless form by defining
\begin{eqnarray}
\alpha = \frac{\tilde{\alpha}}{\mu_0}, \nonumber  \\
\gamma = G  \frac{\tau_M}{\zeta}, \\
\epsilon  = \tau_M \, \tilde{\epsilon}, \nonumber 
\end{eqnarray}
which leads to a dynamical equation
\begin{equation}
\frac{\partial  \phi ({\bf x},\tau)}{\partial t} - \gamma
\frac{\partial \tilde{\phi} ({\bf x},t)}{\partial y} =
\nabla (m(\phi,{\bf x}) \cdot \nabla \mu({\bf x},t))
 - \frac{1}{2} \epsilon (1+\phi({\bf x},t))
\label{p_field_tot}
\end{equation}
where the chemical potential
$\mu = -\phi + \phi^3 - \nabla^2 \phi - \alpha ({\bf x})$.

For slow enough motion of the interface the chemical potential is adjusted
quasi-statically to the interface positions \cite{Bray_1994}, which
themselves are driven by the difference between incoming and outgoing current
${\bf j} \! = \! - m \nabla \mu$. In the sharp interface limit, and for a
slowly moving front, the concentration $\phi$ is approximately constant
in the bulk, changing only at the interface. The chemical potential $\mu$
thus satisfies the equivalent of a Poisson equation 
\begin{equation}
\nabla \cdot \nabla \mu  = \epsilon
\label{poisson_eq}
\end{equation}
in the bulk. At the interface, $\mu$ must obey the Gibbs-Thomson
boundary condition
\begin{equation}
\label{gibbsthompson}
\Delta \phi \; \mu \vert_{int} = \Delta V - \sigma \mathcal{K},
\end{equation}
where $\mathcal{K}$ is the curvature, $\sigma$ the effective surface tension
of the model, the miscibility gap $\Delta \phi = \phi_+ - \phi_-$ and 
$\Delta V = V(\phi_+)-V(\phi_-)$. The quantities $\phi_{\pm}$ are the
{\it equilibrium} values of the phase field, defined by the usual tangent
construction \cite{Bray_1994,Langer_1992}. The interface motion is then
determined by the normal velocity 
\begin{equation}
(\Delta \phi) v_n = - \partial_n \mu \vert_{\pm} - \gamma.
\end{equation}
These conditions are of course completely equivalent to the description
based on capillary pressure and Darcy's law in the context of Eq.\
(\ref{darcy_th}) in Section~\ref{pheno}. 

\subsubsection{Interface equation}

To extract the interface equation from Eq.\ (\ref{pf_gen}), it is
convenient to use a local coordinate system $(u,s)$ close to the
interface \cite{Jasnow_1996,VanSaarloos_1998}. Two-dimensional space
is parametrised by ${\bf x}(u,s) = {\bf X}(s) + u \; {\bf \hat{n}}
(s)$, where ${\bf X}(s)$ is a point of the interface, ${\bf \hat{n}}$
is a unit vector normal to the interface pointing towards the side
where $\phi > 0$, and $s$ is the arc-length coordinate along the
interface \cite{Elder_2001}.  In terms of the phase field, this
corresponds to $\phi(u \! = \! 0,s) = 0$. The time derivative of the
field then becomes
\begin{equation}
\frac{\partial \phi (u,s,t)}{\partial t} = V_n (s) \;
\frac{\partial \phi}{\partial u}
\end{equation}
where $V_n (s)$ is the normal velocity of the interface at position $s$.
If the interface thickness, $\zeta = 1$ in dimensionless units,
is much smaller than the typical radii of curvature of the interface
(the sharp interface limit), the Laplacian term of the chemical potential
may be expanded such that
\begin{equation}
\nabla^2 = \frac{\partial^2}{\partial u^2} + \frac{\partial^2}{\partial s^2}
+ {\cal K} (s) \; \frac{\partial }{\partial u}
\end{equation}
where ${\cal K} (s)$ is the curvature of the interface. For $\alpha=0$,
the one-dimensional kink solution is $\phi (u,s) = \phi_0 (u) =
\tanh (u /\sqrt{2})$, with corrections of order ${\cal O}(\alpha)$
from the finite compressibility of the model. Therefore, to first order,
$\mu \approx -\alpha (u,s)  - {\cal K} (s) \partial \phi_0 (u) / \partial u$.
Still in the sharp interface limit $\zeta {\cal K} \ll 1$, the derivatives
of the kink solutions have properties $\ \partial \phi_0 (u) /
\partial u \approx \Delta \phi \, \delta (u)$ and
\begin{equation}
\sigma = \int du \left( \frac{\partial \phi_0 (u) }{ \partial u} \right)^2
 = 2 \, \frac{\sqrt{2}}{3}
\label{surf_tention}
\end{equation}
where $\Delta \phi \approx 2$ is the miscibility gap and $\sigma$ is the
dimensionless surface tension or, for the fully dimensional expression,
\begin{equation}
\tilde{\sigma} = 2 \; \frac{\sqrt{2}}{3} \; \zeta \phi_e \mu_0.
\label{surf_tension_dimension}
\end{equation}  
Note that this quantity represents an average surface tension associated
with the interface and does not necessarily bear any relationship to
the microscopic surface tension that gives rise to capillary pressure.
Nevertheless, assuming $p_c \sim \sigma / r_0$, with $r_0$ the typical pore
size and using Eq.\ (\ref{alpha-cap-pressure}) yields 
\begin{equation}
\alpha = \frac{\zeta}{r_0},
\label{alpha_cap}
\end{equation}
i.e., the dimensionless parameter $\alpha$ corresponds to the ratio
between the interfacial width and the typical pore size. 

The capillary number $C_a = \eta \tilde{v} / \tilde{\sigma}$,
$\tilde{v}$ denoting the fluid velocity, represents the ratio between
the viscous and capillary forces. Strictly speaking, this cannot be
defined in the phase field model since the fluid is not explicitly
taken into account, only the ratio $\eta/\kappa$. Nevertheless, the
capillary number itself does not play a direct role in the roughening
of the interface, but enters the problems only through the length
scale $\xi_{\times}$. Assuming that the permealibility can be related
to the average pore radius $\kappa \sim r_0^2$, it is easy to show
that,
\begin{equation}
C_a \sim \frac{r_0^2}{\zeta^2} \, \frac{v}{\sigma} \label{c_number},
\end{equation}
where $v$ is the velocity of the {\em interface} -- unlike the fluid velocity
$\tilde v$. For spontaneous imbition, this is easily related to the capillary
pressure, from Eq.\ (\ref{alpha_cap}), to the dimensionless parameter
$\alpha$ to give $C_a = v / (\sigma \alpha^2)$.

For constant mobility $m(\phi,{\bf x}) = 1$, the dynamical phase field
equation may be inverted with the use of Green's function, defined by 
\begin{equation}
\label{gr_def}
\nabla^2  G(x,y|x',y') = - \; \delta (x-x') \; \delta(y-y'),
\end{equation}
for the range $-\infty < x,x' < \infty$, $0 < y,y' < \infty$, with
Dirichlet boundary conditions. The half-plane Green's function
\begin{equation}
\label{gr_form}
G(x,y|x',y') = \frac{1}{4\pi} \ln \; \frac{(x-x')^2 +
(y-y')^2}{(x-x')^2+(y+y')^2}
\end{equation}
must be used, since the presence of an reservoir at position $y=0$
breaks the translational symmetry in $y$
This yields the equation
\begin{widetext}
\begin{equation}
\label{pf_invert}
\int \! d{\bf x}' \; G ({\bf x}|{\bf x}') \;
\left( 
\frac{\partial \phi({\bf x}',t)}{\partial t} - \gamma 
\frac{\partial \phi({\bf x}',t)}{\partial t} 
+ \frac{1}{2} \; \epsilon \; \left( 1 + \phi ({\bf x}',t) \right) \right)
=
\mu ({\bf x},t).
\end{equation}
Multiplying  Eq.\ (\ref{pf_invert}) by $\int \! du \; (\partial \phi_0 (u)
/ \partial u)$ then effectively projects the phase field dynamics onto 
the interface $u=0$. A translation $u \rightarrow u + h(s,t)$ then yields
\begin{equation}
\int \! ds' \; G (s,h(s,t)|s',h(s',t)) \; (V_n (s') - \gamma )
+ \Delta \phi \; \epsilon 
\int \! ds' \int_{0}^{h(s',t)} \! \! \! du' G (s,h(s,t)|s',u') 
= \eta (x,h(x,t)) + \sigma {\cal K}. \label{int_eq_tot}
\end{equation}
\end{widetext}
where  $\eta(x,h) \equiv \int \! dy \; \phi_{0}'(y-h(x,t)) \;
\alpha(x,y) \sim 2 \alpha (x,h)$. 

Without disorder, the advancing liquid front interface remains flat,
${\cal K} = 0$, with position described by $H(t)$ and the value of the
chemical potential at the interface $\mu\vert_{\rm int} = - \alpha$,
c.f.\ Eq.\ (\ref{gibbsthompson}). It is then straightforward to obtain
the dynamical evolution 
\begin{equation}
\frac{d H(t)}{d t} =  \frac{\bar{\alpha}}{2H(t)}
 - \gamma -  \frac{1}{4} \, \epsilon \, H(t).
\label{wash-mod}
\end{equation}

Without gravity or evaporation, the interface then progresses in a
typical Lucas-Washburn behaviour, $H(t) \sim t^{1/2}$ while non-zero
values of $\gamma$ or $\epsilon$ eventually stop the interface at an
equilibrium height $H_p (\gamma,\epsilon)$ given by the zero of the
right-hand side of Eq.~(\ref{wash-mod}) with limiting cases $H_{p}
(\gamma,\epsilon \! = \! 0) = \bar{\alpha} / (2 \gamma)$ and $H_p
(\gamma \! = \! 0,\epsilon) = \sqrt{ 2 \bar{\alpha}/\epsilon}$.

At early times $t \ll t^{*} \equiv \min(\bar{\alpha} \gamma^{-2},
\epsilon^{-1})$, the rise of the interface follows Washburn behaviour,
$H(t)= (\bar{\alpha} t)^{1/2}$.
For $t \gg t^{*}$, the pinning height is approached exponentially. In
cases where gravity is absent, the average interface, of initial
height $H(t \! = \! 0) = 0$, is described at all times by
\begin{equation}
\left(\frac{H(t)}{H_p}\right)^2 =  1-e^{-\epsilon t/2},
\label{mf_evap}
\end{equation}
while for $\epsilon=0$, the rise follows the transcendental equation
\begin{equation}
\frac{H(t)}{H_p} + \ln \left( 1-\frac{H(t)}{H_p} \right)
= -\frac{2}{\bar{\alpha}} \gamma^2 t. 
\label{mf_grav}
\end{equation}

Although both gravity and evaporation pin the interface at a
given height, it must be kept in mind that the physics behind
these two effects is quite different \cite{Dube_2000b}.
Spontaneous imbibition, without external influences, is characterised
by a Laplace equation for the chemical potential $\nabla^2 \mu=0$ in
the bulk (i.e., far away from the interface). When solved with boundary
conditions $\mu (y \! = \! 0) = 0$ and $\mu (y \! = \! H) = -\bar{\alpha}$,
this yields a gradient $\partial \mu \propto -1/H$ in the liquid phase.
The gravity term of Eq.\ (\ref{pf_gen}) thus has the effect of
stopping the interface when $\partial \mu \simeq -\gamma$.

On the other hand, the non-conserving term introduced by evaporation is such
that the chemical potential in the bulk must be a solution of the Poisson
equation $\nabla^2 \mu = \epsilon$ and at pinning height, the gradient in the
chemical potential $\partial \mu = 0$. This simply represents the fact
that the flux of liquid from the reservoir exactly balances the losses
due to evaporation.

The time and length scales coming from gravity and evaporation can
however be very different. Studies of capillary rise with light
organic liquids (for which evaporation is presumably an extremely
weak effect compared to ordinary water) imbibed in filter papers
\cite{Gillespie_1958} have found a pinning height
of the order of $1$ m, with time scales on the order of
days. In contrast, evaporation effects can cause the pinning height to be
of the order of $15 - 50$ cm, with correspondingly much faster
time scales.

When capillary disorder is introduced, the interface equation is best
expressed in terms of the Fourier modes of the interface fluctuations
$h_k$. A straightforward expansion of Eq.\ (\ref{int_eq_tot}) yields
the linear equations
\begin{widetext}
\begin{equation}
\left( \dot{h_k} + \frac{1}{2} \epsilon h_k \right)
\left( 1 
- e^{-2 \vert k \vert H} 
\right) 
+ |k| (\dot H + \gamma) \; h_k \left( 1 + e^{-2 \vert k \vert H} \right) 
 = 
\frac{1}{4} |k| \left(\{ \eta \}_k -  \sigma k^2 h_k \right). 
\label{non-local}
\end{equation}
\end{widetext}
The quenched noise enters in a somewhat non-standard way, by a
Fourier transform of the disorder values at the interface,
$\{ \eta (t)\}_k \equiv \int_{x} \! e^{-ikx} \eta(x,h(x,t))$. 
Immediately apparent is the presence of the modulus of the
wave-vector, $|k|$, arising from the conservation law. Non-linear terms,
similar to those in Eq.~(\ref{imbi_non_linear}) obtained by Ganesan and
Brenner \cite{Ganesan_1998} can also be obtained from Eq.\ (\ref{int_eq_tot}).

\subsubsection{Spontaneous imbibition without pinning}
\label{spon_imbi}
In this section, we specialise to the case of spontaneous imbibition, 
where the porous medium is simply put into contact with a liquid reservoir. In
this setup the interface follows the typical square root behaviour
in time for the average interface height, $H \sim t^{1/2}$. 
It is worth mentioning that similar thermodynamical non-equilibrium
states are found in other systems. An example is the invasion
of a type II superconductor by magnetic field vortices. Indeed,
lattice models similar in spirit to the phase field formalism
have been developed, and it is an interesting question how the
understanding of imbibition (dynamics, roughening, continuum
equations) would find applications in that context 
\cite{Jackson_2000,Nicodemi_2001}.

Disorder enters the model in very different ways, and changes in the local
capillary pressure enter directly into the chemical potential through
the Gibbs-Thomson boundary condition. This term is present independently
of the velocity of the interface. On the other hand, it is easy
to show that changes in the chemical potential $\delta \mu$ from its
constant mobility value $\mu_0$ due to changes in the mobility $\delta m$
can be estimated as 
\begin{equation}
\nabla^2 (\delta \mu) \sim \nabla (\delta m) \nabla \mu_0 \sim 
\nabla( \delta m) v_0, 
\end{equation}
which means that the effects of disorder in the mobility are essentially
proportional to the interfacial velocity, a quite intuitive result.

In the case of spontaneous imbibition, the interface continuously slows
down and the effect of mobility disorder becomes smaller and smaller until
finally disorder in the capillary forces completely controls
the roughening process. In this Subsection, we thus consider a constant
mobility $m(\phi,{\bf x}) = 1$, and refer to the implications of mobility
variations in the next Subsection. 

The limit $\gamma \! = \! \epsilon \! = \! 0$ of Eq.\
(\ref{non-local}) shows that there exists a lateral length scale
separating two different modes of damping of the interfacial
fluctuations. The changes in the chemical potential due to curvature
(Gibbs-Thomson effect) and those due to slowing down of the front
balance at a length scale
\begin{equation}
\label{xi_cross}
\xi_\times \; \simeq \; \left( \frac{\sigma}{v} \right)^{1/2} 
 \, = \, \left( \frac{ \sigma H}{ {\bar \alpha}} \right)^{1/2}.
\end{equation}
The front is smoothed on length scales larger than $\xi_{\times}$
since parts of the interface further away from the reservoir have a
slower velocity. Large fluctuations are constantly suppressed. 

A convenient way to study this case is to adapt the model to the
experimental setup invented by Horv\'ath and Stanley
\cite{Horvath_1995} where a steady state is reached by pulling a paper
sheet at a constant velocity ${\bf v} \! = \! -v {\bf \hat y}$ towards
a reservoir, or (almost) equivalently by monitoring the average
interface height to remain at a constant value. In this case, the
equation of motion reads 
\begin{equation}
\label{horvath_eq}
\partial_t \phi = - \nabla^2 \left[ \nabla^2 \phi + \phi - \phi^3 +
\alpha({\bf x}-{\bf v}t) \right] + {\bf v} \cdot
{\mbox {\boldmath $\nabla$}} \phi.
\end{equation}
The corresponding interface remains at a fixed average height
$H = \bar{\alpha}/2 v$, where a freely rising interface would have 
velocity $v$.

A simple spatial discretisation of Eq.~(\ref{horvath_eq}) on a square
 lattice yields satisfactory results. The Laplace operator is
then discretised in the standard way of finite differences
\begin{equation}
\label{disc_laplace_eq}
( \Delta \phi )_{i,j} = \sum_{| (i,j) \! - \! (i',j')| \! = \! \Delta x}
(\phi_{i',j'} - \phi_{i,j}).
\end{equation}
The grid spacing $\Delta x$ has to be adapted to the variation
in the phase field and should not be larger than $1/3$ of
the interface width.

It would be interesting to apply a  finite element
method to imbibition problems which could resolve the
interface on a finer scale and remain coarse in the bulk
phases. One has however to be careful
to keep track of the quenched disorder field $\alpha(x)$ when rearranging
the mesh. Also, a slightly modified version of
the phase field equations, where the disorder acts only
at the interface and not in the bulk is better suited.
This could be achieved by, e.g., changing the noise term
to $(1 - \phi(x))^2 \alpha(x)$.

For the slow diffusive motion an explicit Euler time
integration is also satisfactory, even if the fourth spatial
derivative requires relatively small time steps $dt \sim 0.01$.
Recent developments on semi-implicit schemes, where the
nonlinear terms are treated explicitly, but the linear ones
implicitly, might prove useful for models of this type \cite{Dziuk_2000}.

\begin{figure}[ht] 
\includegraphics[width=8cm]{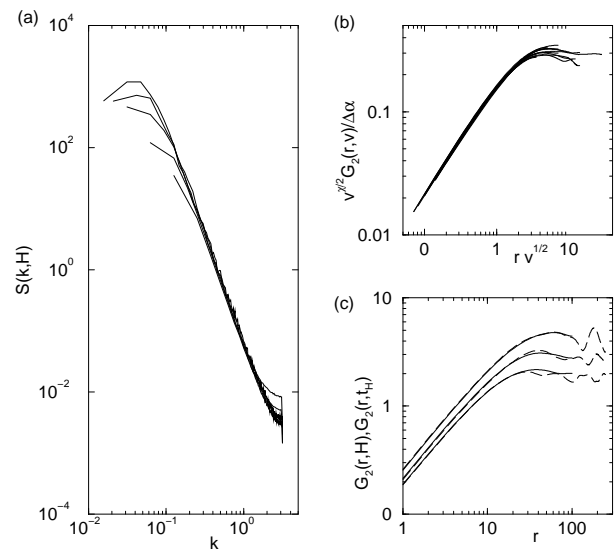}
\caption{(a) Structure factors for systems of height $H \! = \!
50,100,150,200$ and lateral size $L \! = \! 2 H$. Note that for
large system sizes the power spectrum is {\em not} cut off by the system size,
but by some other length scale $\xi_\times < L$. (b) Correlation functions
$G_2 (r,H)$, scaled according to Eq.\ \protect (\ref{g_scaling}) for
various $v \! \sim \! H^{-1/2}$ and $\Delta \alpha$. (c) Comparison of
the correlation functions in the steady state, $G_2 (r,H)$ (solid
lines) for $H \! = \! 25,50,100$ and of the rising case $G_2 (r,t_H)$
(dashed lines), at times $t_H \! = \! H^2/\bar \alpha$.
}
\label{spatial}
\end{figure}

The steady-state structure factor $S(k,H)$ for the stationary setup of
Eq.\ (\ref{horvath_eq}), shown in Fig.\ \ref{spatial}(a), shows that
there is a maximum length scale for the fluctuations, independent of
the  lateral system size, with the global roughness exponent extracted
from the power-law decay $\chi \approx 1.25$. It is slightly
intriguing, that the value is so close to the QEW one, again
(c.f.\ Section \ref{avalanche}).

All the data can be rescaled using $\xi_\times \! \sim \! v^{-1/2} \!
\sim \!(H/\bar \alpha)^{1/2}$. For the second moment height difference
correlation function (defined and schematically shown in Fig.\
\ref{gx_scheme} in Section \ref{kinrough}) one obtains the scaling
function
\begin{equation}
\label{g_scaling}
G_2(r,H) = \Delta \alpha \; v^{-\chi/2} \; g \left(
r \, v^{1/2} \right).
\end{equation}
The dependence of the interface velocity $v$ (and equivalently on the
average height $H$) follows from the above analysis. The relatively
simple scaling dependence on the noise strength $\Delta \alpha$ is
found numerically and shown in Fig.\ \ref{spatial}(b), but a
convincing explanation is to our knowledge missing
\cite{Dube_1999,Dube_2000b}.

The scaling function $g(u)$ is constant for $u \gg 1$ and $g(u) \sim
u^{\chi_{\rm loc}}$ if $u \ll 1$, with $\chi_{\rm loc} \approx 1$
\cite{Dube_1999,Dube_2000b,Krug_1997,Lopez_1997}.

The same scaling picture applies in the freely rising case, where the
average interface height $H(t) = (\bar \alpha t)^{1/2}$ provided that
a dynamical correlation length $\xi_\times \! \sim \! 
(t/\bar{\alpha})^{1/4}$ is used. There appears a complete equivalence
between the interfacial fluctuations at an instantaneous height $H(t)$
and the saturated fluctuations of a stationary interface. The length
scale $\xi_\times$ is thus {\em conceptually different} from the intrinsic
time dependent lateral correlation length $\xi_t$ commonly found in
models of kinetic roughening (see Section \ref{kinrough}).
Here $\xi_{\times}$ merely fixes the
maximum range of correlated roughness and can only occur if the ``natural''
dynamical exponent $z < 4$, so that the interface fluctuations can
always catch up instantly with the available area of correlation.
This also implies a dynamical behaviour of the total width of the
freely rising interface $W \! \sim \!  \xi_\times^\chi \! \sim \!
t^{\chi/4} = t^{\beta}$, with $\beta \approx 0.31$
\cite{Dube_1999,Dube_2000b}. It cannot be overemphasised that the
exponent $\beta$ found here is conceptually completely different from
the usual exponent in traditional descriptions of kinetic roughening
introduced earlier in Eq.\ (\ref{wt_scaling}).

\begin{figure}[ht]
\includegraphics[width=8cm]{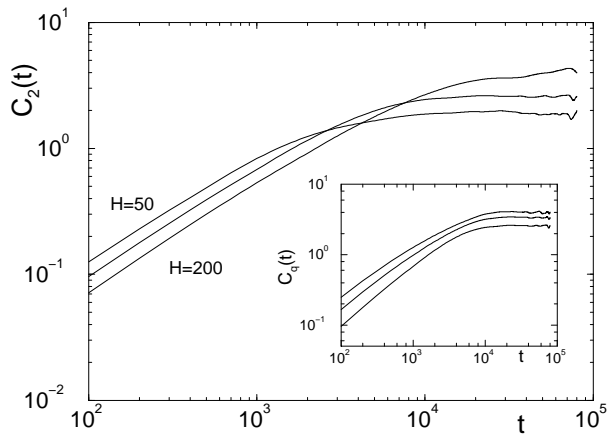}
\caption{Steady state temporal correlation functions for
systems of size $L_x = 400$, at heights $H=50,100,200$. Saturation level
and overall roughness increase with $H$, whilst short time differences
and typical velocities decrease. The inset shows the correlation functions
of moments $q=2,4,6$ (from bottom to top) for $H=100$, which decrease in
their logarithmic slopes. 
}
\label{temporal}
\end{figure}

The temporal correlations in the steady state, obtained from Eq.\
(\ref{horvath_eq}), can be analysed through the $q^{\rm th}$ order
height difference correlation functions $C_q(t,H) = \langle \vert
h(x,t+s) \! - \! h(x,s) \vert^q \rangle^{1/q}$.
For small time differences $t$ there appear logarithmic slopes (that have
been interpreted as ``effective'' exponents $\beta_q$) which are
independent of $H$ and {\em decrease} with $q$,
as shown in Fig.\ \ref{temporal} and its inset. Note that the exponent
$\beta_2 \approx 0.85$ is fundamentally different from the growth exponent
associated with the width of a freely rising front. Standard concepts of multi-scaling
are not easily recovered in this system, even if the only reason is the
restriction in the system sizes simulated.

\begin{figure}[ht]
\includegraphics[width=8cm]{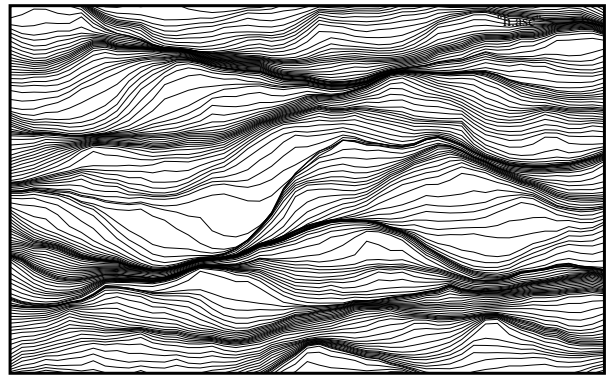}
\caption{An example of an interface in the phase field model propagating
by avalanches. Configurations at equidistant times are plotted
together. Regions with ``avalanche sweeps'' become visible (low
density of lines), as well as pinning zones (high density).}
\label{avalanches_fig}
\end{figure}

It seems however certain that the reason for the decrease of $\beta_q$
is linked to the propagation of the front by avalanches, which in the
QEW equation (\ref{qew_eq}) leads to similar statistics in temporal
correlations \cite{Leschhorn_1994}. In Figure \ref{avalanches_fig} we
show a few dozen interface configurations at equidistant times which
apparently propagate by effective avalanches.
Zones where interfaces get blocked for a longer time appear as thick dark
lines which are separated by regions through which the interface
sweeps in a fast avalanche.

\begin{figure}
\includegraphics[width=8cm]{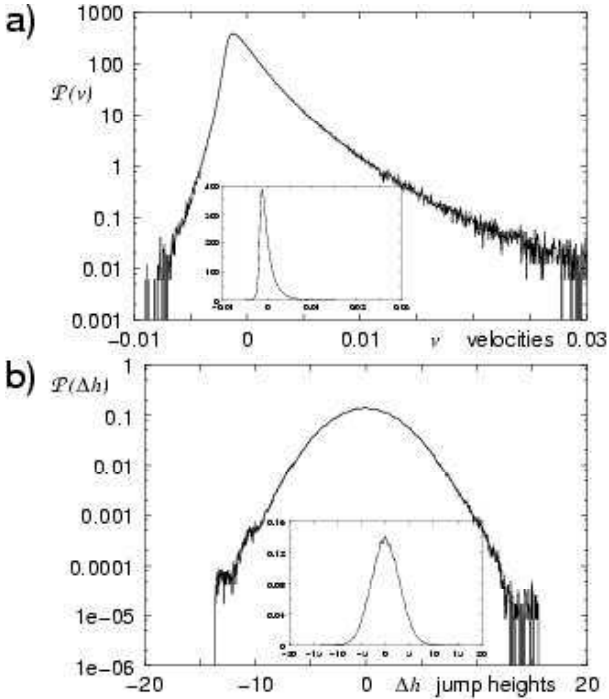}
\caption{a) probability density function of the interface velocity in
a numerical simulation of the phase field model
Eq.~(\ref{horvath_eq}). b) Distribution of height differences taken at
two different times. The insets show the same data on a linear scale.}
\label{velocity_distr}
\end{figure}

The precise nature of the decrease of $\beta_q$ has not been studied
thoroughly. An answer to the origins of such behaviour may lie in the
{\em distribution functions} of the fluctuating variables. In
Fig.~\ref{velocity_distr}a) we show the distribution of the normal
interface velocity in the phase field model. It is kept in a steady
state by shifting the system towards the reservoir at constant
velocity ${\bf v} = - v {\bf e}_y$, as in the experiments of
Horv\'ath and Stanley \cite{Horvath_1995} expressed by the modified
phase field equation (\ref{horvath_eq}).

The velocity distribution is related to the distribution of
differences $h(x,t+\tau)-h(x,t)$ in the height profile in the limit of
small time steps $\tau \to 0$. For large $\tau \to \infty$, when both
profiles are essentially uncorrelated the shape of the height
difference distribution crosses over from that of Figure
Fig.~\ref{velocity_distr}a) to that of \ref{velocity_distr}b). It
seems plausible that the characteristics of this crossover become more
apparent from the probability distributions themselves than from the
behaviour of their moments $\langle | h(x,t+\tau)-h(x,t) |^q
\rangle^{1/q}$. The behaviour of the velocity distributions gives rise
to analogies with other systems (with ``local'' Langevin equations)
that exhibit interesting probability distributions
\cite{Bramwell_1998}. The deviations from Gaussian distributions
relate to the existence of correlations, and make the question
pertinent why self-averaging fails.

\begin{figure}
\includegraphics[width=8cm]{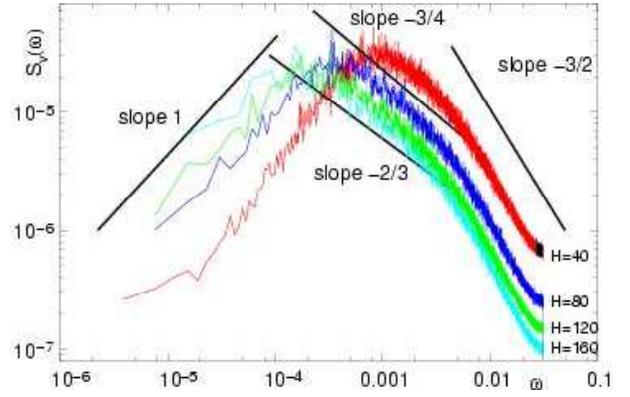}
\caption{Power spectra for the interface velocities in the steady state of
Eq.~(\ref{horvath_eq}). At low frequencies $S_v(\omega) \sim \omega$, at
high frequencies $S_v(\omega) = \omega^{1 - 2 \beta_2^{\rm global}}$ with
$\beta_2^{\rm global} \approx \chi \approx 5/4$ indicating a dynamical
exponent $z \approx 1$. The behaviour in the intermediate frequency
range is not clear. Data were obtained in systems with size $L \! = \! 500$ 
at average interface heights $H = 40$, $80$, $120$, and $160$ respectively.}
\label{velpower_fig}
\end{figure}

Anomalous roughness with a global roughness exponent $\xi \approx 1.25$
has been found in the spatial structure factor or power spectrum $S(k,t)$.
For the second moments of temporal  fluctuations or the exponent $\beta_2$
a similar phenomenon is found: Figure~\ref{velpower_fig} shows the
velocity power spectra
\begin{equation}
S_v(\omega) = \langle | v(x,\omega) |^2 \rangle
\end{equation}
obtained from simulations of steady state distributions
of Eq.~(\ref{horvath_eq}) at different average heights $H = 40$, $80$, $120$,
and $160$. The overbar denotes a spatial average over the
system, the brackets and ensemble average, and $v(x,\omega)$
is the Fourier transform of $v(x,t) = \partial_t h(x,t)$ which leads
to the obvious identity $S_v(\omega) = \omega^2 S_h(\omega)$.
From the power spectra one can read a global temporal
exponent $\beta_2^{\rm global}$: A possible power law behaviour goes
like
\begin{equation}
S_v(\omega) = \omega^{1 - 2 \beta_2^{\rm global}}.
\end{equation}
It cannot be overemphasised that this is the global counterpart to the
local exponent defined via $C_{q = 2}(t,H)$ in Figure~\ref{temporal}, and
{\bf not} the exponent related to the early time increase in interface
roughness when starting from an initially flat configuration.

The crossover observed in the velocity distributions,
Fig.~\ref{velocity_distr}, will of course prevent a ``pure'' power law in
$S_v(\omega)$, but we can nevertheless observe three regions giving
rise to ``effective'' exponents. (i) For low frequencies,
long times, the fluctuations are stationary, $S_v(\omega) \sim \omega$
and therefore $\beta_2 = 0$. (ii) Fluctuations at short times
and high frequencies, on the time scale of avalanche duration, are
``turbulent'' in the sense of \cite{Krug_1994}: We find
$S_v(\omega) \sim \omega^{-3/2}$ and thus $\beta_2^{\mathrm global} = 5/4$.
It certainly would be of interest to examine these in greater detail and
pin down the connection to the velocity distributions shown in
Fig.~\ref{velocity_distr}. (iii) Last we can identify
an intermediate regime with an apparent behaviour between
$S_v(\omega) \sim \omega^{-3/4}$ and $\omega^{-2/3}$, which would
correspond to $\beta_2^{\rm global} = 7/8$ or $5/6$.
We would however suggest not to put too much emphasis on the power
law nature of this regime, which probably is caused by an interplay of
several different mechanisms during the crossover from
avalanche propagation to the stationary fluctuations at
long time scales.

The different regimes appearing in $S_v(\omega)$ exhibit different
scaling behaviour with respect to the average interface velocity
$|{\bf v}|$. The stationary regime with $S_v(\omega) \sim \omega$
ends at $\omega_0 \sim |{\bf v}|^{\beta_2^{\rm global}}$. A possible
interpretation is to take the typical duration of avalanches
$\tau_{\rm ava} \simeq \Delta h_{\rm ava}/v_{\rm ava}$ from their
vertical extent $\Delta h_{\rm ava} \sim \xi_\times^\chi$ and their
``sweeping velocity'' $v_{\rm ava} sim |{\bf v}|^{-\beta_2^{\rm global}/2}$.
The velocity $v_{\rm ava}$ also appears in the large frequency tail where
$S_v(\omega) \sim \omega^{-3/2}$ with an amplitude $\propto v_{\rm ava}^2$.
The lower bound of this tail, $\omega_1$, at the crossover to the
intermediate regime seems to be independent of the average velocity
$|{\bf v}|$. Also these findings certainly need to be examined more
thoroughly.

Interestingly we find the exponent $\beta_2^{\mathrm global} = 5/4$ in the
short time regime to be the same as the global roughness
exponent $\chi$. Assuming that these short time fluctuations
follow a standard scaling picture we can combine them to
$z = \chi/beta \approx 1$ which means linear or ballistic propagation
of fluctuations along the interface. At first sight this is
plausible: We are dealing with the advancement of the
front in the course of single avalanches, between encountering
regions of stronger pinning. Here the front can propagate
freely with some typical velocity, and it is easy to imagine
that fluctuations will be carried along the interface with
a constant velocity during that process. Again, this would
be an interesting topic for further studies. We have not pursued
an analogous direction, by computing the power spectrum of the
interface velocity (recall Fig.~\ref{velocity_distr}). In the
context of microscopic models - related to e.g.
Barkhausen noise - with avalanches that are easier to define
than in imbibition, the $S(\omega_{v,int})$ can be related
to the exponents describing the avalanches which in turn 
relate to those that describe the underlying critical point itself
\cite{Kuntz_2000}.

It is worth remembering here
both with respect to the correlation functions $C_q$ and the 
$\beta$-exponents, that the deviations from ordinary kinetic
roughening stem from the memory or history effects in imbibition.
In local models the central question in this respect concerns
whether the local fluctuation $\delta h(x,t) \equiv h(x,t) - \bar{h} (t)$
defines effectively a Markovian process. Imbibition is different,
since, among others, clearly the dynamics measured by
the two-point functions $C_q$ depends on both 
the fluctuating field $\delta h(x,t)$ and also on $\bar{h}$,
the distance to the reservoir. Secondly, it is clear 
though only on a qualitative level that the existence of $\xi_\times$
implies memory effects that are valid also beyond the particular
details of spontaneous imbibition.
 
In any case, the model results shown in Figure~\ref{temporal} reproduce several
experimental features found by Horv\'ath and Stanley \cite{Horvath_1995}:
First, the late time saturation level of $C_q$ increases with $H$ indicating
larger overall roughness, and second, $C_q(t,H)$ at fixed small $t$ decreases
with $H$, indicating a faster intrinsic avalanche velocity for a smaller $H$.

However, the picture developed so far contradicts strongly with the 
exponent identity, Eq.\ (\ref{exp_ident}) proposed by Lam and
Horv\'ath \cite{Dube_2001}. This identity is derived under the
assumption that the dynamics are controlled by a single time scale
which can be obtained by two means. For a moving interface, a width
$w$ is reached after a time $t_1 \sim w^{(\Omega+1) / (\kappa
\Omega)}$. On the other hand, the form of the time correlation
function, Eq.\ (\ref{LH_ct_scaling}) suggests a time scale $t_2 \sim
v^{-(\theta_t + \kappa)/\beta} \sim w^{(\theta_t+\kappa)/\kappa
\beta}$, where $\theta_t = 0.38$ and $\beta = 0.63$ are found.
Assuming that $t_1 \propto t_2$, the exponent identity is
obtained. This is trivially true in usual kinetic roughening but wrong
for spontaneous imbibition since $t_1$ and $t_2$ describe two distinct
physical time scales. It takes a time $t_1$ for the correlation length
to have value $\xi_v (t_1)$. This in turns controls the width as $w
\sim \xi_v^{\chi}$. On the other hand, $t_2$ is the relaxation time of
the fluctuations when the interface is kept at a fixed height. In this
case, the correlation length $\xi_v$ is a predetermined constant and
the saturation within this zone is obtained when the spatial extent of
the correlations equals $\xi_v$. We emphasise that the growth of the
width in spontaneous imbibition is controlled by the increase of
$\xi_{\times}$, the zone available for correlated fluctuations, and
not from the intrinsic dynamical fluctuations of the interface. 

In fact, the correlation function, Eq.~(\ref{LH_ge_scaling}), clearly
defines a length scale $\xi_v \sim v^{-\gamma}$, where $\gamma \equiv
(\theta_l + \kappa)/\alpha \simeq 0.4$, which then implies $G_2(r)
\sim \xi_v^{\chi} g(r/\xi_v)$. Thus, $\xi_v \sim v^{-0.4}$ is
analogous to the $\xi_{\times} \sim v^{-1/2}$ discussed for the ideal
case of spontaneous imbibition with Darcy-Washburn behaviour. Equation
(\ref{LH_ge_scaling}) also defines a {\em global} roughness exponent,
$\chi = \kappa \alpha /(\kappa + \theta_l)=1.25$, again very similar
to the results obtained from the phase field model.

The scenario proposed by Lam and Horv\a'ath could only occur if the
fluctuations were slower than the increase of $\xi_{\times}$, which
implies $t_2 \gg t_1$. Recall the point that for $z < 4$, in the case
of the phase field model, this is never true. Indeed, the numerical
data indicate $t_2 \sim w^{2.8}$ and $t_1 \sim w^{3.25}$ which shows
this assumption to be false asymptotically.

\subsubsection{Spontaneous imbibition with pinning}
\label{imbi_pinning}
In presence of the pinning effects of gravity or evaporation,
numerical simulations of the phase field model show that the behaviour
of the average interface height is well described by the
transcendental equation, Eq.\ (\ref{mf_evap}) and Eq.\
(\ref{mf_grav}). The actual pinning heights deviate only slightly 
from the predictions due to the presence of disorder
\cite{Dube_2001}.

It is convenient to analyse the effects of gravity and evaporation
separately. In the absence of evaporation, the linearised equation of
the fluctuations Eq.\ (\ref{non-local}), immediately shows the
existence of a length scale $\xi_{\rm g} (t)$ restricting the range of
the spatial fluctuations and evolving in time as
\begin{equation}
\label{xi_g}
\xi_{\rm g}(t) = \frac{1}{2} \,
\left( \frac{\sigma}{ 2(\dot{H} + \gamma)} \right)^{1/2} = \
\frac{1}{2} \, \left( \frac{\sigma H(t) }{\bar{\alpha}} \right)^{1/2},
\end{equation}
where $H(t)$ is given by the solution of Eq.\ (\ref{mf_grav}) (minor
corrections of order ${\cal O}(\xi_{\rm g}/H)$ are also expected). The
length scale $\xi_{\rm g}(t)$ is thus analogous to the length scale
$\xi_{\times} (t)$ in spontaneous imbibition without external
influences, but it does not increase in time according to a simple
power law any more. At pinning, this length scale becomes
\begin{equation}
\xi_{\rm g} (H_{\rm p}) = \frac{1}{2}\left( \frac{\sigma}{2 \gamma}
\right)^{1/2} = \,  
\xi_{\times} (H_{\rm p}).
\label{grav-pinned}
\end{equation}

As in spontaneous imbibition without gravity, dynamical scaling
relations can be established by assuming a single correlation length
$\xi_{\rm g} (t) \sim \gamma^{-1/2} \, (\sigma H/H_{\rm p})^{1/2}$,
where, from Eq.\ (\ref{mf_grav}), the quantity $H / H_{\rm p}$ is a
function of $\gamma^2 t$ only. The temporal behaviour of the width of
the interface $W(t)$ can then be scaled as
\begin{equation}
\label{sc_width}
W(t) = \gamma^{-\chi/2} w(\gamma^2 t)
\end{equation}
where $w(x) \sim x^{\chi/4}$ for $x \ll 1$ and tends to a constant
for large arguments. Numerically, the global roughness exponent
$\chi \approx 1.25$ has the same value as for spontaneous imbibition
without external influences \cite{Dube_2001}.

\begin{figure}
\includegraphics[width=8cm]{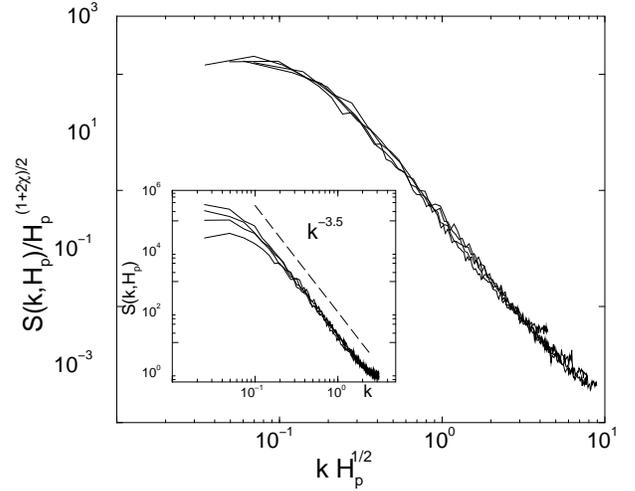}
\caption{A log-log plot of structure factors of the pinned
interface in presence of gravity.  The inset shows the value
of $S(k,H_{\rm p})$ for pinning heights $H_{\rm p} = 20, 40,60$ and $80$.
The main figure shows the collapse of the data under the assumption of a
correlation length $\xi_{\rm g} (H_{\rm p}) \sim H_{\rm p}^{1/2}$.
The dashed line indicates a roughness exponent of $\chi=1.25$.
All quantities are taken in the dimensionless units of
Eq.\ (\ref{pf_gen}).
}
\label{strf_g}
\end{figure}

The scaling picture at pinning is confirmed from the structure factor.
Figure \ref{strf_g} shows the structure factor for systems with different
pinning heights (arising from different values of the constant $\gamma$). The
structure factor decays as $k^{-3.5}$ at large values of wave-vector,
consistent with the value $\chi = 1.25$ and the curves can be collapsed
on the common scaling form by setting $\xi_{\rm g} (H_{\rm p}) \sim
\gamma^{-1/2} \sim H_{\rm p}^{1/2}$:
\begin{equation}
S(k,H_{\rm p}) = (\xi_{\rm g}(H_{\rm p}))^{1+2\chi} \, s (k \,
\xi_{\rm g} (H_{\rm p})).
\label{sk_scaling_pinning_g}
\end{equation}
The scaling function $s(k\xi)$ is the same as in the freely rising case and
thus relates to the scaling function $g$ in Eq.\ (\ref{g_scaling}).

As for pure spontaneous imbibition the concept of a dynamical exponent $z$
describing the propagation of fluctuations along the interface
does not apply to imbibition with gravity. The fluctuations
always catch up to the available zone of correlated roughness,
correlation lengths cannot exceed $\xi_{\rm g}$.

Dynamical roughening of the interface in the  presence of evaporation 
(and absence of gravity) is the most complex case. Although the
average interface dynamics is well described by the mean-field equation, 
the extra term in $\epsilon (1-e^{-2|k|H})$ means that a single
correlation length $\xi = \xi(t,H(t))$ cannot be unambiguously
identified.

The situation becomes clear only in the pinned interface limit, where a
simple correlation length can be defined as
\begin{equation}
\sigma \, \xi_{\rm e}^{-3} = 2 \, \epsilon \,
(1 - e^{-2 H_{\rm p}/\xi_{\rm e}}).
\label{xi_e}
\end{equation}
In the limit $\xi_{\rm e} \ll H_{\rm p}$, $\xi_{\rm e} \sim
(\sigma/\epsilon)^{1/3}$ while in the opposite limit
$\xi_{\rm e} \gg H_{\rm p}$, $\xi_{\rm e} \sim
(\sigma^2 / (\bar{\alpha} \epsilon) )^{1/4}$.
The first case corresponds to weak evaporation, defined by
$\epsilon \ll \bar{\alpha}^3 /\sigma^2$,
while the second case is that of strong evaporation,
$\epsilon \gg \bar{\alpha}^3 /\sigma^2$. 
The value of the roughness exponents is 
established from the decay of the structure factor,
with again the result $\chi = 1.25$ \cite{Dube_2001}.

\begin{figure}
\includegraphics[width=8cm]{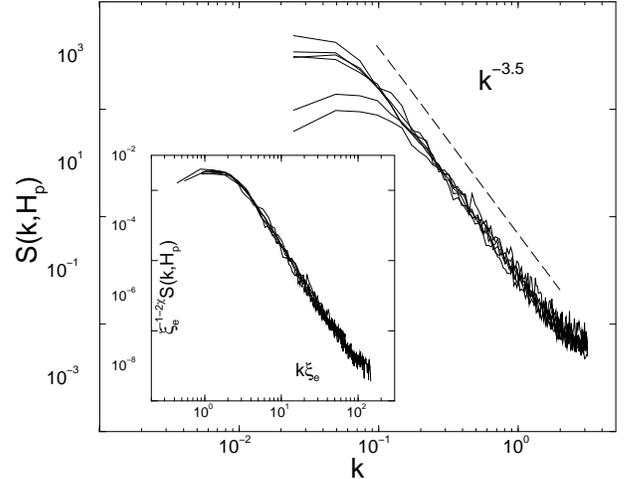}
\caption{A log-log plot of structure factors of
the pinned interface in the presence
of evaporation. The main figure shows the value of $S(k,H_{\rm p})$ for
pinning heights (from bottom to top) :
$H_{\rm p} \simeq 20$ ($\bar{\alpha}=0.2, \epsilon=10^{-3}$);
$H_{\rm p} \simeq 28$ ($\bar{\alpha}=0.2, \epsilon=5 \times 10^{-4}$);
$H_{\rm p} \simeq 45$ ($\bar{\alpha}=0.1, \epsilon=10^{-4}$);
$H_{\rm p} \simeq 63$ ($\bar{\alpha}=0.2, \epsilon=10^{-4}$);
$H_{\rm p} \simeq 77$ ($\bar{\alpha}=0.3, \epsilon=10^{-4}$);
$H_{\rm p} \simeq 109$ ($\bar{\alpha}=0.3, \epsilon=5 \times 10^{-5}$).
The structure factors for $\epsilon \sim 10^{-4}$
all collapse on the same curve, independent of the value of $\bar{\alpha}$.
A complete collapse of the data can be accomplished by solving
Eq.\ (\ref{xi_e}) numerically, as shown in the inset. The dashed line indicates
a roughness exponent of $\chi=1.25$.
Again, all quantities in the dimensionless units of
Eq.\ (\ref{pf_gen}).
}
\label{strf_e}
\end{figure}

The prediction for the pinned correlation length can be checked by
considering the scaling behaviour of the pinned structure factors
$S(k,H_{\rm p})$ as shown in Fig.\ \ref{strf_e}. The data with the highest
evaporation rate can all be collapsed assuming a correlation length
$\xi_{\rm e} \sim (\bar{\alpha} \epsilon)^{-1/4}$, while for lower evaporation
$\xi_{\rm e} \sim \epsilon^{-1/3}$. The crossover occurs roughly for parameters
$\bar{\alpha} = 0.2$ and $\epsilon = 10^{-4}$. The complete set of data
can be reduced to a common scaling form by solving Eq.\ (\ref{xi_e})
numerically (using a value of $\sigma = 2 \sqrt{2}/3$). As a result the
structure factor at pinning has a scaling form similar to Eq.\
(\ref{sk_scaling_pinning_g}), although care must be taken in identifying
the correlation length.


To summarise, the main result of the analysis presented here is that the
fluctuations of the interface are intimately coupled to the average position,
through the length scale $\xi_{\times}$. Both $\xi_{\times}$ and $H$ should
be studied simultaneously. Experimentally, the existence and dynamical
behaviour of this length scale can be inferred from the saturation of the
spatial and temporal correlation functions and should be easier to observe
than the precise values of the scaling exponents on scales $r < \xi_{\times}$.

\subsubsection{Constant flow}
\label{constantflow}
Imbibition with a constant flow rate can easily be performed, e.g., in
Hele-Shaw cells with disorder
and can be modelled within the phase field approach by
a simple modification of the boundary conditions. So far we have had
for the average velocity of the interface $2 v = | \langle \grad \mu
|_{y=0} \rangle |$ (the factor 2 coming from the miscibility gap). If
now the disorder is in the mobility, its effect on the interface
dynamics is proportional to the flow influx, so a priori it cannot be
neglected. This adds a term to the interface equation, of the form
\begin{equation}
\int d{\bf x}' \; G ({\bf x}|{\bf x}') \;
\left( 
\nabla \cdot m({\bf x}) \nabla \mu \right). 
\label{bulk_green_mob}
\end{equation}

The usual projection technique becomes quite involved in this case. 
However, for a linear analysis, it is sufficient to relate the 
gradient of the chemical potential to the average interface
velocity $\grad \mu \sim v$, which yields 
\begin{equation}
\label{linear_cap_mob}
\dot{h_k} = - \left( v +  \sigma k^2 \right) |k| h_k + 
\frac{1}{4} \, |k| \, \{ \eta \}_k + v \, \{ m \}_k
\end{equation}
where $\{ m \}_k \approx 2 \int_{x} e^{-ikx} m(x,h(x,t))$ represents the
noise induced by the mobility. In contrast to the capillary disorder, 
represented by $|k| \{ \eta \}_k$, 
this noise is non-conserved, but with a strength proportional to the
mean velocity of the interface. 

This equation, also derived in \cite{Paune_2002} and 
partly in \cite{Hernandez_Machado_2001}
shows that mobility and capillary disorder influence the
fluctuations on different length scales. Small length scales are dominated
by capillary disorder while the effect of mobility disorder becomes 
dominant on large scales. This is evident and natural, since the latter
reflects fluctuations in the local fluid flow in the bulk, which obviously
have an effect beyond local pore randomness.
A crucial point however is that the crossover
length-scale $\xi_{\rm mob}$ between the two regimes is velocity dependent,
in terms of dimensional quantities
\begin{equation}
\xi_{\rm mob} \sim \frac{\kappa^2}{v \eta} \, \frac{\delta p_c}{\delta \kappa}.
\label{EqXimob}
\end{equation}
In the simple model of Paun\'e and Casademunt, where both permeability
and capillary pressure are related to an effective pore radius $r_0$,
$\xi_{\rm mob} \sim r_0/C_a$ \cite{Paune_2002}.

\begin{figure}
\includegraphics[width=8cm]{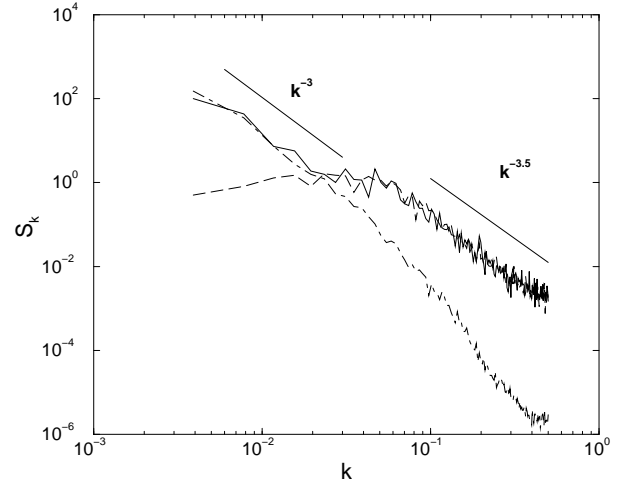}
\caption{Comparison of roughness spectra (structure factors) for capillary and
mobility disorder. The dashed line represents capillary disorder only,
the dot-dashed line mobility disorder only and the full line 
to both sources of disorder present. The simulation is done at $v=0.005$,
taken at an average height $H=50$. 
Beyond a length scale $\xi_{\rm mob}$ (cf.\
Eq.~(\ref{EqXimob})) mobility disorder, with $\chi \approx 1$ 
gives the dominant contribution to
roughness, on smaller scales disorder in capillarity, with $\chi=1.25$ 
is more important.}
\label{cap_mob_disorder}
\end{figure}

\begin{figure}
\includegraphics[width=8cm]{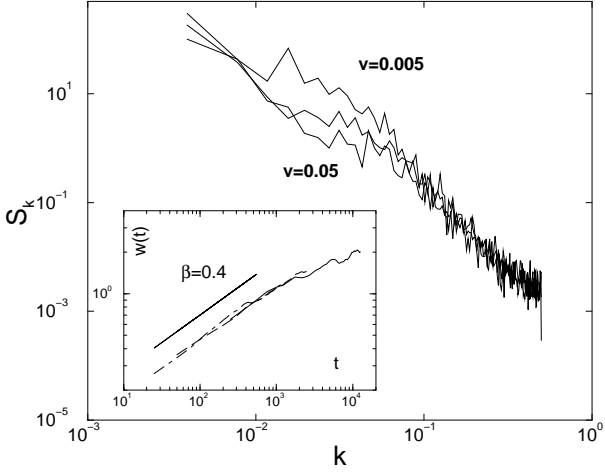}
\caption{Comparison structure factors for both capillary and mobility
disorder at different velocities. The simulations are done for three
different velocities, $v \! = \! 0.005, 0.01$ and $0.05$ with the
structure factors taken at the same average height $H \! = \! 50$. A
clear separation between the two roughening regimes occurs only if
$\xi_{\rm mob} \sim 1/v \rightarrow \infty$. Beyond a length scale
$\xi_{\rm mob}$ (cf.\ Eq.~(\ref{EqXimob})) mobility disorder, with
$\chi \approx 1$ gives the dominant contribution to roughness, on
smaller scales disorder in capillarity, with $\chi=1.25$ is more
important.}
\label{imbi_both}
\end{figure}

The length scale $\xi_{\rm mob}$ can be seen in Figure~\ref{cap_mob_disorder},
where the roughness spectra of three model setups are plotted on top of each
other: (i) capillary disorder only, which gives the main contribution to
roughness on short scales $< \xi_{\rm mob}$, (ii) mobility disorder only, which
roughens the opposite regime, and (iii) including both, which gives an envelope
of the other two curves. 
The comparison between the structure factors 
corresponding to different average interface velocity is shown on 
Fig~\ref{imbi_both}. An effective exponent $\beta \simeq 0.4$ can be
defined, but it is strongly influenced by cross-over between the capillary
and mobility regimes of disorder. 

\begin{figure}
\includegraphics[width=8cm]{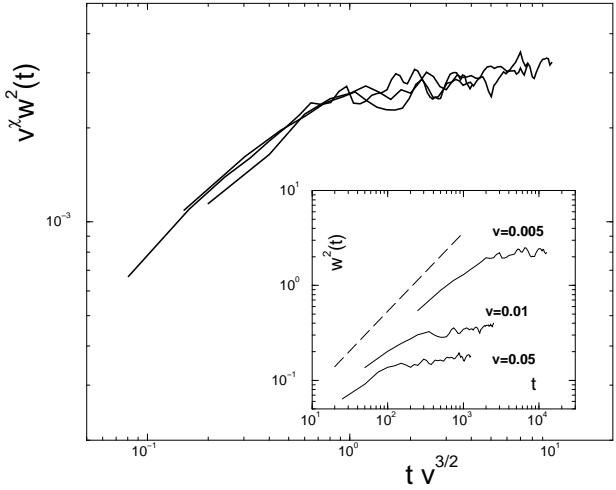}
\caption{Growth of the width for interfaces moving at constant
velocity $v=0.005, 0.01$ and $0.05$. The inset shows an initial growth
exponent $\beta \approx 0.42$ followed by a weak logarithmic
regime. The main figure shows the rescaling of the data using Eq.\
(\ref{cap_kst_scaling}) and a crossover time $t_{\times} \sim
v^{-3/2}$. The roughness exponent $\chi=1.25$. 
}
\label{w_cap_disorder}
\end{figure}

\begin{figure}
\includegraphics[width=8cm]{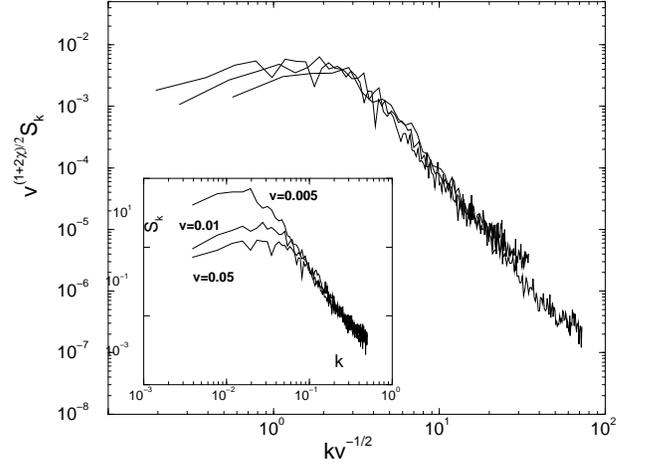}
\caption{Structure factor at saturation for interfaces moving at constant
velocity $v=0.005, 0.01$ and $0.05$ taken in the logarithmic regime for
the growth of the width (nearly saturated regime). 
The inset of the Figure shows the structure
factors, the faster velocity have the smaller range of correlated roughness.
The main Figure shows the rescaled data using the correlation length 
$\xi_{\times} \sim v^{-1/2}$ and the roughness exponent $\chi = 1.25$.
}
\label{strf_cap_disorder}
\end{figure}

The behaviour of the interfacial fluctuations is a lot clearer at {\em
low capillary number}, where only capillary disorder is relevant . The
intrinsic length scale $\xi_{\times} = (\sigma /v)^{1/2}$ is still
present but it is now static, and does not play an explicit role in
the dynamics of the fluctuations. Figures~\ref{w_cap_disorder} and
\ref{strf_cap_disorder} show the numerical results obtained from a
simulation of the phase field model with constant flow, $m({\bf x}) =
1$ and without pinning ($\gamma \! = \! \epsilon \! = \! 0$). The
structure factor at saturation shows that the short wavelength
exponent $\chi \approx 1.25$ with a much smoother roughness at large
length scales. The dynamical behaviour of the width shows first a
steep increase, with exponent $\beta \approx 0.42$ which is indicative
of a dynamical exponent $z = \chi/\beta = 3$. At later times, the
width flattens out, to an effective exponent (only defined on the
small window available) $\beta \approx 0.1$, more representative of
logarithmic roughening than true power-law behaviour, a trend
confirmed by the form of the structure function.

The dynamics is thus controlled by two correlation lengths, 
$\xi_1 \! \sim \! vt$, which dominates the dynamics for wave-vectors
$k \ll \xi_{\times}$, and $\xi_3 \sim (\sigma t)^{1/3}$, which controls
the scale $k \gg \xi_{\times}^{-1}$. The time development of the width 
is determined by the $\xi_3$ term as long as $\xi_3 \ll \xi_{\times}$
while the slower $\xi_1$ is only effective at times when $\xi_1 \ll
\xi_{\times}$:
\begin{eqnarray}
w (t) & \sim & \xi_{3}^{\chi} (t) \sim t^{\chi/3}
\mbox{\hspace{2.0cm}} t \ll t_{\times}, \nonumber \\
w (t) & \sim & w(t_{\times}) + {\cal O}(\log(t/t_{\times}))
\mbox{\hspace{1.0cm}} t \gg t_{\times}.
\label{cap_kst_scaling}
\end{eqnarray}
The crossover time between the two dynamical regimes is $t_{\times} \sim
\sigma^{1/2} / v^{3/2}$ and the width after saturation of the rough zone
$w(t_{\times}) \sim \xi_{\times}^{\chi}$.

\subsubsection{Columnar Disorder} 
\label{imbi_columnar}
\begin{figure}
\includegraphics[width=8cm]{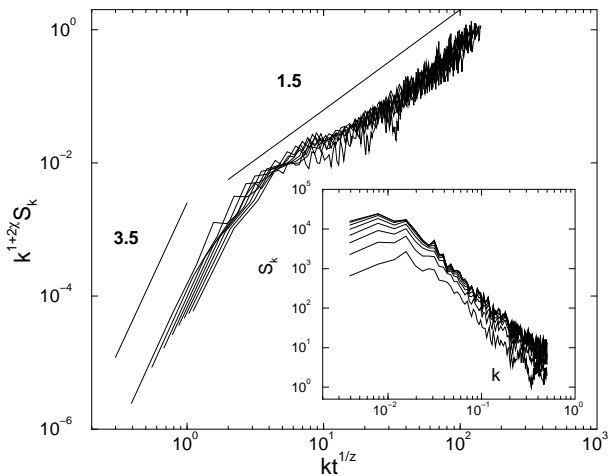}
\caption{Structure factor obtained from the phase field model with
columnar capillary disorder and constant flow rate. Intrinsic anomalous
roughening \protect \cite{Lopez_1997} is visible in the inset. The scaling plot
(main graph) indicates a roughness exponent $\chi \approx 1.25$ and
dynamical exponent $z \approx 2$. The scaling scaling plot indicates an
anomalous exponent $\theta_a \approx 0.75$, similar to the value found
experimentally \cite{Soriano_2002b}. The simulations were done with a 
imposed velocity $v=6.25 \times 10^{-4}$ 
and capillary disorder only.}
\label{col_dis}
\end{figure}

Recent experiments have also dealt with the case of imbibition
with columnar disorder \cite{Soriano_2002a,Soriano_2002b}, defined by 
\begin{equation}
\langle \alpha({\bf x}) \alpha({\bf x'})\rangle  -  
\bar \alpha^2  =  (\Delta \alpha)^2 \delta(x \! - \! x').
\label{colnoise}
\end{equation}
An example is shown in Figure \ref{sor01b} (Section~\ref{introughfflow}). 
The main characteristic of
this type of noise is that the dimension perpendicular to the reservoir
does not come into play, it corresponds to imbibition on ``stripes'' of
different properties. Thus, at the level of the interface equations,
the position of the interface does not enter the quenched noise. The
linear interface equation is thus 
\begin{equation}
\label{linear_column}
\frac{ \partial h_k(t)}{\partial t} = - \left( v +  \sigma k^2 \right)
\; |k| \; h_k + \frac{1}{4} \; |k| \; \eta_k 
\end{equation}
where now $\langle \eta_k \eta_{k'} \rangle =  (\Delta \alpha)^2
\delta_{k,- k'}$. With this simple nature of the disorder, the
roughening process can be studied analytically at the linear level. At
saturation, the structure factor
\begin{equation}
S_k \sim \frac{1}{( v + \sigma k^{2} )^2}
\end{equation}
indicates a short-wavelength roughness exponent $\chi = 3/2$ and a
flat interface at large distances. However, due to the development of
fingers in the interface, separated by large slopes, the linear
approximation of Eq.~(\ref{linear_column}) breaks down and
experiments (Figure~\ref{sor01c}), as well as simulations of the phase
field model (Figure~\ref{col_dis}), show instead intrinsic anomalous
roughening \cite{Lopez_1997}. In this case, there is a time-dependent
increasing amplitude of the structure factor, characterised by an
exponent $\theta_a$, due to steepening of local slopes on the interface.
\begin{equation}
\label{strf_anom}
S_k(t) \sim
\Biggl\{
\begin{array}{ll}
\! \xi^{2\theta_a} \; k^{-2 (\chi+\theta_a) - 1} & \mbox{for }k \gg \xi^{-1} \\
\! \xi^{2 \chi + 1} & \mbox{for }k \ll \xi^{-1}
\end{array}
\end{equation}
and $\xi = t^{1/z}$.
Indeed, the experimental distribution of slopes shows a power 
law tail \cite{Soriano_2002a,Soriano_2002b}, similar to what is
observed for roughening in fractal structures \cite{Asikainen_2002}.
At this point it becomes clear that the linear interface
equation fails to appropriately describe this case.


\section{Experiments on imbibition}
\label{Experiment}
In the following we summarise the current experimental situation in
imbibition with regard to the theoretical ideas presented earlier. The
main goal of this section is to give some experimental substance to 
theoretical ideas presented earlier. Beyond this we want to understand
where further interesting avenues exist, and to outline which of the
recent advances in models have been confirmed.
From the existing works one can see some clear directions for
further investigations.

The introduction presented already some experimental complications 
associated with both spontaneous and forced imbibition. 
Since our approach is predominantly based on  an interfacial approach,
where the advancing front of liquid has an effective 
surface tension, it is important to discuss in which 
cases the simplifications of the theoretical models are expected
to be valid.

To recapitulate the viewpoint of a statistical physicist interested
in the front roughening, the main issue is whether
an interface description actually makes sense at 
all. In our case this is considered to be true if the 
large-scale geometry of the boundary between the ``dry''
and ``wetted'' parts of the system can be described
with the aid of self-affine concepts. Exceptions can arise
if the saturation of the imbibing fluid is a very smooth function,
in the coarse-grained sense. Then the transport dynamics are determined 
on the level of individual pores, and coarse-graining to an interface
is not possible. When an interfacial description is possible, 
pore-scale phenomena can be embedded in the
noise contained in the interfacial equation of motion.
A second question to be asked is whether 
the Darcy-Washburn behaviour is valid (recall that it is based on a
idealised picture of fluid flow), even in the presence of a well-defined
imbibition front. Failure of simple Darcy-flow
concepts can stem from fluid-matrix interactions, from hydrodynamical
reasons (inertial and transient effects, as Bosanquet-flow),
and from contact-line physics, like the change of the contact
angle over time \cite{Poulin_1997,Siebold_2000}.

In this section, when appropriate, we also point out analogies between
kinetic roughening in imbibition and other experimental systems, 
(see \cite{Meakin_1998,Barabasi_1995,HalpinHealy_1995}
for early reviews of the subject). 
We first discuss the specific problem of paper, or fibre
networks, and then continue with results on interface dynamics for 
constant flow rate (forced fluid flow)
imbibition, followed with the spontaneous case.

Before embarking on this task, it is worth keeping in mind
some basic facts of roughening in nature. The theoretical
models are always a ``coarse-grained'' description of the experimental
system considered, that is
the micro-structure enters only via an effective description,
most often encoded in the noise terms. Several length scales are also
usually present and many cross-over phenomena can occur. In real systems,
as well as in simulations, one encounters a typical structural 
length-scale $l_c$ (e.g., the bead size
in a Hele-Shaw cell) so that real scaling behaviour is established
only for $l \gg l_c$. Cross-over behaviours are also 
due to the finite size $L$ of system itself which 
enters the scaling function of any measure of roughness or correlations.
Finally the boundary conditions can induce boundary effects
by suppressing or enhancing local dynamics. It is customary
to neglect these by considering only a part of the system
(to use a window in space), but there is of course no guarantee
that this procedure works a priori. In particular, it is often
assumed that one needs to subtract the 'tilt' (linear trend) from
the height profile $h(x,t)$ to avoid long-wavelength effects
(see e.g. \cite{Schmittbuhl_1995} for a discussion). 
For all these three reasons
it is clear that one should take any exponent values measured
for a fixed $l/l_c$ with a grain of salt. Indeed, the effect of
boundaries on the behaviour of systems exhibiting criticality
is a very complex topic by itself, with many facets, e.g., for
directed percolation \cite{Lauritsen_1998} and the KPZ universality
class \cite{Krug_1994b}.

A second important point is that true scaling behaviour
is observed only if good averaging, in the ``thermal'' sense is 
obtained \cite{Myllys_2001}. The interface must see several different
realisations of disorder before any scaling behaviour can be observed.
This is intuitively clear even in a simulation, where
one run just considers a particular noise history and representative
scaling is established only within a statistical ensemble. 
There are two ways around this problem. The simplest, brute force approach,
is simply to
do a large enough number of experimental runs with different noise
configurations. This is however difficult with imbibition since, analogous to
fracture experiments, it is often impossible to repeat the 
experiment very many times. The other option is to consider long
enough experimental runs, so that the interface fluctuations
reach a steady state \cite{Horvath_1995,Myllys_2001}. This is again
very difficult in imbibition, due to the slow time scales involved. 
Also, in the theoretically interesting case of spontaneous imbibition
the front slows down constantly and there is no steady state.
In cases where thermal noise becomes irrelevant (slow,
avalanche- or activation-like dynamics) the situation is
in practical terms even more precarious than in usual kinetic
roughening when it comes to obtain reliable data.

\subsubsection{Difficulties inherent to paper imbibition: a case
study}
In many statistical physics experiments on imbibition the medium of
choice has been either a Hele-Shaw cell filled with (transparent)
beads up to a certain volume fraction or, for obvious simple practical
reasons, paper. The first approach is perhaps less subject to
accidental experimental complications. To both assist with the
interpretation of actual scaling results and to highlight the
difficulties that could arise in various media we now discuss the
specific  case of paper and its impact on studies of
imbibition. Sheets have been used to study striped media
\cite{Zik_1998} and {\em radial} propagation \cite{Medina_2001}, and
two fronts can be made to collide \cite{Nagamine_1996}, in addition to
the papers cited below in later sections.

Experiments have demonstrated that clear, Washburn-like behaviour can
only arise if there is not much interaction (swelling, prewetting)
between the fibre networks and the penetrating liquid
\cite{Gillespie_1958}. Ordinary paper is made from wood fibres and,
in many cases, chemical additives and filler materials like talc and
clay, arranged in a disordered structure \cite{Niskanen_1998}. There
is a wide distribution of pore sizes (this can be simulated by
computer models \cite{Niskanen_1994}, or studied via, e.g., X-ray
tomography \cite{Samuelsen_2001}) with a high effective
tortuosity. The actual mass distribution per unit area varies (due to
a phenomenon called flocculation \cite{Niskanen_1998}) and the
resulting structure has on the top of this power-law correlations up to
a cut-off of a few average fibre lengths \cite{Provatas_1996}.

The surface of paper sheets is extremely uneven and the concept of
surface pore is itself ill-defined. This gives rise to problems in
defining static and dynamical contact angles as is typical of
inhomogeneous substrates and implies that there will be large
fluctuations in the capillary pressures. In commercial papers, there
is also an anisotropy between the ``machine-direction'' (MD, parallel
to the web) and ``cross-direction'' (CD, perpendicular to the web),
which may also influence the liquid penetration in the structure. For
industrial papers there are often residues of chemicals, which may
induce drastic changes in the effective viscosity or surface tension
of the invading liquid. 

In paper, the whole fibre structure may be modified by
swelling of the fibres in contact with the liquid
\cite{Bristow_1971}. Cellulose fibres show a great affinity to water and
can absorb large quantities in millisecond time-scales, giving rise to
concomitant changes in fibre volume and pore structure. By nuclear
magnetic resonance (NMR)
it is possible to observe the simultaneous intake of water into pores 
between fibres but also {\em inside} the fibres themselves.
However, this is not the case for many organic fluids and oils. 
Any intuitive picture of fibres (or the network) as  capillary tubes is
thus often false: the pore
structure is highly non-trivial and in some cases time--dependent
(see also \cite{Thompson_2002,Koponen_1998}).
If swelling occurs, the volume to be filled with liquid increases 
and the flow resistance of the pore network changes, which leads possibly to
non-Washburn behaviour. 

There can also be an exchange of liquid between the inside of a fibre
and the ``surface'' pores, important in particular since the
imbibition takes place ``along'' the sheet. This complicates
enormously the fluid flow since there are no well defined
``structures'' (either the pores, or the fibres) responsible for the
capillary forces \cite{Aspler_1987,Pezron_1995,Knackstedt_2001}. The
fluid may very well undergo imbibition as such only on the rough paper
surface. This is a mechanism that has been studied itself by various
authors in other contexts \cite{Bico_2001,Rye_1996,Mann_1995}, but not
for such rough surfaces. All this makes it questionable as to whether it
is possible to easily find a representative network description
\cite{Lenormand_1988,Schoelkopf_2000a}. Mathematically, the problems
amount to defining the essential effective properties of a pore for
the imbibition process. Classical applications of imbibition to study
the flow properties of porous media hope to answer this 
unequivocally, but correlations in the pore-scale properties may
however render the starting point useless, since both the typical
scale of pores and the effective correlations need to be
mathematically specified  in any model of a ``typical'' pore
\cite{Sok_2002,Knackstedt_2001}.


\subsection{Forced flow imbibition}
Simultaneously to the development of the theoretical picture of
``local'' kinetic roughening several groups attempted to measure the
properties of a fluid-gas interface in Hele-Shaw cells
\cite{Rubio_1989,Family_1992,He_1992}. These were typically done in
planar cells of thickness between one and a few bead diameters. The
general outcome is that no really convincing link could be established
with such ``local'' descriptions. Quite recently this has been taken
up by the group of Ort\'\i{}n, Hern\'andez-Machado, and co-workers, using
specially prepared substrates with controlled disorder
\cite{Hernandez_Machado_2001,Soriano_2002a,Soriano_2002b,Soriano_2002c}.
We outline below the current experimental situation.

\subsubsection{Interface roughening}
\label{introughfflow}
The early works \cite{Rubio_1989,Family_1992,He_1992} measured
typically the temporal exponent $\beta$ and the roughness exponent
$\chi$ by using digitised pictures of interfaces and two-point
correlation functions  (c.f.\ Eqs.\ (\ref{G_vs._S}) and
(\ref{Ct})). This method has many natural limitations: the typical
resolution available for spatial structure was of the order of 250
points of $h(x,t)$ and the two-point correlation function cannot
measure roughness exponents larger than 1. Representative values for
the exponents would be $\beta = 0.65$ and $\chi =0.80 \dots
0.9$. These are substantially higher than standard KPZ exponents
$\beta_{\rm KPZ} = 1/3$ and $\chi_{\rm KPZ} = 1/2$.

\begin{figure}
\includegraphics[width=8cm]{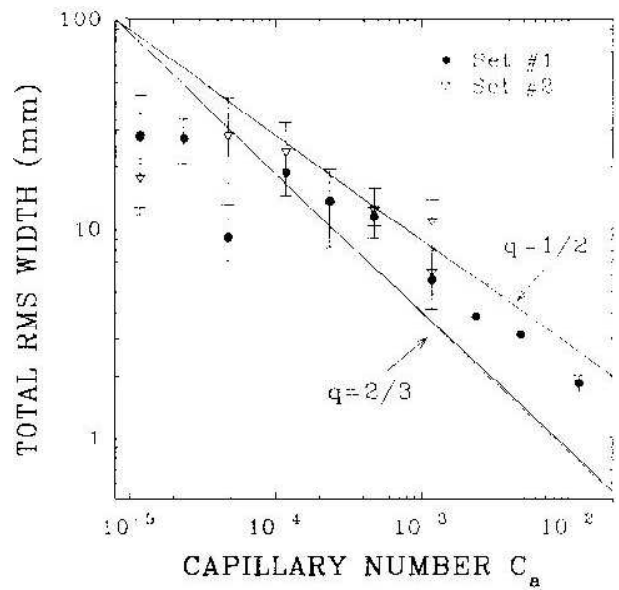}
\caption{Interface width vs.\ the capillary number $C_{\rm a}$ in a
Hele-Shaw cell with a constant flow rate. (Experiment by He et al.\
\cite{He_1992}). The lines show scaling predictions (see text).}
\label{he2}
\end{figure}

\begin{figure}
\includegraphics[width=8cm]{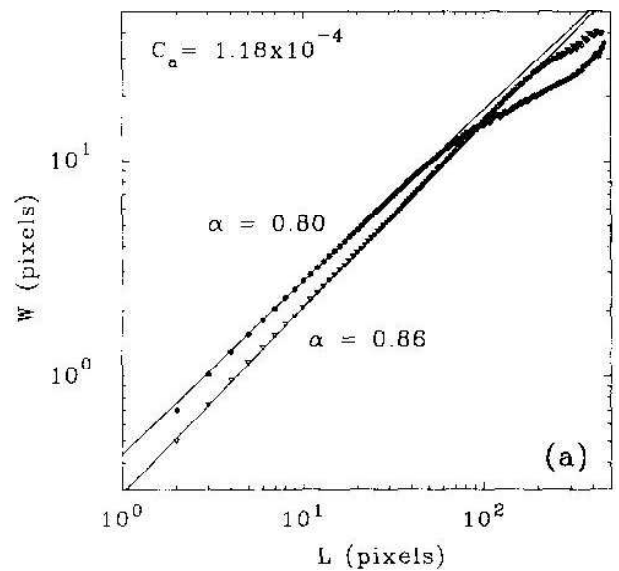}
\caption{Two examples of the local width (called $W$ here) for the
experiment of He et al.\ \cite{He_1992}. For $L$ small the effective
roughness exponent is about 0.8, whereas for larger scales the scaling
changes.}
\label{he3a}
\end{figure}

He et al.\ \cite{He_1992} observed the roughness in a Hele-Shaw cell
with capillary numbers $ 10^{-5} < C_{\rm a} < 10^{-2}$. They determined a
large roughness exponent $\chi \sim 0.8 - 0.9$ (as illustrated
in Fig.\ \ref{he2}, from \cite{He_1992}) and a
decrease of the global width with increasing capillary number with a
possible saturation of the width at very low capillary number (Fig.\
\ref{he3a} \cite{He_1992}). 
The roughness exponent was obtained from the two-point
correlation function and is probably indicative of a super-rough ($\chi
> 1$) interface. They also made the important point that the dynamics
is non-local and introduced a simple scaling argument to relate the
total width of the interface to the capillary number $C_{\rm a}$
\begin{equation}
w \sim C_{\rm a}^{-q}
\label{He-scaling}
\end{equation}
where $q = 2\chi/(2\chi+1)$ in 2D. This argument is obtained by
balancing the local (pore level) capillary forces to the global
viscous pressure drop driving the interface and agrees qualitatively
with the experimental results at large capillary numbers, although it
cannot fit with the plateau at low capillary numbers.  

However, this argument does not take into account any large scale curvature
of the  interface, from which the length scale $\xi_{\times} \sim
C_{\rm a}^{-1/2}$ in Eq.~(\ref{xi_cross_intro}) arises. The experimental
results can then also be
explained  with the theoretical picture developed in the preceding
section. Since the experimental capillary number is small, we assume
that only capillary disorder is important and that the mobility has a
negligible effect. At very small capillary numbers, the length of the
cell $L \ll \xi_{\times}$ and so the width is independent of $C_{\rm a}$. As
the capillary number increases, $\xi_{\times} < L$ and, neglecting the
small logarithmic correction at large scales, $w \sim
\xi_{\times}^{\chi} \sim C_{\rm a}^{-\chi/2}$. A roughness exponent $1.0 <
\chi < 1.25$ yields $ 0.5 < 0.625$ which is also in qualitative
agreement with the experimental results.

\begin{figure}
\includegraphics[width=8cm]{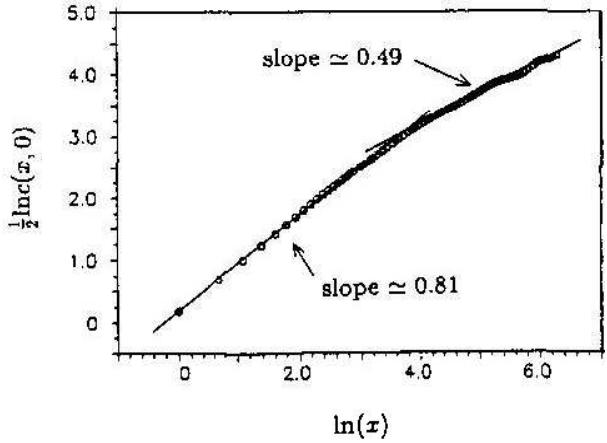}
\caption{The spatial two-point correlation function in a forced fluid
flow experiment by Horv\'ath et al.\ \cite{Horvath_1991a}. Notice the
presence of two scales in the data, with a larger exponent at small
scales.}
\label{ja9101063}
\end{figure}

At large capillary numbers, Horv\'ath et al.\ noted that there seemed
to be a cross-over in $\chi$ to values $\sim 0.5$ \cite{Horvath_1991a} at
large length scales (a clear example is depicted in Fig.\ \ref{ja9101063},
from the same reference) which would be consistent 
with a regime where either mobility disorder or the ``thermalisation''
of the quenched disorder becomes relevant. The measured dynamical
exponent $\beta = 0.65$ is also indicative of a smaller value of $z$.


\begin{figure}
\includegraphics[width=8cm]{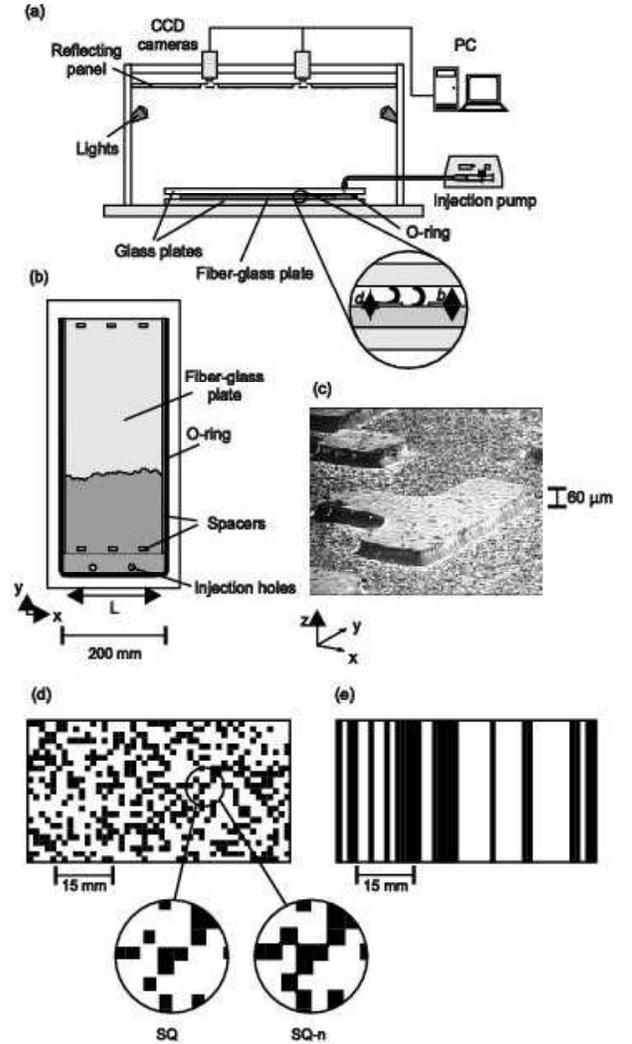}
\caption{Sketch of the experimental setup of \cite{Soriano_2002a}.
a) and b) present views of the experimental apparatus,
c) a SEM image of the copper islands on the fibre--glass plate, and
d) and e) present views of the disorder pattern for ``point-disorder''
and ``columnar disorder'', such that copper islands are denoted
by the black regions.
}
\label{sor01a}
\end{figure}

\begin{figure}
\includegraphics[width=8cm]{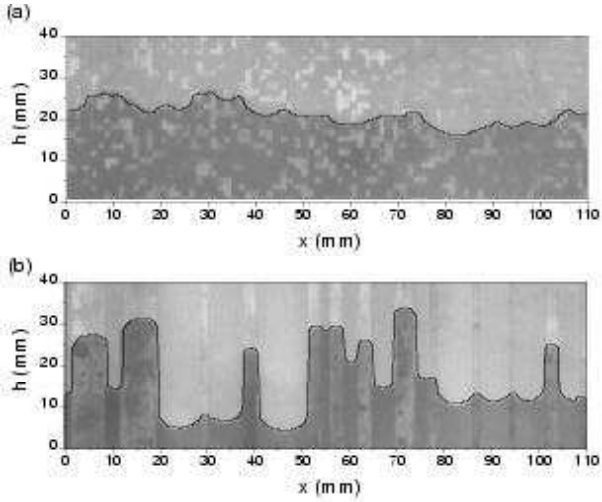}
\caption{Interface profiles for examples of disorder,
point-like in a), columnar in b), with the original
images included. Notice in particular the scale of the
vertical ``discontinuities'' in b).
From \cite{Soriano_2002a}.
}
\label{sor01b}
\end{figure}

Lately the forced flow case has been studied intensively both
theoretically (see the preceding section) and experimentally by the
group of Ort\'\i{}n, Hern\'andez-Machado and co-workers
\cite{Hernandez_Machado_2001,Soriano_2002a,Soriano_2002b,Soriano_2002c}. 
The central idea, different from earlier approaches, was to use
Hele-Shaw cells with a pattern of copper tracks or islands placed on a
substrate of fibre glass. An illustration of the experiment is shown
in Fig. \ref{sor01a}, together with the way of introducing randomness
into the samples.  Here such disorder is given simply by the
fluctuations in the cell gap. A definite advantage of this setup is
that the disorder is under better qualitative control. In particular
it allowed for the first time the study of ``columnar disorder'',
where the gap value is quenched and does not vary in the direction of
interface propagation,  for the first time (see
Eq.~(\ref{colnoise})). On the other hand this very interesting work
also highlights the standard experimental difficulties in this field:
the need for many samples (disorder averaging) and the problems in
attaining reasonable scaling windows for the possibly critical
quantities are difficult to circumvent. The in practice 
achievable length-scales are obvious in Fig. \ref{sor01b},
and one should note in particular the case with columnar 
disorder.


\begin{figure}
\includegraphics[width=8cm]{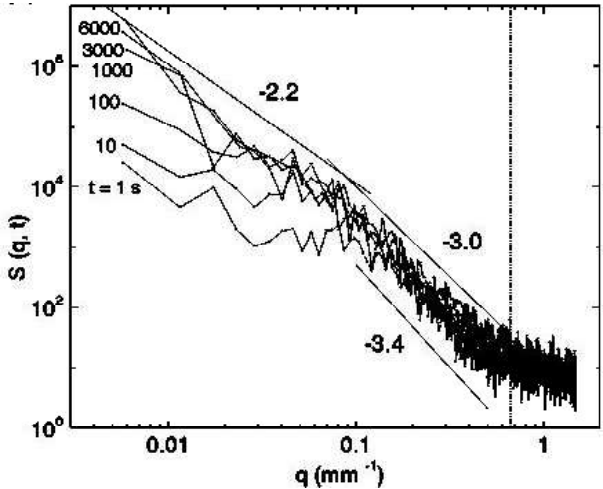}
\caption{Results from Ref.~\cite{Soriano_2002a} for the power-spectra
of the interfaces at different times approaching a steady state under
point disorder as in Fig.~\ref{sor01b}(a).}
\label{sor01new}
\end{figure}

Hele-Shaw cells with a pattern of square islands (square disorder)
is similar to using cells with randomly packed beads. Intuitively 
one may perhaps expect a better qualitative picture of the porosity
(gap) fluctuations than in such earlier setups. For capillary 
numbers in the range $1.33 < C_{\rm a} <17.0$, the cross-over phenomenon
already observed by Horv\'ath et al.\ \cite{Horvath_1991a}, together
with exponent $\beta \sim 0.5$ was also present. The structure factor
of Fig.\ \ref{sor01new} clearly shows a short-range exponent
$\chi_{\rm SR} \sim 1.0$. At large scales, it appears that the roughness
exponent changes continuously as a function of velocity, but this
behaviour should be contrasted to the theoretical results of Section
\ref{Theory}. From the phase-field simulations (see
Figs.~\ref{cap_mob_disorder} and \ref{imbi_both}), it is known
that roughening due to capillary forces gives rise to a large
roughness exponent but only up to a length scale $\xi_{\times}$. At
larger length scales, mobility disorder dominates, but there can be a
very long crossover between the two regimes, particularly at large
capillary numbers (i.e., for a large interface velocity). The
experimental data of Fig.~\ref{sor01new} are then more indicative of
a well-defined long-range roughness exponent $\chi_{\rm LR} \sim 1.0$ and
of a cross-over behaviour, than a true velocity-dependent roughness
exponent.

\begin{figure}
\includegraphics[width=8cm]{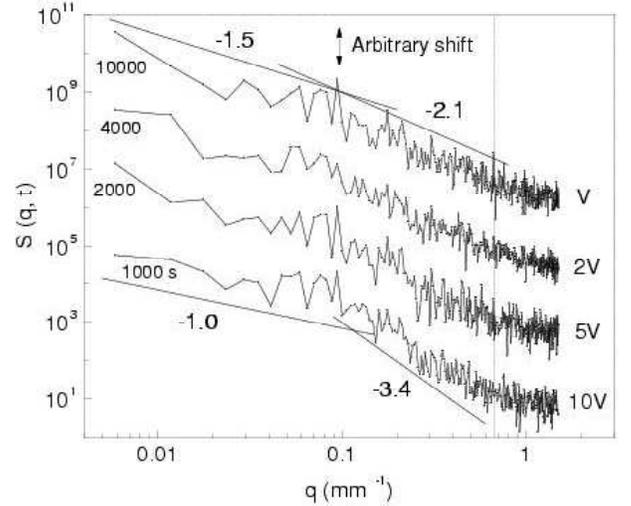}
\caption{Results from \cite{Soriano_2002a} for the power-spectra
of the interfaces at four different imposed velocities and columnar
disorder. The curves are shifted for clarity. The vertical line gives the
scale corresponding to the typical width of the columnar tracks, and
the other lines are fits to determine roughness exponents. Notice
how a large velocity seems to imply in particular a larger 
short-range roughness exponent (recall that the structure
factor decays as $k^{-(1+2\chi)}$).
}
\label{sor01c}
\end{figure}

\begin{figure}
\includegraphics[width=8cm]{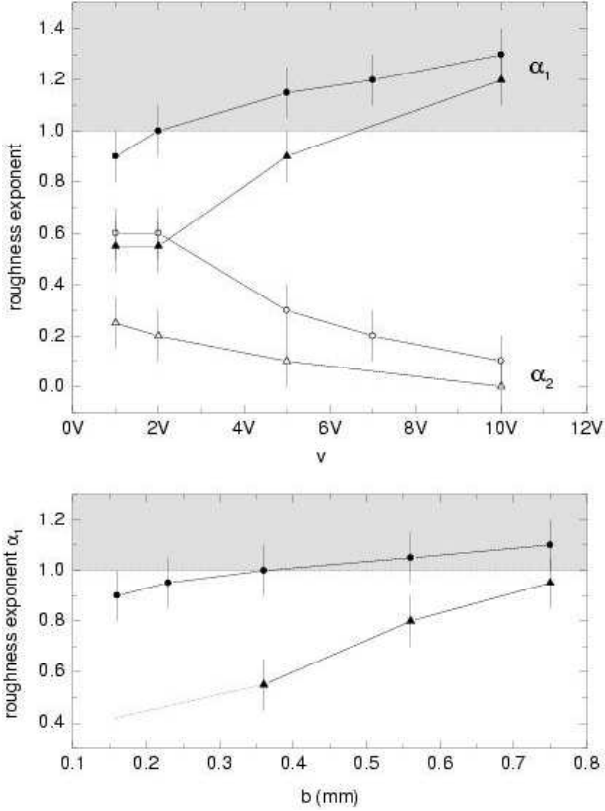}
\caption{Results from \cite{Soriano_2002a} on the roughness
exponents, where the type of disorder is varied (circles: point-like,
triangles: columnar). The top one shows variation with
imposed velocity, and the bottom one with the gap between the
plates between which the advancing liquid is confined.
Solid and open markers correspond to short and long range
exponents, respectively. For $\chi>1$ (or $\alpha$ in the notation
of the figure, from \cite{Soriano_2002a}) denoting ``super-roughness''
the region is shaded in grey.
}
\label{sor01d}
\end{figure}

\begin{figure}
\includegraphics[width=8cm]{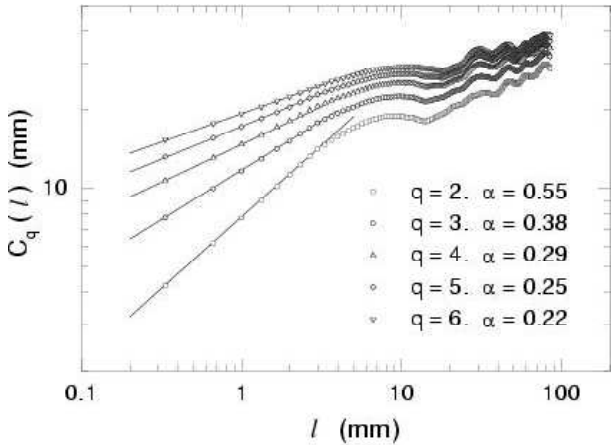}
\caption{$q$-th order two-point correlation functions of the
interfaces in the experiment of Soriano et al.\ show at saturation
some type of effective multiscaling as their logarithmic slope (taken
as $\chi_q$) is clearly $q$-dependent. Here the disorder used was of
columnar type \cite{Soriano_2002b}. The long-range scaling is not
reliable due to poor statistics. 
}
\label{sor01e}
\end{figure}

\begin{figure}
\includegraphics[width=8cm]{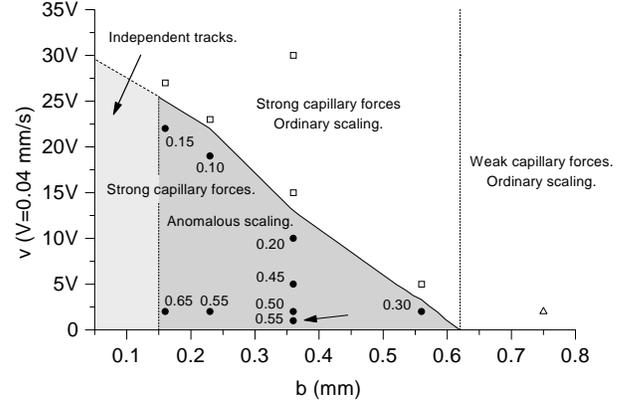}
\caption{An experimental phase diagram for anomalous scaling
in the setup of Soriano et al.\ \cite{Soriano_2002b}. The symbols 
indicate the points explored. The numbers next to the symbols
tell the difference between the ``global'' and ``local''
roughness exponents, $\chi$ and $\chi_{\rm loc}$. The effect
is strongest for slow interface progression and small values
of the gap --- which renders the disorder strong and would
perhaps indicate a simultaneous decrease of the effective
interface tension.
}
\label{sor01g}
\end{figure}

For columnar disorder, analysed theoretically in Section
\ref{constantflow}, the papers brought up the idea of a phase diagram,
which depends on the strength of the capillary forces (gap in the cell
in the experiments being the control parameter) and the forced flow
velocity. If the capillary forces are of major importance, {\em
anomalous scaling} is found \cite{Lopez_1997,Soriano_2002b}. This
results from another length-scale, the ``lag'' between neighbouring
columns due to the difference in local mobility. The usual signs of
anomalous behaviour could be extracted: correlation functions for the
spatial fluctuation would exhibit different scaling exponents $\chi_q$
depending on the order (or moment) of the correlations measured, the
local and global exponents for the roughness would differ, and finally
the global width would in addition exhibit ``super-rough'' behaviour,
$\chi>1$ \cite{Lopez_1997}. The scaling of the structure factor of the
interfaces also pointed out the possibility of several $\chi$ values,
as shown in Fig.~\ref{sor01c} for the experiments:
for short scales the roughness exponent would exhibit such anomalous,
super-rough scaling ($\chi \approx 1.2$) while on more macroscopic
scales the same cross-over behaviour as for square disorder would be
observed. Again, this may be a crossover effect between capillary and
mobility disorder.
For the other exponents $\beta \approx 0.5$ and $z \approx 2.0$ were
obtained (see Fig.~\ref{sor01d}).
Again, higher order correlation functions show different
effective exponents, a possible indicator for multi-scaling.
Fig.~\ref{sor01e} gives evidence for this, for interfaces with
columnar disorder and at saturation.
A distinction between the various phases for multi-scaling
is illustrated in Fig.~\ref{sor01g}.
Such signs are
similar to what is seen in the phase-field description of interface
dynamics (though the actual exponent values may not agree), and
indicates that the presence of non-local dynamics invalidates simple
scaling.

\subsubsection{Avalanches and depinning}
\label{SectAval}
\begin{figure}
\includegraphics[width=8cm]{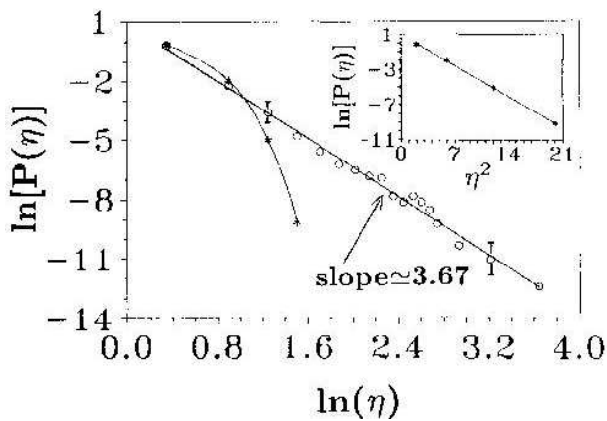}
\caption{The effective noise distribution
$P(\eta)$, or equivalently the distribution of
the local interface velocities in a 2d forced
fluid flow imbibition (open circles). Clearly, there is a wide range
of local rates of advancement. For comparison
the data include the result from the RSOS model (stars),
that belongs to the KPZ universality class
(see text, figure from \cite{Horvath_1991a}).
}
\label{horvath2}
\end{figure}

All the early works in particular in Hele-Shaw cells pointed out more
or less directly the difficulty in observing the slow, avalanche-like
motion for small $C_{\rm a}$, due to the lack of self-averaging. Further
experiments then concentrated on the study of avalanches and noise
properties. Horv\'ath et al.\ studied the {\em noise distribution}
$P(\eta(x,t))$ \cite{Horvath_1991b}.
For {\em local interface equations}, this can be obtained directly
from the local increments of the interface height ($v(x,t) \sim
\eta(x,t)$), as can easily be seen from, e.g., the slope-averaged KPZ
equation. This quantity, turned out to have a power-law tail which is
in qualitative agreement with the fact that no standard exponents were
observed either. Fig.~\ref{horvath2} shows the
power-law-like shape of $P(\eta)$).

\begin{figure}
\includegraphics[width=8cm]{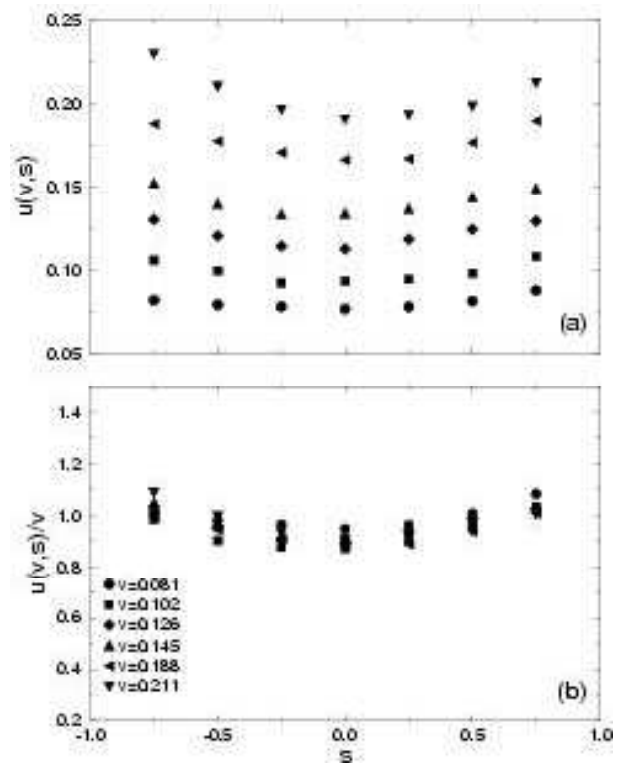}
\caption{a) Dependence of the local velocity $u(v,s)$ on the local
slope $s$. The average velocity is given in bead diameters/second. 
In b) the same data is represented after rescaling
by the velocities \cite{Albert_1998}. The parabolic shape of
$u(v,s)$ would imply the presence of a KPZ-like nonlinearity in the
effective dynamics.}
\label{alb98}
\end{figure}

In later papers by Dougherty and co-workers, the
avalanche-like motion of the interface was studied in more detail
\cite{Dougherty_1998,Albert_1998}
by looking at the average interface velocity as a function of the
local average slope $| \nabla h |$ (inside a measurement window). It
would be first assumed that, for a fixed average interface velocity,
the function $\langle v \rangle$ is parabolic in the interface
gradient. This is a consequence of the non-linear, KPZ-like terms in
the interface equation which imply $\langle v (\Delta h)\rangle
\propto \lambda (\Delta h)^2$ in the ensemble-averaged sense. As the
velocity of the interface approaches zero, depinning models
predict that the quantity $\langle v (\Delta h) \rangle $ can either
vanish, which indicates the isotropic depinning universality class
(QEW-like), or not, meaning that the depinning is anisotropic
(QKPZ-like) if one considers local interface equations. The parabolic
dependence of the velocity on the slope was indeed obtained
experimentally, and the results further indicated that the depinning
process was isotropic \cite{Albert_1998}, consistent with the value
$\chi = 1.25$ obtained from the phase field simulations
(see Fig.~\ref{alb98}).

\begin{figure}
\includegraphics[width=8cm]{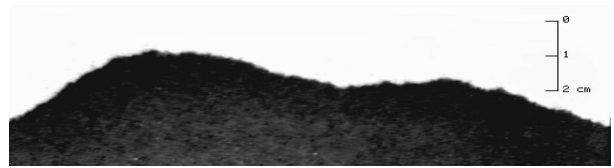}
\caption{Typical interface in a forced fluid flow 
experiment by Dougherty and Carle \cite{Dougherty_1998}.
The velocity was 13.7 $\mu m/s$, which should be compared
with the scale in the figure.
}
\label{dough1}
\end{figure}

\begin{figure}
\includegraphics[width=8cm]{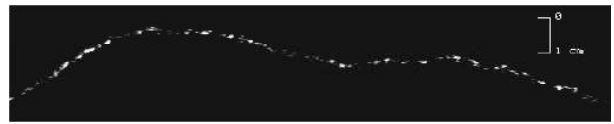}
\caption{The contrasted image (from \cite{Dougherty_1998})
shows where the interface has
moved during a 4 $s$ interval. As can be deduced, the motion
takes place rather uniformly in small patches,
without the presence of a wide variety of sizes of
``avalanches''. This might be taken to indicate
the presence of a correlation length, due to
the non-locality of the dynamics.
}
\label{dough3}
\end{figure}

\begin{figure}
\includegraphics[width=8cm]{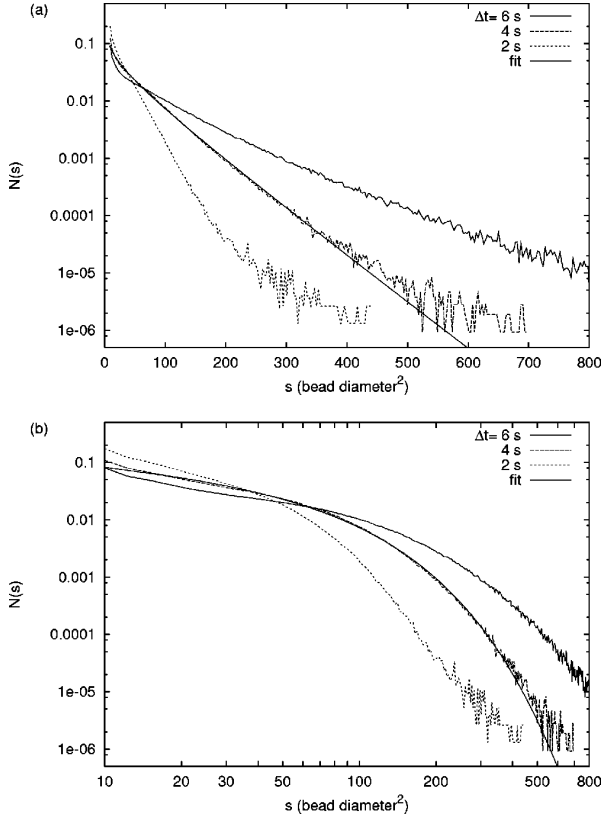}
\caption{The distribution of avalanche sizes
in the experiment of ref.~\cite{Dougherty_1998}.
The tail resembles more an exponential than a power-law.
}
\label{dough4}
\end{figure}

Meanwhile in
another paper, using the same setup, it was established that
the avalanche statistics are not quite akin to normal depinning
models: the area distribution was close to an exponential one
and not a power-law, which also was manifest in the fact that
the avalanches spread mostly laterally \cite{Dougherty_1998}.
One constructs from images of interfaces (Fig.~\ref{dough1})
typical avalanches by using time steps (Fig.~\ref{dough3}) to
recover the areas, locally, swept by the interface.
Clearly, the data in Fig.~\ref{dough4} are more in line with
an exponential distribution.

This is in clear distinction from critical behaviour found in
self-organised systems (e.g., SOC sandpiles or QEW). 
Criticality in these systems is measured in terms of the order
parameter (average velocity) going to zero. An interface 
is then expected to exhibit avalanche-like motion with a wide
distribution of scales {\em both} perpendicular and parallel to
the orientation of the interface. Another landmark would
be a power-law distribution of avalanche sizes that also
is a signature of the lack of an intrinsic scale.
These experimental results imply at least
indirectly that the non-locality of the interface dynamics, which
provides a definite length scale for the avalanche dynamics, 
was of importance, as one would expect in particular in the
limit $v \rightarrow 0$. They also highlight the difference
to other scenarios (theoretical and experimental) where the
details of microscopic physics upon coarse-graining yield 
a ``local'' (KPZ) interface equation \cite{Manneville_1996,Cuerno_2001,Myllys_2001}.

In this respect we also would like to point out the possibility of
measuring directly (in model media) quantities that relate to the
non-local properties as pressure fluctuations and saturation
\cite{vanderMarck_1997}. These could be perhaps {\em combined} with
various analysis of the interface dynamics in future experiments.


\subsection{Spontaneous imbibition}

So far, both the theoretical and experimental developments
in imbibition highlight the specific features of fluid invasion 
together with quenched
roughness, and it is important to outline what in particular one
should look for in the case of an interface that spontaneously
slows down. Some of the specific ideas are:

\begin{itemize}
\item
a length-scale, as $\xi_\times$, separating regimes of effectively
local and non-local interface dynamics,

\item multi-scaling, or anomalous scaling, absence of simple
scale invariance,

\item the interface velocity and its variations as the interface
approaches an equilibrium height,

\item scaling functions for fluctuations, including the properties
of pinned interfaces.
\end{itemize}

Next we overview both the older experiments and the more recent
ones, while trying to relate them to the background established.

\subsubsection{Pinned interfaces}

A set of statistical physics imbibition experiments were done by Buldyrev 
et al.\ \cite{Buldyrev_1992a,Barabasi_1992,Buldyrev_1992b} and Family
et al.\ \cite{Family_1992}. The first one was performed with a dye
solution in a vertical capillary rise setup. The rising front moved
from the reservoir, until eventually, the dye front stopped, due to
gravity and/or evaporation (no dynamical measurements were done), and
the roughness of the pinned interface was measured. The main
experimental finding was a roughness exponent $\chi = 0.63$,
consistent with the DPD/KPZ model. It should be noted, however, that
the length scale of the scaling behaviour was extremely small: For a
of total lateral extent $L = 40$ cm, $C(l) \sim l^{\chi}$ only for
length scales $l$ smaller than $l_{\rm max} \approx 1$ cm, with a
cross-over to a constant or to logarithmic behaviour at larger
$l$. This scaling region is only a few times larger than the average
fibre length in paper, so the result should be taken with some care. A
similar experiment was done in a three dimensional sponge-like
material \cite{Buldyrev_1992b}. The stopped interface yielded a
roughness exponent $\chi^{(2d)} \approx 0.5$, again consistent with
the $2d$ DPD model.

\begin{figure}
\includegraphics[width=8cm]{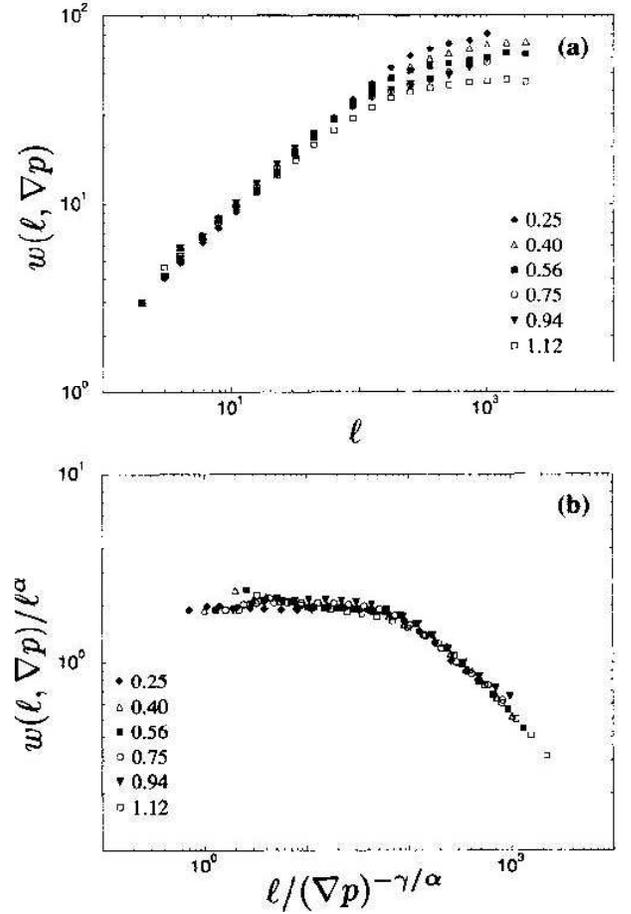}

\caption{The scaling of the interface width for a pinned
interface as a function of the evaporation
rate, related to the equilibrium height of the interface. Both
the ``raw data'' and the scaling used by Amaral et al.\ are 
depicted \cite{Amaral_1994}.
}
\label{amaral4}
\end{figure}

Amaral et al.\ \cite{Amaral_1994} studied the role of evaporation 
induced pinning, by controlling the pinning height via the
evaporation rate (presumably by modifying the humidity
during the experiment). The 
result is that the width of the pinned interface is related to the
pinning height $h_p$ (and thus to the evaporation strength) through a 
novel exponent $\gamma$, as $w_{sat} \sim h_p^\gamma$. The experimental
value was found to be $\gamma = 0.49$, which could also be related to
a modified version of the DPD model \cite{Amaral_1994} as shown
in Fig.~\ref{amaral4} together with the scaling ansatz used.
This result can also be related to the length scale $\xi_\times (h)$:
if the role of evaporation is to stop the
interface, with little or no influence
on the statistical fluctuations, then
$w(t) \sim \xi_\times^{\chi} (h_p) \sim h_p^{\chi/2}$ and
$\gamma = \chi/2$.

Kumar and Juma
\cite{Kumar_1996} also performed an experiment in presence of varying
evaporation conditions, with the claim that the roughness exponent depended
strongly on the evaporation rate, i.e., $\chi
= \chi (\epsilon)$. 

The last results in this respect are
those by the Mexican group \cite{Balankin_2000,Balankin_2003}.
Here, the pinned interfaces in paper-ink wetting experiments
were analysed. The main idea was to compare the front roughness
(after usual digitisation procedures from a digital picture)
between samples of different size. The pinned interface
width vs.\ sample width was stated \cite{Balankin_2000}
to follow an exponent of the order of 0.75 or 0.63 $\dots$ 0.64
depending on the paper grade (type of sample). It is worth
noting that the Washburn-behaviour was absent, i.e., the
measured interface heights vs.\ time increased much slower.
In such conditions it is no surprise that the interfaces
possess lots of overhangs. Such a result is, in any case,
qualitatively in accordance with the fact that Washburn-scaling
is not observed - this must imply, that the pinning height
can, as again was observed, vary with the sample size as well.
However the later interpretation \cite{Balankin_2003} that
the actual exponent value would relate to the correlations
in porosity and thus on the local ``force'' acting on the 
interface must be questioned. Unlike in many rocks etc.\
the pore space in paper is constrained by the finite thickness
(of the order of 0.1 mm) and a typical pore thus has a size
of a few tens of $\mu$ m$^3$, while the pore space is not
expected to have fractal properties on any sizeable scale.
There is no evidence that pores are correlated in paper.
(Note that ref.\ \cite{Provatas_1996} discusses {\em areal mass}
fluctuations and the correlations there in, not pores.)

These results highlight
the fact that the slow dynamics close to pinning are very
difficult to understand. Indeed, Delker, Pengra and Wong, and later
Lago and Araujo, have demonstrated in Hele-Shaw cells that
the approach to pinning can follow a power-law, similarly
to depinning transitions at the critical point in general 
\cite{Delker_1996,Lago_2001a,Lago_2001b}. This would follow
after a Washburn-like temporal regime has been observed.
With pinning allowed or included,
Washburnian dynamics would imply an exponential law,
also in contradiction with such experiments.

Recall that at the critical point of an absorbing state phase
transition, or a depinning transition, the velocity often follows a
power-law decay in time. The actual exponents of the experiments are
perhaps not very surprisingly close to any standard depinning
models. The attained values found by Delker and Wong for the decay
exponent of the interface velocity were in excess of unity, and
dependent on many factors. Indeed, similar power-law behaviour has been
reproduced for a contact line inside a  capillary tube. The Hele-Shaw
one may of course also simply demonstrate how local evaporation or
slow surface flows lead to creep-like motion of the interface
\cite{Cachile_1996}. Other experiments on spontaneous imbibition in
such systems have addressed saturation and cross-overs to fingering
flow \cite{Rodriguez_2001,Hayashi_2001}. These results do not seem to 
be in line with any of the standard local models, which does not seem
to be too surprising, either.

\subsubsection{Moving interfaces}
In analogy with the pinned interfaces, the results on
moving ones are also quite varied.
Family et al.\  performed an experiment in a horizontal
capillary setup with water, with the position of the air-water
interface recorded both temporally and spatially
\cite{Family_1992}. The main results were an average
interface progression $\bar{h} \sim t^{\delta}$,  with $\delta = 0.7$
(faster than the Washburn one)
and a self-affine interface described by a Family-Vicsek scaling
relation, and characterised by the exponents $\beta = 0.38$ and $\chi
= 0.76$. Concerning $h(t) \sim  t^{0.7}$ it may be so that
the details of the setup were at play: a reservoir was placed
underneath the paper used, to combat evaporation. This may
have caused condensation, or prewetting, thus
increasing the velocity of the front. It is of course no news
that water penetration in paper can be non--Washburn-like.
The spatial scaling regime was rather
small, for distances below $l_{max} \approx 2$ cm for a $40$ cm wide
sheet of paper.

\begin{figure}
\includegraphics[width=8cm]{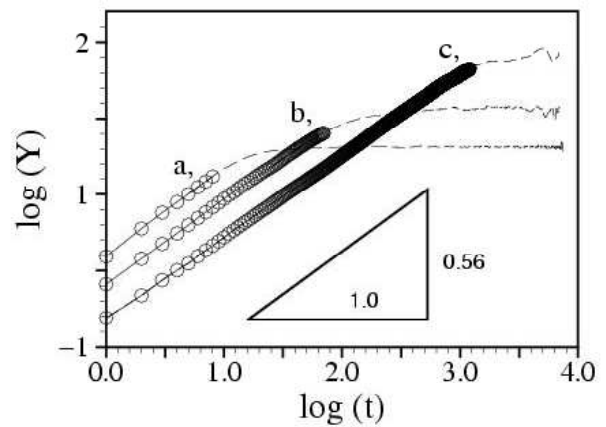}
\caption{The temporal two-point correlation function from the
stationary height experiment of Horv\a'ath and Stanley,
\cite{Horvath_1995}. The apparent roughening exponent is about 0.56
regardless of the speed ($V=4.03\times 10^{-2}$, $1.32\times 10^{-2}$
and $1.40\times 10^{-3}$cm/sec). Saturation takes place the sooner the
larger the velocity.}
\label{hor95}
\end{figure}

The temporal scaling of the interface was studied in detail by
Horv\'ath and Stanley \cite{Horvath_1995}. A paper sheet was moved
so as to keep the interface always at a fixed distance $h$ above a
reservoir. A power law behaviour for the time correlation
function $C_2(t)$ was established, with
$\sim t^{\beta}$, $\beta = 0.56$. The experimental result is
shown in Fig.~\ref{hor95}. The velocity $v$ at which the paper 
must be moved towards the reservoir
varied as $v \sim h^{-\Omega}$, $\Omega = 1.6$.
This implies that the interface propagated slower than $t^{1/2}$,
$h(t) \sim t^{1/(\Omega+1)}$, from $v = dh/dt$. No spatial scaling results
were reported. The value of $\beta$ was independent of $h$,
in contrast to the saturation time of $C_2 (t)$. The correlation function
was brought into a scaling picture,
\begin{equation}
C_2(t) \sim v^{-\Theta_L} f (t v^{(\Theta_t+\Theta_L)/\beta} ),
\label{horvath_scaling}
\end{equation}
where $f(y)$ is a scaling function such that $f(y) \sim y^{\beta}$ for
$y \ll 1$ and $f(y) \sim const$ for $y \gg 1$. The values of the
independent exponents were $\Theta_L=0.48$ and $\Theta_T=0.37$.
One can again try to rewrite this in terms of $\xi_\times$ by
relating the lateral length $\xi_\times$ to the velocity at $h$, so that
\begin{equation}
C^2_2 (\tau \rightarrow \infty) \sim \int_{1/\xi_\times (h)}^{1/a} 
\frac{dk}{k^{2\chi+1}} 
\end{equation}
which yields $C_2(\tau \! \rightarrow \! \infty) \sim h^{\chi/2} \sim
v^{-\chi/2}$. As long as $\xi_\times < L$, the correlation function
$C_2(t)$ is {\em independent} of the total width of the
system. Horv\'ath and Stanley used only paper of one, fixed, size.

A similar velocity result was obtained by Kwon et al.\ \cite{Kwon_1996}. 
A paper towel was set on an inclined glass plate, to follow
the capillary rise of a dye solution. This gave
$\bar{h} (t) \sim t^{0.37}$. Two scaling regimes
were present in the saturated width; on small length scales ($\leq 2$
cm), $\chi = 0.67$ while on larger length scales (up to $20$ cm) $\chi
\approx 0.2$. Within the simple Family-Vicsek picture they obtained
$\beta = 0.24$ on the short length-scale regime. It is conceivable
again that the cross-over is simply due to non-local effects.

The only experiment so far on horizontal front roughening was
performed by Zik et al.\ \cite{Zik_1997}. They obtained rough
interfaces only with what was called ``highly anisotropic'' paper,
with $\chi = 0.4$. For isotropic paper, the roughness was at best
logarithmic. It is remarkable, that the scaling for the anisotropic
paper was observed through a large range of length scales, not only up
to a few fibre lengths.

\begin{figure}
\includegraphics[width=8cm]{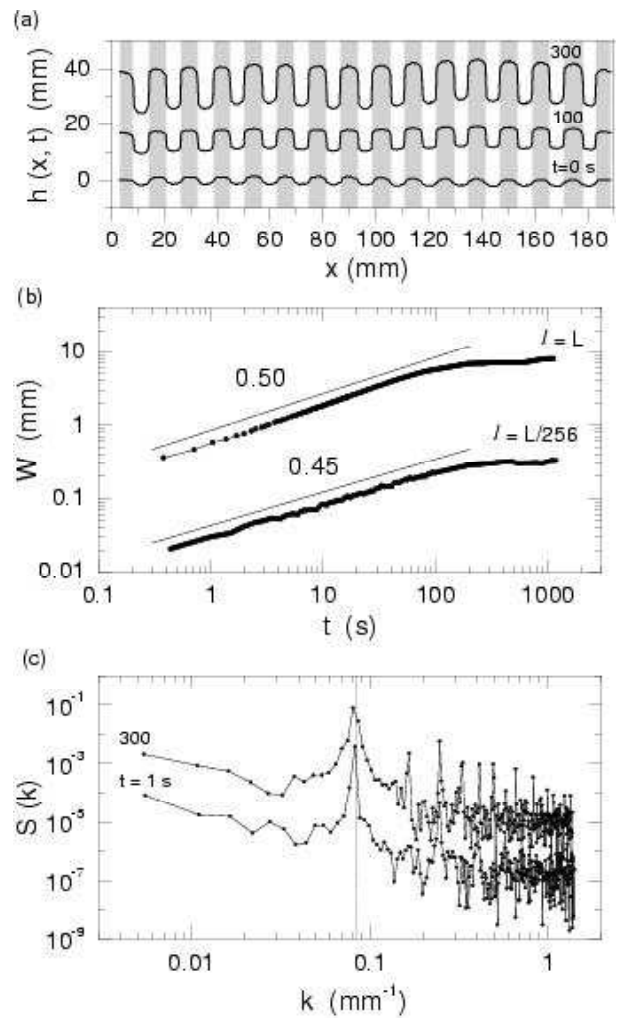}
\caption{Interfaces with a regular spacing of the tracks (or
disorder columns) 6 $mm$ wide. a) snapshots at different
times, b) local width $w(l,t)$ for two values of $l$, 
and the power spectrum at saturation. The vertical line indicates 
wavelength of the pattern. From \cite{Soriano_2002b}.
}
\label{sor01f}
\end{figure}

There are two recent, intriguing experiments on spontaneous
imbibition that have shown promise with respect to agreement
with theory. First, Soriano et al.\ also used the Hele-Shaw
setup to explore the effects of $v=0$ by preparing the interface
by finite-velocity transients. This means that the interface was first
forced, as in a fixed flow rate experiment, until a certain height,
and then let to relax to $v=0$ \cite{Soriano_2002b}. It was
established, with columnar disorder, that the local progression {\em
inside} the copper tracks (with preferential wetting properties)
followed Washburn-like dynamics. This setup and, possibly, experiments
on regular assemblies of tracks would certainly be useful to
investigate the cross-over between local and non-local dynamics
(again, $\xi_\times$) and the role of annealed (thermal) disorder by
repeating the test in the same fixed geometry, as illustrated
in Fig.~\ref{sor01f} where the effect becomes visible in the
local width $w(l,t)$ and the saturation power-spectrum.

\begin{figure}
\includegraphics[width=8cm]{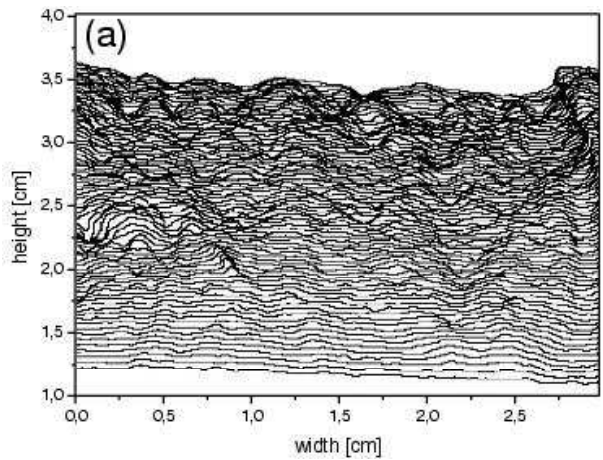}
\caption{A rising spontaneous imbibition front in an
experiment with two rough, parallel glass plates.
The snapshots are separated by 10 seconds \cite{Geromichalos_2002}.
}
\label{gero2a}
\end{figure}

\begin{figure}
\includegraphics[width=8cm]{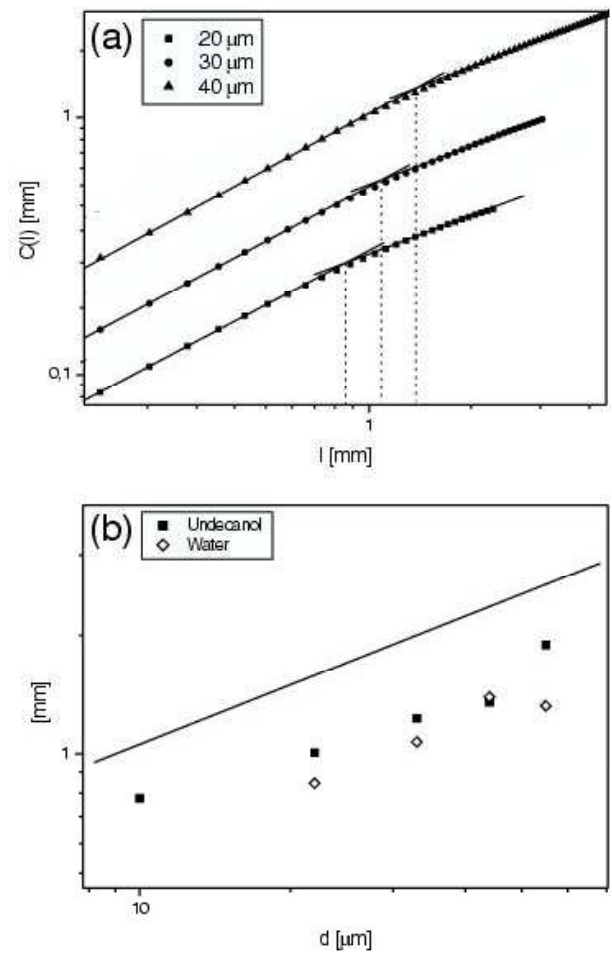}
\caption{In a) the spatial correlation functions of
the experiments of Geromichalos et al.\ \cite{Geromichalos_2002}
are demonstrated,
with varying separations of the two glass plates. The
presence of two scaling regimes is apparent, as is a
cross-over length-scale that decreases with the separation.
In b) the length-scale is shown, both
water and undecanol, as a function of the gap
width. The square-root scaling
of Eq.~(\ref{geroeq}) is roughly reproduced by the data.
}
\label{gero3}
\end{figure}

In a recent paper Geromichalos et al.\ explicitly demonstrate
the existence of a scale-length similar to $\xi_\times$ by studying
rising liquid fronts between two rough glass 
plates \cite{Geromichalos_2002}; we reproduce examples in
Fig. \ref{gero2a}. First, the
Washburn-scaling of the velocity was established, although with
standard prefactor-problems: the dynamic contact angle is 
unknown quantitatively. Measurements of the spatial
two-point correlation function indicated the presence of
two kinds of scaling, with $\chi_{\rm SR}\approx 0.8$ and
$\chi_{\rm LR} \approx 0.6$ (see Figure \ref{gero3}).
Maybe coincidentally, the values and the existence of the
cross-over are similar to what has been established in a
number of forced-flow experiments discussed above
\cite{Horvath_1991a,He_1992,Soriano_2002b}. 

\begin{figure}
\includegraphics[width=8cm]{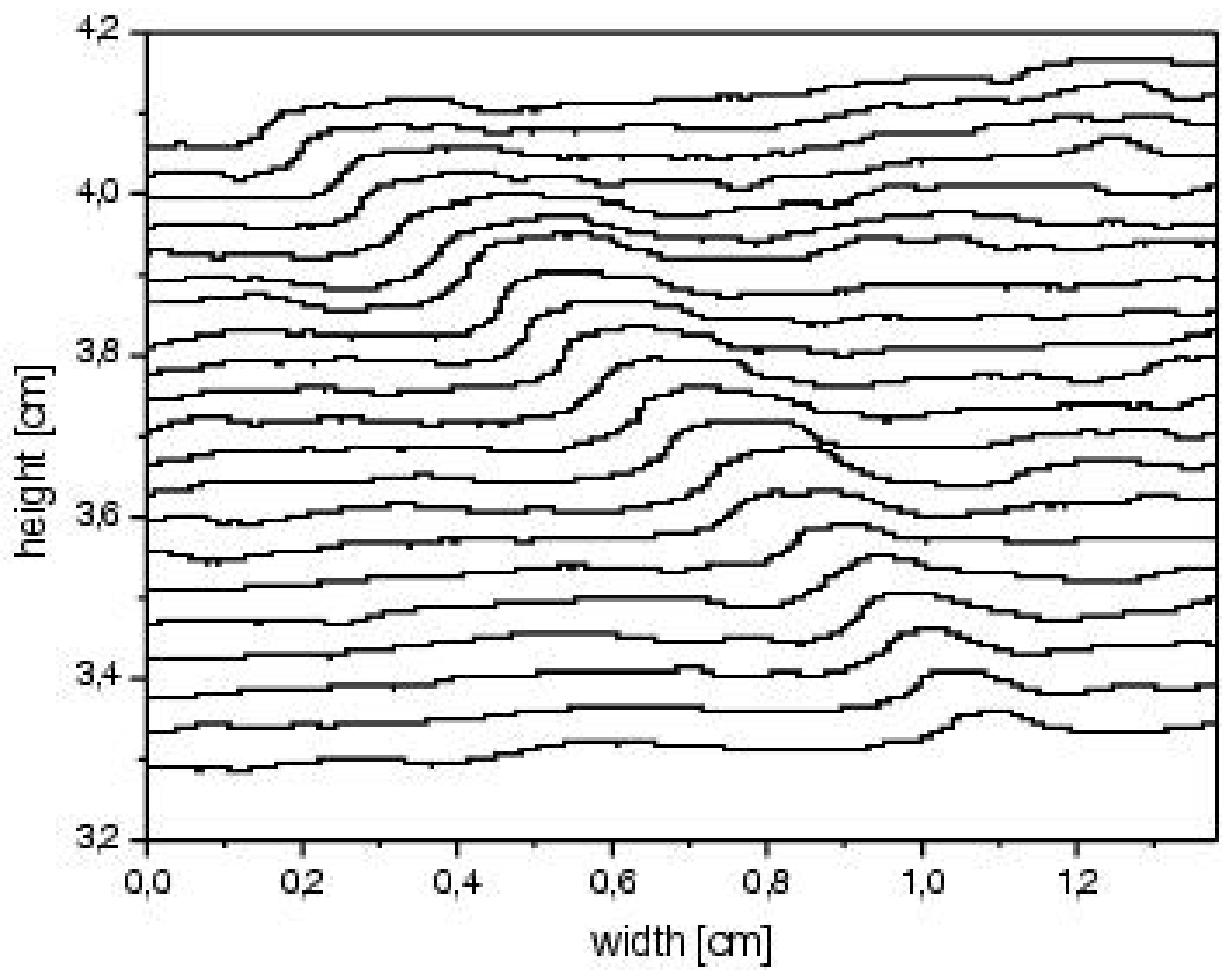}
\caption{A detailed look of some fronts (with undecanol),
showing the presence of coherent features or deterministically moving
``kinks'' in the front dynamics (\cite{Geromichalos_2002}).
}
\label{gero4}
\end{figure}

The cross-over scaling should follow, as a function of the
cell gap $d$ the law
\begin{equation}
\xi_\times \sim \sqrt{d}
\label{geroeq}
\end{equation}
which was in line with the results for dodecanol and water
as the rising liquid.
The exponents differ from most of the models except for
perhaps that of Ganesan and Brenner \cite{Ganesan_1998}, 
and also one should note that $\chi_{\rm LR}$
is numerically close to the quenched KPZ value of 0.63. Unfortunately the
published results do not consider the other exponents, as e.g.\ $\beta$
(unpublished data may imply, that $\beta \sim 0.5 \dots 0.6$
\cite{Geromichalos_2003})
nor the structure factor or higher order correlation functions
that could be used to study multi-scaling. The authors pointed
out that fronts sometimes had clearly visible dynamical structures or
``kinks'' one of which is shown in Figure \ref{gero4}.
This would be
certainly worth further study as would be the repeatability of the
measurements and correlations between runs using the same glass plates
several times.


\subsection{Imbibition, wicking, and liquid transport in living organisms}
\label{biosection}

Transport of water and solubles, and water entering parts of an organism are
essential processes of its life. We finish this Section by giving a (very)
short overview on recent studies of imbibition and water transport in plants,
focussing on those which by their methods or subject come close to the physics
aspects, as presented in this Review.

In a plant water flows through capillaries in the stem, roots, branches, and
leaves, in vessels forming a fractal structure. Perhaps not too surprising
from a Statistical Physics point of view, scaling behaviour can be observed in
this structure for various quantities, among them the vessel diameters in
relation to the biomass, the ratios of biomasses between roots and leaves,
between stem and leaves, etc. Some simple arguments indicate that the plant
structure, as it is found, optimises the supply of cells with respect to
resource use \cite{Enquist_2003}.

One has to be careful to apply concepts of physics to biological systems,
e.g., Darcy's law connecting pressure gradient to flux, because living
organisms may, and generally do {\em actively} influence all kinds of
processes --- so the exchange of water and solubles between a cell and its
surroundings. Interestingly, the term {\em imbibition} is used for the soaking
of water into seeds and grains, a mainly passive process and therefore in line
with the models and experiments presented above. Imbibition to seeds starts
germination through different mechanisms. Anybody is familiar with the
swelling of a wet grain, which breaks the outer hull and thus enables or eases
the seedling's growth. But water entering the grain also starts physiological
and developmental processes necessary for the early development of the embryo 
\cite{Acevedo_2002,Dominguez_2002,Leubner-Metzger_2002,Ouyang_2002}. It is
influenced among other parameters crucially by temperature and water supply
\cite{Booth_1999,Wuebker_2001}. Surface properties influence wettability and
imbibition \cite{Bet_2003}. Besides these agricultural issues knowledge about
grain imbibition has great importance in food processing and food storage
\cite{Gruwel_2002,Jansen_1994,Sacande_2001}.

Nuclear magnetic resonance allows for a direct measurement of water flow
inside the grain. Generally, the hydration of a seed is a multistage process. 
For example in soy beans the water concentration keeps increasing on the 
time scale of days with water entering different compartments one after 
another and reaching the embryonic axis relatively late \cite{Pietrzak_2002}. 
Similar results are found for cereals in detailed studies, so for barley
\cite{Gruwel_2001,McEntyre_1998}, oats \cite{Hou_1997}, maize
\cite{McDonald_1994}.

It is of considerable interest for agricultural applications to
understand the interaction of a seed and its surrounding soil during
imbibition. It would allow for optimal preparation of the soil
structure (grain size distribution) and its humidity and give an
estimate measure for successful germination. Bruckler has built a
model based on a Darcy ansatz for imbibition of maize grains and was
able to determine essential parameters such as, e.g., grain wall
permeability by direct measurements of single grains which were kept
in well defined conditions \cite{Bruckler_1983}. Subsequent tests in
seed beds yielded good correspondence of observations and model
predictions under variation of soil moisture and soil grain size, but
difficulties in estimating the role of temperature variation
\cite{Boiffin_1983}. This is in fact not too surprising, as
temperature has substantial influence on physiological properties. As
stated above, {\em active} processes in living organisms involve many
details, often too complicated to be captured by simple thermodynamics
or generally physical reasoning.



\section{Conclusions}
\label{conclusion}
\subsection{Overview}
In this Article we have reviewed various aspects of front roughening 
in imbibition. To us, with a statistical
physics viewpoint, it is striking 
that imbibition appears in many different fields,  
from the most abstract statistical physics of rough interfaces 
to detailed engineering studies in, e.g., paper fabrication or oil recovery.
Many years have passed since the original work of Washburn 
\cite{Washburn_1921}, but perhaps it can be stated that 
the general understanding of imbibition has advanced relatively little. 

This work has aimed at collecting the knowledge accumulated so far, 
grouping it into fields, 
and stating the most urgent and interesting open questions and advances
from the theorist's viewpoint. By far 
the most work has been done in numerous detailed technical applications, 
as the references to engineering literature on the subject that we have 
gathered should bear witness to. 

Imbibition has attracted a good deal of interest among statistical physicists,
in great part 
due to the presence of experiments on interface roughening that (still)
remain unexplained. As usual in this context, it is of particular interest 
to look for the simplest possible model that captures 
the essential ingredients of the problem. 
This however implies the presence of the usual dichotomy between theory
and experiment:
The simplified approaches of statistical mechanics predict scaling behaviour
but have difficulties to make 
quantitative predictions
and there 
is often no satisfactory connection between experiment and theory. This is 
true in part for imbibition and it remains
to be seen how quickly the distance between theory and experiment 
can be overcome. This also holds for the comparisons between
numerical simulation models that contain microscopic details - as outlined
in the Introduction - and more coarse-grained ones like the phase field 
approach.
We next summarise the main theoretical issues 
that have been resolved so
far and then point to a number of topics and issues
that remain open and deserve further attention.

Theorists, in evaluating experiments and model
simulations, have quite uncritically been looking for power laws in 
surface roughness without looking at the global picture. In our opinion, 
this has in many cases obfuscated 
the view for the essential features of imbibition. For example, it is 
of primary importance to observe and quantify 
the lateral length-scale $\xi_\times$ 
(c.f.\ Eqns.~(\ref{xi_cross_intro}) and (\ref{xi_cross})),
coming from the 
interplay of surface tension and pressure gradient in the liquid bulk. 
We note that this comes
out of theoretical considerations \cite{Dube_1999,Dube_2000a,Dube_2000b},
and that there is recent experimental proof for its existence
in the case of spontaneous imbibition \cite{Geromichalos_2002}, 
while forced fluid flow in Hele-Shaw cells yielded indications 
even earlier \cite{He_1992}.

Imbibition problems are of a broad general 
theoretical interest since
they present a field in which usual ``non-local'' interface models
(here referring to equations of type $\partial_t h(x,t) = \int_{x'} K(x',
x) \; h(x',t)$) fail to describe the problem properly. As noted in
the theoretical section one may then proceed systematically by
studying the origins of the effective noise terms, e.g., on the
basis of an assumption of Stokes' flow everywhere in the bulk.
This is not only useful for the particular case of roughening in
imbibition but will also advance the 
understanding of the permeability
properties, multi-phase flow and the role of
capillary effects; for disordered porous 
media these are all very challenging problems
(see among others 
\cite{Sahimi_1993,Hulin_1994,vanGenabeek_1996,Blunt_2001,Miller_1998,Dullien_1998}).

The fact that quantitative understanding of kinetic roughening should
also involve considering the solutions to these problems again point to the
complications present in imbibition phenomena. We might then conclude
that this merits a statistical physics approach, in which one eschews
detailed considerations in favour of basic symmetry and scaling
principles, but it is also important to realise that imbibition
represents an important meeting point for statistical physics and
actual applications to disordered media.

\subsection{Future issues for theorists}

The primary importance of imbition points to a number of directions
for theorists' future work. We now outline briefly three separate
topics, that stress quite different aspects. 

{\bf Interface roughness:} from the viewpoint of general interface 
dynamics it would of course be most interesting to develop models
that correctly describe the experimental results on interface
roughness properties. One of the fundamental questions remaining
open is of course the value (or possible values) of the roughness
exponent $\chi$ in a ``most clear-cut'' imbibition experiment.
In the theory section the idea of ``dynamical noise'' was brought
forward to explain some of the scaling regimes of the phase field-type
models corresponding to constant fluid flow setups. This kind of
reasoning is an analogy of the quenched-annealed cross-over in
ordinary interface depinning (QEW or QKPZ), which results  from the
fact that the noise becomes decorrelated due to the finite
velocity. This highlights the point that, as in local models, it is
the noise and the symmetries of the effective interface dynamics that
dictate the ``critical'' exponents - however with the proviso that the
scaling range might be limited due to the intrinsic nature of
imbibition.

In this sense, attempts to quantitatively describe a variable-gap Hele-Shaw 
cell give rise to some optimism that the quenched features of the 
noise could be 
described, even quantitatively, in some simple cases. 
Lest it be forgotten we should emphasise that the theoretical and 
numerical results presented in
Section~\ref{Theory} are inspired by simple experiments
and computational ease: no work exists so far on imbibition-like models
in, say, three dimensions. In the context of various numerical
(lattice Boltzmann or network) models this is a challenging task.
It is clear that the presence of an extra
dimension makes analytical work more challenging, and perhaps presents
new scenarios as in the case of anisotropic depinning \cite{Tang_1995}.
It is also rather obvious that this would be of experimental and
practical interest.

{\bf Washburn behaviour:} 
it is clear that in many instances of 
spontaneous imbibition Washburn-like scaling
is not observed, and other (velocity) laws are obtained.
This can ensue from the dynamic response of the medium (swelling),
from microscopic physics at the meniscus level (surfactant dynamics,
inertial flow in transients), and from the fact that the liquid
does not obey Stokes' flow conditions in general. Also, for e.g. very small
capillary numbers it is possible
that precursors (film flow) or the particular, peculiar pore structure
play a role. We believe that several of these effects 
merit separate studies, particularly in connection to their experimental  
effects on the roughness. The statistical physics viewpoint would
be that such microscopic phenomena give rise to yet another length-scale:
beyond this the effect of liquid conservation should dominate as usual.
Now the question is to understand the dependence on time and
on quantifiable measures of microscopic structure, of such scales.

The {\em coarse-grained} description of these effects can also be 
improved; the phase field models described earlier do not do yet a
very good job on the actual fluid dynamics level. Generalisations to
account for this are of course possible. The scaling away from the
micro-scale towards an effective large-scale description presents in
the context of imbibition several challenges. At least three can be
listed. In spontaneous dynamics, the time-scales change continuously
as the interface slows down (recall that even without an asymptotic
equilibrium $\partial_t h \sim 1/h$). Second, as noted the dynamics on
the pore level can be very slow (coming from surfactant diffusion or
induced dissipation at the contact line) or very fast
(Bosanquet-flow). Third, the correlations in the pore structure and
their size distribution can resist attempts to average, without even
mentioning time-dependent structural changes. In particular, also in
relation to the actual kinetic roughening properties, it is
challenging to describe the dynamics of imbibition when the interface
advancement becomes slow, and stick-slip motion typical of general
depinning/pinning transitions takes place. In the experimental Section
and in the Introduction
we pointed out evidences for deviations from Washburn-like approaches
to a pinning height at late stages. These can in some cases arise
from a change in the morphology of the interface from a compact
(but rough) front, perhaps, and indeed involve further scales.
Likewise, one should note that it should be evidently possible to construct
phase-field -style theories that would allow to handle the
smoothening and dispersion of the front e.g. in the presence
of initial wetting fluid saturations.

{\bf Complete stochastic description:} another promising avenue for
future work is to concentrate on issues other than those dealing with
traditional ``scaling exponents''. The dynamics of an interface in the
presence of external (thermal) and quenched noise is of course a
stochastic process and can be characterised in many ways. In the
statistical physics community, much attention has been recently paid
to concepts like persistence, or equivalently, first-return properties
\cite{Majumdar_1999,Krug_1997b,Kallabis_1999,Majumdar_2001,Merikoski_2003,Majumdar_2003}. 
For example, the local interface height $h(x,t)$ (after subtracting
the average) and the local velocity of the interface $v(x,t)$ are both
stochastic variables. As in the simple cases of Brownian motion or
ordinary random walks, it is then possible to explore the time-series
of $h$, or e.g. return-to-origin statistics. In usual kinetic
roughening models, which are most often in the steady-state, it
follows that this defines a {\em persistence} exponent, simply related
to the exponent $\beta$ of the two-point temporal correlation function
if the dynamics is Markovian. The probability distributions or {\em
scaling functions} of various statistical and fluctuating quantities,
such as the velocity distribution $P(v,L)$ (see
Fig.~\ref{velocity_distr}) or the roughness distribution in depinning
problems of the QEW type $P(w,F)$ \cite{Rosso_2003} can are also
similar alternatives descriptions. There has also been some recent
interest in  distributions such as ``$P(v)$'' as general signatures of
non-equilibrium behaviour and the possible universality therein
\cite{Bramwell_1998}.

It has to be stressed however that the
non-local dynamics of imbibition does not necessarily
maintain local scale-invariance, and thus many of
these questions have to be re-evaluated. 
Also, one probably cannot expect 
analytical solutions as in the case of KPZ interfaces with their
deep connections to random matrices \cite{Prahofer_2000}. 

The scaling exponents of ordinary local interface equations are also
manifest in many other properties. Examples are found in
non-equilibrium steady states, as in the SOC ensemble
\cite{Alava_2002}, and in case the system is driven, e.g., in an
oscillatory manner. In this case, one runs into concepts such as
hysteresis, aging, and the general effect of ``AC'' driving fields
\cite{Nattermann_2001,Glatz_2003}. In the context of imbibition it is
clear that analogous scenarios are easy to set up in experiments by,
e.g., varying the injection rate in a forced fluid flow test. 

\subsection{Experimental and practical implications}
To finish the discussion on possible directions for
further work we outline some suggestions for experimentalists.
Kinetic roughening and related questions should be studied in systems,
where the micro-structure is under control --- which is certainly not the
case for the ``classical'' imbibition of water into paper. The often-used
micro-channel networks are perhaps too idealised for such purposes. First
of all, the average flow and its influence on the lateral length scale 
$\xi_\times$ should be addressed.
Roughness exponents should be measured very carefully, not by a mere bona fide
fit to various moments of height differences $\langle | h(x+\xi,t)-h(x,t) |^q
\rangle^{1/q}$, but also by analysing the structure factor and the
probability distribution functions of these quantities. 

The predictions of the phase field model can be applied directly also
to the length-scales. For the $\xi_\times$, Eq.~(\ref{cross}) implies
that a knowledge of the most important fluid parameters and a
characterisation of the porous medium can be used to estimate its
range and effects on scaling. Consider e.g.\ a Hele-Shaw cell,
analogous to \cite{He_1992}. Using water and beads of size $ 0.4 \, 
\mbox{mm}$,
and varying the flow velocity between $10^{-6}$ to $10^{-3} \, \mbox{m/s}$
yields capillary numbers in the range 10$^{-5}$ to 10$^{-2}$. Assuming
now that the typical pore size is of the same order of magnitude as
the beads, this yields (with a surface tension of $\sim 72.5 \, 
\mbox{mJ/m}^2$) $\xi_\times$-values of about $1.5 \, \mbox{mm}$ for $C_a =
10^{-2}$. In other words, a rather small length-scale if compared to
typical kinetic roughening experiments. For ordinary paper the pores
are smaller by only one order of magnitude. This shows that the
roughness often seen in paper-based imbibition experiments may
directly be due to the structural disorder which is {\em discrete} on
the sub-millimetre scales (recall that a typical fibre length is up to
2 to 4 $mm$) and does not follow any real ``continuum description''.

A central question for the dynamics is interface propagation by
avalanches close to pinning. The complicated behaviour of simultaneous
avalanches, in a system with non-local dynamics, is not understood
well theoretically as already Fig.~\ref{velocity_distr} above
indicates. Possibly it may eventually become feasible to explore the
details of fluid dynamical fluctuations (pressure), as
e.g. Ref.~\cite{vanderMarck_1997} hints. The question of pore-scale
physics in this context and its relation to interface dynamics
would seem to merit attention \cite{Lenormand_1984}.

Last we recall that quantitatively by far the largest amount of work
has been done in detailed studies of phenomena relevant to all kinds
of applications. Our understanding is that this will remain so, since
many of the phenomena listed in the experimental Section III can be
examined more thoroughly: Pre-wetting layers along and inside pores,
the influence of surfactants and solubles, partial saturation, the
flow and transport in the bulk, and in particular behaviour with
non-Newtonian  liquids. Advances in various experimental
techniques (X-ray tomography for pore structure determination, NMR
with high enough temporal resolution for dynamical measurements)
should also provide with new results. As mentioned in the Introduction,
such methods are very close to getting to the level of imagining
imbibition fronts with sufficient accuracy, and comparisons can then
be made to the local pore structure including cross-correlations.

\subsection{Last outlook}
In this Review we have collected experimental and theoretical studies
of imbibition and tried to place them in a common framework of
understanding the phenomenon. This can not be gained without including
the lateral length-scale $\xi_\times$ appearing in the interface
fluctuations, which is a central point in our theory Section. Also the
scaling behaviour of interface fluctuations has to be studied
carefully. Different methods have to be compared for consistency, in
particular most valuable information can be obtained from the
structure factor $S(k,t)$ and the probability distribution of height
and velocity fluctuations.

To finish this work we mention a few directions that should
be interesting to follow or which may very well become more important
in the future than what one might conclude from our exposition of the
field here. First of all, we hope that ideas about the nature of
systems where disorder and global conservation laws combine can
benefit from the phase field lessons. One particular example could be
the description of vortex phases in superconductors and the
coarse-grained theories thereof 
\cite{Giamarchi_2002,Jackson_2000,Moreira_2002,Zapperi_2001,Miguel_2003,
Zapperi_2003}.

Second, we have not discussed the complications that ensue in the
presence of more than two phases (three phase imbibition) except
briefly in the context of surfactants. These systems are of obvious
interest in some oil recovery scenarios, but have so far been analysed
mainly numerically in the literature 
\cite{Fenwick_1998,Mani_1998,Hashemi_1999,Laroche_1999,Dijke_2003a,Dijke_2003b}. Here the pore-scale complications get even more manifold, and it is
obvious that coarse-graining these into any kind of continuum theory
is a challenge. Similar scenarios exist if non-Newtonian fluids 
are involved \cite{Kondic_1998,Tsakiroglou_2003}: the rate-dependence of the
liquid(s) will make itself manifest already inside single pores,
and if one for instance considers the slow-down in spontaneous
penetration, it is evident that the {\em average} flow of the
wetting liquid is hard to quantify. This area is however apt for
much future development given the natural role of surfactants
in many practical applications, and the fact that e.g. in petroleum
industry -related scenarios even the basic constituents - like crude -
have very non-ideal properties \cite{Morrow_2001}.

Another aspect not mentioned is the presence of simultaneous transport
processes (e.g. heat) during imbibition (see \cite{Medina_2003} for
the effect of thermal gradients). This can be complicated
further, if some of the phases involved undergo chemical
reactions (see \cite{DeWit_2004} and references therein). 
The fingering properties of such processes have been investigated in
Hele-Shaw cells, but imbibition -related conditions have not been
considered in general. One may for instance consider a porous medium, where 
a more viscous fluid displaces another, while at the 
front chemical reactions take place. These may involve
or even produce a surfactant, giving rise to a coupling
to the contact angle.

Third, we believe also that a sufficient understanding of imbibition
will prove to be of interest in a variety of technological
applications. Micro- and nano-machinery and chemically structured
surfaces should provide ample examples of situations of practical and
industrial interest \cite{Giordano_2001,Rauscher_2003},
as imbibition into carbon nanotubes \cite{Supple_2003}. In such
contexts the continuum description may fail due to the small scales
that necessitate an atomistic treatment; consider for instance the
physics of contact lines for a start. It should on the other hand be
of interest to be able control imbibition properties (including
dynamics) together with the permeability as hinted many times earlier,
here.

\noindent
{\bf Acknowledgements:}\\
MJA would like to thank the Centre of Excellence program of the
Academy of Finland for support and the SMC centre at La Sapienza,
Rome, for hospitality. MD would like to thank the Canada Research Chair
on Value Added Paper and Joe Aspler, Fran\,cois Drolet and Lyne Cormier
for interesting discussions. MR was supported by SFB 611 ({\em Singul\"are
Ph\"anomene in mathematischen Modellen}) of Deutsche
Forschungsgemeinschaft and thanks Helsinki University of Technology
for its hospitality. All authors wish to thank Ken R.\ Elder, Sami Majaniemi,
and Tapio Ala-Nissil\"a for collaboration on several works on
imbibition theory. Also Janoinen Lohi is gratefully acknowedged for a
long-standing and very fruitful collaboration.

\end{document}